 \documentclass[authoryear]{elsarticle}

\usepackage{amssymb,amsmath}
\usepackage{rotating}
\usepackage{lineno}
\usepackage{url}
\usepackage{longtable}
\usepackage{rotating}
\usepackage{geometry}
\journal{Earth-Science Reviews}

\begin{document}

\begin{frontmatter}

\title{Discussion on climate oscillations: CMIP5 general circulation models versus a semi-empirical harmonic model based on astronomical cycles}

\author{Nicola Scafetta $^{1,2}$}
 \address{$^{1}$ Active Cavity Radiometer Irradiance Monitor (ACRIM) Lab, Coronado, CA 92118, USA.}
 \address{$^{2}$ Duke University, Durham, NC 27708, USA.}

\begin{abstract}
Power spectra of global surface temperature (GST) records (available since 1850) reveal major periodicities at about 9.1, 10-11, 19-22 and 59-62 years. Equivalent  oscillations are  found in numerous multisecular paleoclimatic records.  The Coupled Model Intercomparison Project 5 (CMIP5) general circulation models (GCMs), to be used in the IPCC Fifth Assessment Report (AR5, 2013), are analyzed  and found not  able to reconstruct this  variability. In particular, from 2000 to 2013.5 a GST plateau is observed while the GCMs predicted a  warming rate of about 2 $^oC/century$.  In contrast,  the hypothesis that the climate is regulated by specific natural oscillations  more accurately fits the GST records at multiple time scales. For example, a quasi 60-year natural oscillation  simultaneously explains  the  1850-1880, 1910-1940 and 1970-2000 warming periods,   the 1880-1910 and 1940-1970 cooling periods and the post 2000 GST plateau.   This hypothesis  implies that about 50\% of the $\sim0.5$ $^oC$ global surface warming observed from 1970 to 2000 was due to natural oscillations of the climate system, not to anthropogenic forcing as modeled by the CMIP3 and CMIP5 GCMs. Consequently, the climate sensitivity to $CO_2$ doubling should be reduced by half, for example from the  2.0-4.5 $^oC$ range (as claimed by the IPCC, 2007) to  1.0-2.3 $^oC$ with a likely median of $\sim1.5$ $^oC$ instead of $\sim3.0$ $^oC$. Also  modern paleoclimatic temperature reconstructions showing a larger preindustrial variability than the hockey-stick shaped temperature reconstructions developed in early 2000 imply a weaker anthropogenic effect and a stronger solar contribution to climatic changes.   The observed natural oscillations could be driven by astronomical forcings. The $\sim9.1$ year oscillation appears to be a combination of  long soli-lunar tidal oscillations, while quasi 10-11, 20 and 60 year oscillations are typically found among major solar and heliospheric oscillations driven mostly by Jupiter and Saturn movements. Solar models based on heliospheric oscillations also predict quasi secular (e.g. $\sim115$ years) and millennial (e.g. $\sim983$ years) solar oscillations, which hindcast observed climate oscillations during the Holocene.  Herein I propose a semi-empirical climate model made of six specific astronomical oscillations as constructors of the natural climate variability spanning from the decadal to the millennial scales plus a 50\% attenuated radiative warming component deduced from the GCM mean simulation as a measure of  the anthropogenic plus volcano contribution to climatic changes. The semi-empirical model  reconstructs the 1850-2013 GST patterns significantly better than any CMIP5 GCM simulation. Under the same CMIP5 anthropogenic emission scenarios, the model projects a possible 2000-2100 average warming ranging from about 0.3 $^oC$ to 1.8 $^oC$. This range   is significantly below the original CMIP5 GCM ensemble mean  projections spanning from about 1 $^oC$ to 4 $^oC$.   Future research should investigate  space-climate coupling mechanisms in order to develop more advanced analytical and semi-empirical climate models. The HadCRUT3 and HadCRUT4, UAH MSU, RSS MSU, GISS and NCDC GST reconstructions and 162 CMIP5 GCM GST simulations are analyzed. \\ . \\
 Please cite this article as: Scafetta, N., (2013). Discussion on climate oscillations: CMIP5 general circulation models versus a semi-empirical harmonic model based on astronomical cycles. \emph{Earth-Science Reviews} 126, 321-357. \\ 
\url{http://dx.doi.org/10.1016/j.earscirev.2013.08.008}
\end{abstract}


\end{frontmatter}

\twocolumn

\section{Introduction}

Natural cyclical variability has been observed in geophysical systems at many time scales from a few hours to several hundred thousand and million years. The physical origin of many climatic oscillations  has often been found in astronomical mechanisms  \citep[e.g:][]{House}. Persistent quasi decadal, bidecadal, 60-years, 80-90 years, 115-years, 1000-years and other oscillations  have been found  in global and regional temperature records, in the Atlantic Multidecadal Oscillation (AMO), in North Atlantic Oscillation (NAO), in the Pacific Decadal Oscillation  (PDO), in global sea level rise indexes, monsoon records, Greenland temperatures  and  in many other climate proxy records covering centuries and millennia. Similar oscillations have been found also in luni-solar tidal cycles, solar and historical aurora long records  and in planetary oscillations   \citep[e.g.:][]{Agnihotri,Bond,Chylek,Chylek2011,Chylek2012,Cook,Currie,Davis,Gray2004,Hoyt,Humlum,Keelinga,
Keelinga,Klyashtorin,Knudsen,Kobashi,Jevrejeva,Mazzarella,Ogurtsov,Qian,Steinhilber,Schulz,
Sinha,Stockton,Yadava,Scafetta2010,Scafetta2012a,Scafetta2012c,Scafetta2013a,Scafetta2013,
Scafetta2013c,Scafetta2013al,ScafettaW2013a,ScafettaW2013b}.   These results suggest an astronomical origin of the observed decadal to millennial climatic oscillations.

Figure 1 shows  power spectral analyses \citep{Press} of all available global surface temperature (GST) records (HadCRUT4, HadCRUT3,  GISS and NCDC). These graphs  show prominent power spectral peaks at about 9.1, 10-11, 19-22 and 59-62 year periods since 1850. Similar frequencies are found  in major astronomical records, which are highlighted in Figure 1A with red boxes  \citep{Scafetta2010,Scafetta2012a,Scafetta2012b,Scafetta2012c,Scafetta2012d,Scafetta2013a,
Scafetta2013al,ScafettaW2013a}.

\begin{figure*}[!t]
\begin{center}
 \includegraphics[angle=0,width=0.75\textwidth]{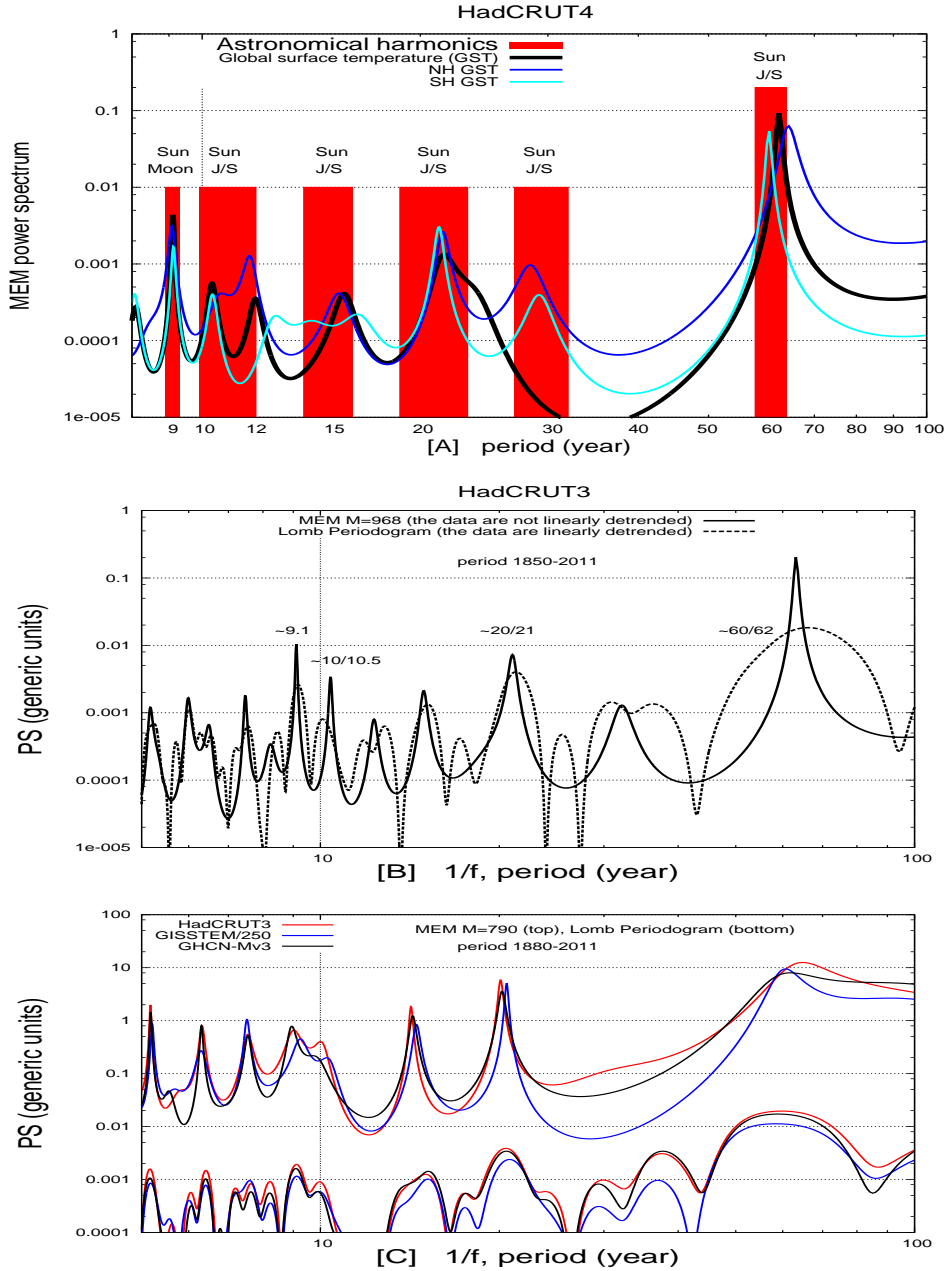}
\end{center}
 \caption{[A] Power spectra  of the HadCRUT4 GST (1850-2012) (black) and of the Northern Hemisphere and Southern Hemisphere  using the Maximum entropy Method (MEM); red boxes represents major astronomical oscillations associated to a decadal soli-lunar tidal cycle (about 9.1 years), and to the major heliospheric harmonics associated to Jupiter and Saturn gravitational and electromagnetic effects and to solar cycles (about 10-12, 15, 19-22, 59-63 years)     [B] MEM and periodogram for HadCRUT3 (1850-2011). [C] MEM and  periodogram for HadCRUT3, GISS and NCDC GSTs from 1880 to 2011. See \cite{Scafetta2010,Scafetta2012a,Scafetta2012b,Scafetta2012c,Scafetta2012d,Scafetta2013a} for details.  }
\end{figure*}

Theoretical climate models should ideally be able to predict the observed GST oscillations. In case of significant mismatches, solutions requiring minor adjustments of the models (for example, tweaking the forcings) may be proposed. However,  some of the basic physical assumptions of the models may also be flawed. In the latter case new mechanisms need to be identified in order to upgrade the models. Because of non-reducible uncertainties \citep{Curry}, alternative semi-empirical modeling strategies should also be developed and considered parallel to the analytic GCM methodology.

\cite{Scafetta2012b} tested all Coupled  Model Intercomparison Project 3 (CMIP3) GCMs used by the \cite{IPCC} and found that those models poorly reconstruct the  observed decadal and multidecadal GST oscillations  at about 9.1, 10-11, 20 and 60 year periods since 1850.   Herein the ability of the  GCMs of the Coupled Model Intercomparison Project 5 (CMIP5) to reproduce the historical global surface temperature patterns  since 1850 is tested as  well. As an alternate, I    conjecture that a significant component of the observed GST variations can be more efficiently modeled using  a set of astronomically induced harmonics of solar and lunar origin whose mechanisms are not included in the GCMs yet.

A GCM failure to reproduce  large natural multidecadal oscillations (periods, amplitudes and phases)  has  relevant theoretical implications. For example, Figure 9.5a and b published by the U. N. Intergovernmental Panel on Climate Change  \citep{IPCC} (\url{http://www.ipcc.ch/publications_and_data/ar4/wg1/en/figure-9-5.html}) show GCM simulations obtained with all adopted natural and anthropogenic forcings (figure 9.5a) and with natural (solar and volcano) forcings alone (figure 9.5b), respectively. The results depicted by this figure were used to claim  that more than 90\% of the warming observed since 1900 (about $0.8$ $^oC$) and practically 100\% of the warming observed since 1970 (about $0.5$ $^oC$) could only be explained by anthropogenic forcing (see figure in 9.5a). The reasoning was that  when only  solar and volcano forcings alone were used the GCMs predicted  a slight cooling since 1970 (see figure 9.5b). The theory emerging from these computer simulations is commonly known as the \textit{anthropogenic global warming theory} (AGWT). However,  if the warming observed since 1970 could be  produced  by a non-modeled multidecadal natural oscillation linked to  major ocean and/or astronomical-solar oscillations, then the AGWT should be questioned and/or  significantly revised.

\begin{figure*}[!t]
\begin{center}
 \includegraphics[angle=0,width=0.8\textwidth]{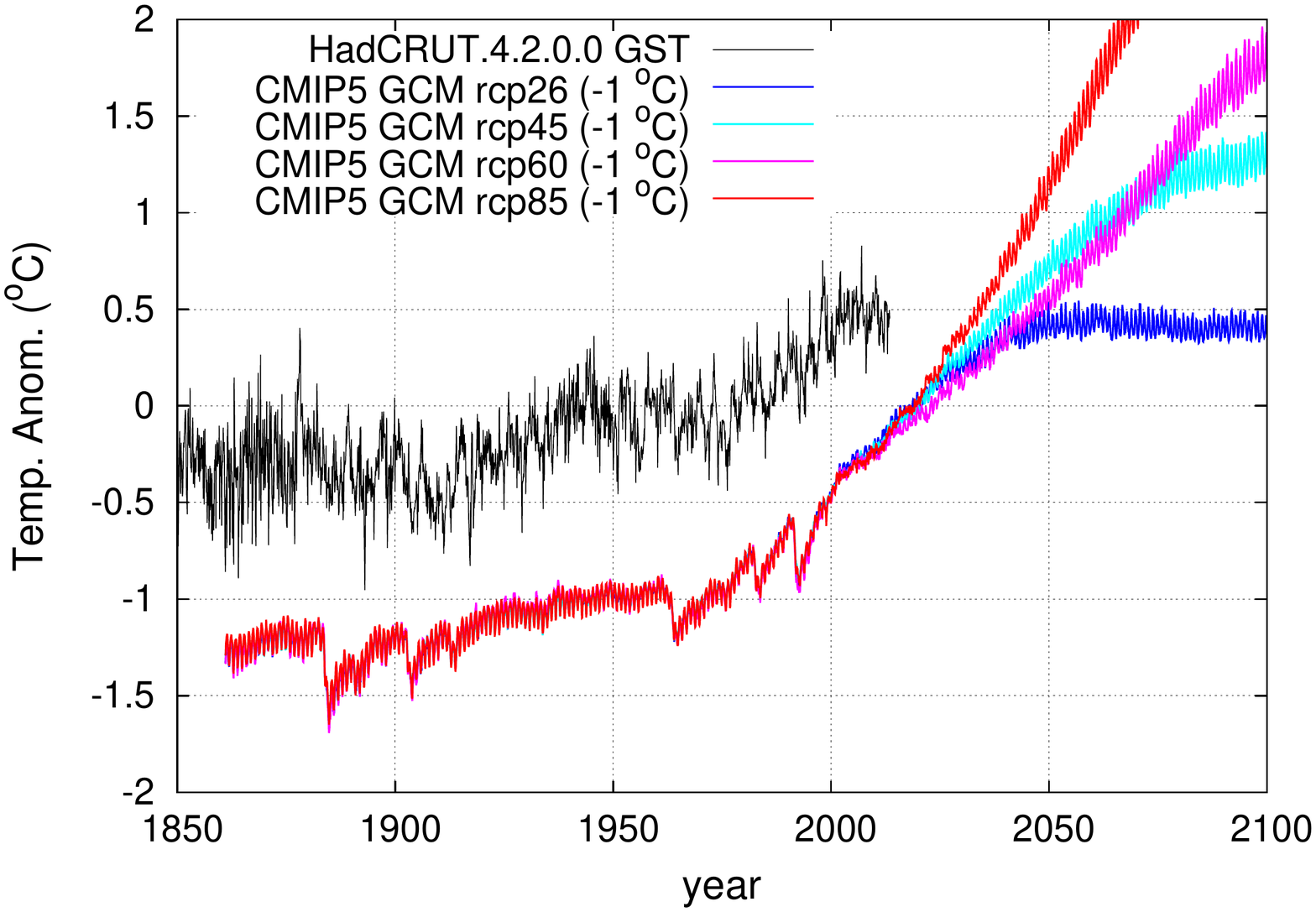}
\end{center}
 \caption{HadCRUT4 global surface temperature record (Jan/1850 - Jun/2013) (black). Four CMIP5 GCM ensemble means based on the historical forcings and four alternative 21st century projections.}
\end{figure*}

The AGWT may be erroneous because  of large  uncertainties in the forcings (in particular in the aerosol forcing, \url{http://www.ipcc.ch/publications_and_data/ar4/wg1/en/figure-spm-2.html}) and in the   equilibrium climate sensitivity to radiative forcing. The GCMs used by the \cite{IPCC} claim that a doubling of $CO_2$ atmospheric concentration  induces  a warming between 2 and 4.5 $^oC$ with a total range spanning   between 1 and 9 $^oC$  (see \cite{Forest} and \url{http://www.ipcc.ch/publications_and_data/ar4/wg1/en/box-10-2-figure-1.html}). The equilibrium climate sensitivity   derives from  the direct $CO_2$ greenhouse warming effect  plus the warming contribution  its  climatic feedbacks. For example, the direct $CO_2$ warming would be significantly enhanced by a water vapor feedback since water vapor is a greenhouse gas (GHG) as well. However, the strength and the nature of the various feedbacks is still quite uncertain  while the $CO_2$ greenhouse properties  are experimentally determined and are undisputed: without feedbacks a doubling of $CO_2$ amounts to a forcing of about 3.7 $W/m^2$ that should cause a warming of about $1~^oC$ \citep{Rahmstorf}. The strength of the feedbacks is  estimated in various ways and also calculated using the GCMs themselves. In the latter case, the calculated climate sensitivity value is a simple byproduct of the physical mechanisms and of the parameters currently implemented in the GCMs such as   those related to the parametrization of the cloud formation and water vapor feedbacks \citep{Hansen1988}. If the modeled feedback mechanisms are erroneous, then   the modeled climate sensitivity would be inaccurate.

Indeed, some authors  have pointed out that observational data  indicate that positive and negative climate feedbacks to $CO_2$ variations  compensate each other, leaving a net equilibrium climate sensitivity to $CO_2$ doubling ranging between 0.5 $^oC$ and 1.3 $^oC$ \citep{Lindzen2009,Lindzen2011,Spencer2010,Spencer2011}. \cite{Chylek2008} found a climate sensitivity between 1.3 $^oC$ and 2.3 $^oC$ due to doubling of atmospheric concentration of $CO_2$,  \cite{Ring} found a climate sensitivity between 1.5 $^oC$ to 2.0 $^oC$ and  \cite{Lewis}  found a range from 1.2 $^oC$ to 2.2 $^oC$ (median 1.6 $^oC$). The model proposed by \cite{Scafetta2010,Scafetta2012b,Scafetta2013a}, which herein will be reviewed and updated, implies  a climate sensitivity  from about 0.9 $^oC$ to 2.0 $^oC$ (median 1.35 $^oC$). Moreover, despite some controversy about the tropospheric records \citep{Thorne},   NOAA balloon measurements do not show the GCM-predicted  $CO_2$ induced hot-spot maximum trend in the tropical region  at an altitude of about 10 km \citep{Douglass,Singer}. \cite{Vonder} showed data that could severely question the existence of a  strong GCM global water vapor feedback to anthropogenic GHGs. These findings  suggest that current GCMs  severely overestimate the climatic effect of the anthropogenic GHG forcing.

As an alternative, I   propose a semi-empirical model composed of six major specific astronomically-deduced  oscillations spanning from the decadal to the millennial scales \citep{Scafetta2010,Scafetta2012c}.
Climatic oscillations can be to a first approximation empirically modeled in particular if they  have an astronomical physical origin  even if their specific microscopic physical mechanisms remain unknown. Astronomically based harmonic constituent models have been used to predict ocean tides since ancient times \citep{Ptolemy,Kepler,Ehret}. The oscillations that will be used are found among solar, lunar and planetary harmonics. Climatic effects due to volcano activity and anthropogenic emissions,  and the chaotic internal variability of the climate  are modeled to a first approximation by properly attenuating the GCM outputs.

It will be shown that the proposed semi-empirical model reconstructs and hindcasts the 1850-2013 climatic patterns significantly better than any CMIP5 GCM simulation and their ensemble mean, and may provide more reliable projections for the 21st century under the same emission scenarios. The finding would suggest that important astronomical forcings of the climate system and the climatic feedbacks to them are still missing and possibly  not yet known. The result reinforces \cite{Scafetta2010} where it was found that power spectra of global temperature records are  more coherent to solar/astronomical gravitational and electromagnetic  oscillations and  to soli-lunar tidal long-scale oscillations than to power spectra deduced from current GCM simulations. Therefore,  the author proposes that future research should incorporate additional astronomical mechanism of climate change.

\begin{figure*}[!t]
\begin{center}
 \includegraphics[angle=+90,width=0.86\textwidth]{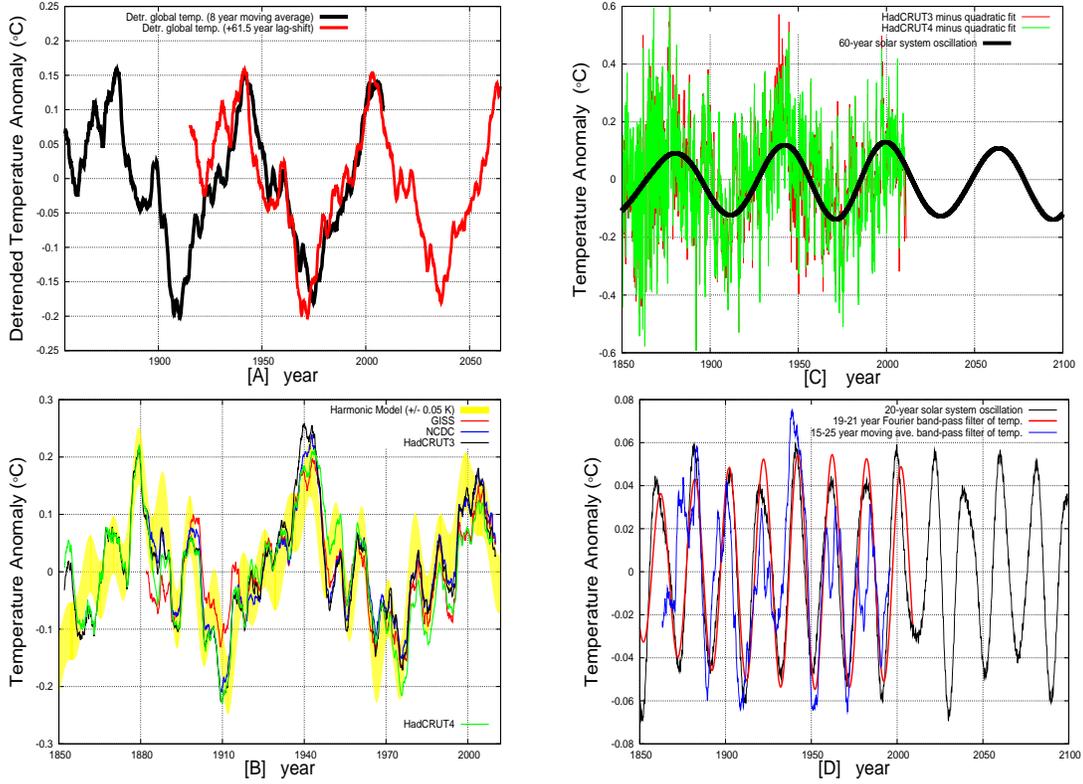}
\end{center}
 \caption{[A]  8-year moving average of the detrended
HadCRUT4 GST plotted against itself with a 61.5-year lag shift (red). The quadratic fitting trend
applied is $f(t)=0.0000298*(t-1850)^2-0.384$.   [B] Four GST records (HadCRUT3, HadCRUT4, GISS and NCDC) after being detrended of their upward trend, and smoothed with a 49-month moving average algorithm against the 4-frequency harmonic model (yellow area) proposed in \cite{Scafetta2010,Scafetta2012a,Scafetta2012c}.  [C] and [D] show the HadCRUT records detrended of their warming trend and band-pass filtered to highlight their quasi bi-decadal oscillation. The two black oscillating curves with periods of about 20 years and 60 years  are two major astronomical oscillations of the solar system induced by Jupiter and Saturn \citep{Scafetta2010}.  }
\end{figure*}

\section{Simple analysis of GST and the CMIP5 ensemble mean simulations}

All CMIP5 GCM simulations  studied were downloaded from Climate Explorer (\url{http://climexp.knmi.nl}, for details see also \url{http://cmip-pcmdi.llnl.gov/cmip5}). These records are 162 monthly resolved temperature-at-surface (tas) historical simulations from 48 models plus numerous  simulations for the 21st century. These are classified under four alternative representative concentration pathway (RCP) emission scenarios labeled as:  RCP 8.5 (rcp85), business-as-usual emission scenario; RCP 6.0 (rcp60), lower emission scenario; RCP 4.5 (rcp45), stabilization emission scenario; RCP 2.6 (rcp26), strong mitigation emission scenario. The RCP number indicates the rising radiative forcing pathway level (in $W/m^2$) from 2000 to 2100.

Figure 2 compares the HadCRUT4 GST (\url{http://www.cru.uea.ac.uk/}) \citep{Morice} from Jan/1850 to Jun/2013 against the CMIP5 model ensemble mean simulations under  the four alternative 21st century emission scenarios. The curves are plotted using a common 1900-2000 baseline, with the GCM curves  downshifted by 1 $^oC$  for visual clarity.

The GST warmed  about 0.80-0.85 $^oC$  since 1850. The GST also presents complex dynamical patterns  dominated by a quasi 60-year oscillation revolving around an upward trend:  1850-1880, 1910-1940 and 1970-2000 were warming periods, and 1880-1910 and 1940-1970 were cooling periods. Since 2000 the GST has been fairly steady.

\begin{table}[!t]
\centering{}%
\begin{tabular}{|c|c|c|}
\hline
period & GST-trend & GCM-trend\tabularnewline
\hline
\hline
1860-1880 & +1.01$\pm$0.24 & +0.50$\pm$0.06\tabularnewline
\hline
1880-1910 & -0.56$\pm$0.09 & +0.28$\pm$0.07\tabularnewline
\hline
1910-1940 & +1.33$\pm$0.08 & +0.72$\pm$0.03\tabularnewline
\hline
1940-1960 & -0.47$\pm$0.16 & +0.18$\pm$0.04\tabularnewline
\hline
1940-1970 & -0.26$\pm$0.09 & -0.33$\pm$0.04\tabularnewline
\hline
1970-2000 & +1.68$\pm$0.08 & +1.50$\pm$0.05\tabularnewline
\hline
2000-2013.5 & +0.35$\pm$0.22 & +1.88$\pm$0.04\tabularnewline
\hline
\end{tabular}\caption{Comparison of 30-year period trends in $^{o}C/century$ between the
HadCRUT4 GST and the CMIP5 GCM ensemble mean simulation depicted in
Figure 1.}
\end{table}

Figure 3  highlights the GST decadal and multidecadal patterns.
Figure 3A shows the HadCRUT4 GST smoothed and detrended of its quadratic polynomial fit and plotted against itself with a lag-shift of 61.5 years. The figure demonstrates that the GST modulation from 1880 to 1940 is very similar to the modulation from 1940 to 2000. This autocorrelation pattern  highlights the existence of a possible quasi 60-year oscillation. Additional modulations induced by other patterns may exist.  For example, the climate system may also be modulated by a 80-90 year oscillation that has been detected in long solar and climatic proxy records  \citep{Knudsen,Ogurtsov,ScafettaW2013a}.   However, a 80-90 year oscillation cannot be separated from a 60-year oscillation using the current GST records, since a 163-year long record is too short.

\begin{figure*}[!t]
\begin{center}
 \includegraphics[angle=0,width=0.8\textwidth]{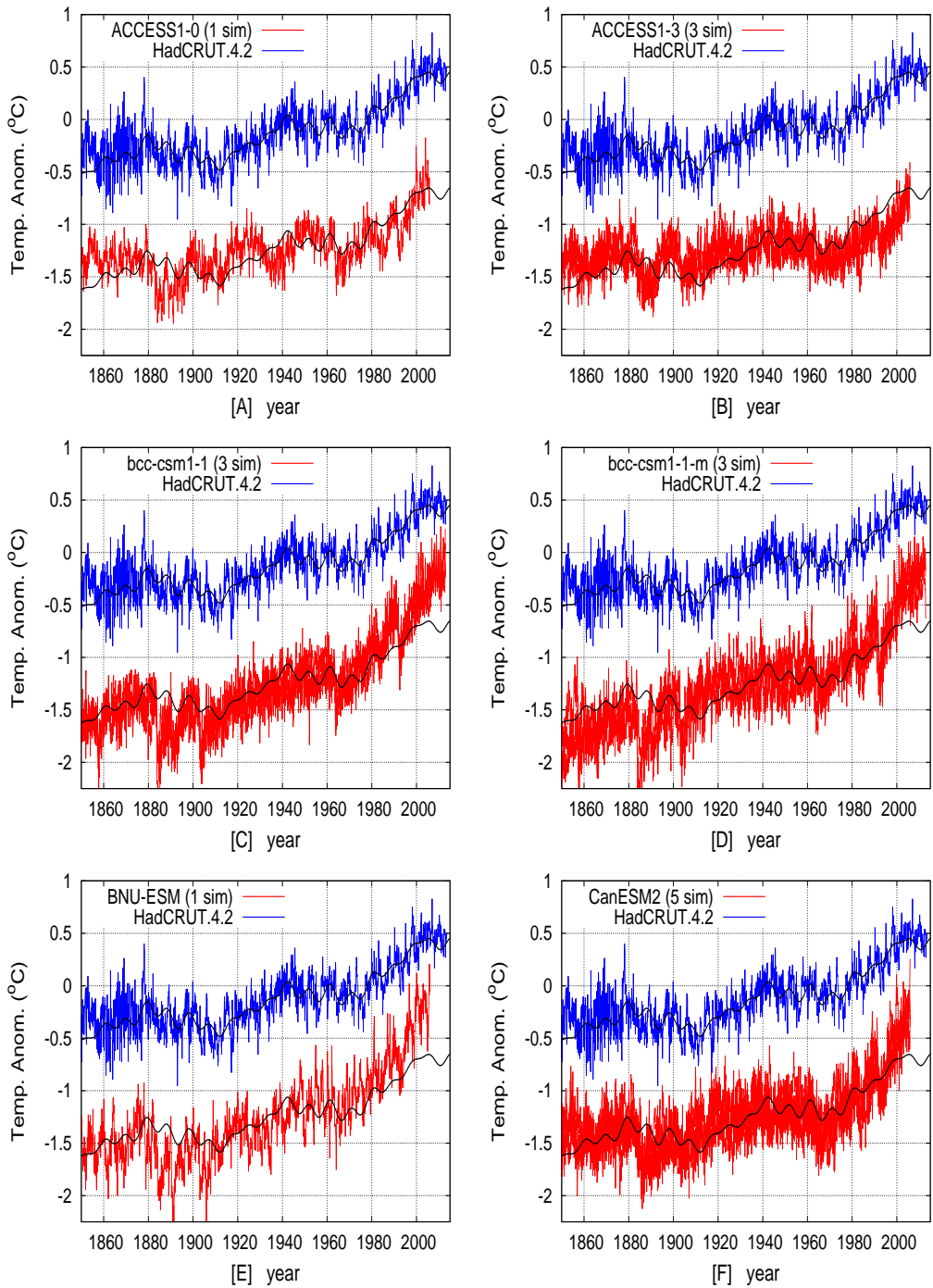}
\end{center}
 \caption{HadCRUT4 GST (blue) vs. CMIP5 GCM simulations (red). Eq. \ref{eqf} (black).  The number of individual available simulations simultaneously plotted is in parenthesis. }
\end{figure*}

Figure 3B highlights both the decadal and the multidecadal GST oscillations by showing four global surface temperature records (HadCRUT3, HadCRUT4, GISS and NCDC) after detrending the upward quadratic trend and smoothing the data  with a 49-month moving average algorithm. The harmonic model (yellow) proposed in \cite{Scafetta2010,Scafetta2012a,Scafetta2012b}  approximately reproduces the decadal and multidecadal patterns observed in the detrended GST curves.

Figure 3C shows the HadCRUT3 and HadCRUT4 records detrended of their warming trend as above. Figure 3D shows two band-pass filtered curves of HadCRUT3 (similar pattern is observed for the HadCRUT4 record) to highlight its quasi bi-decadal oscillations. The two black oscillating curves with periods of about 20 years and 60 years  are two major astronomical oscillations of the solar system. These can be easily observed, for example, in the speed of the Sun relative to the barycenter of the solar system and in the beats of the gravitational tides induced by Jupiter and Saturn \citep{Scafetta2010,Scafetta2012a,Scafetta2012c,Scafetta2013a,ScafettaW2013a}. The observed climatic oscillations appear synchronized with the two depicted astronomical oscillations. Similar results are obtained with the alternative GST records.

\begin{figure*}[!t]
\begin{center}
 \includegraphics[angle=0,width=0.8\textwidth]{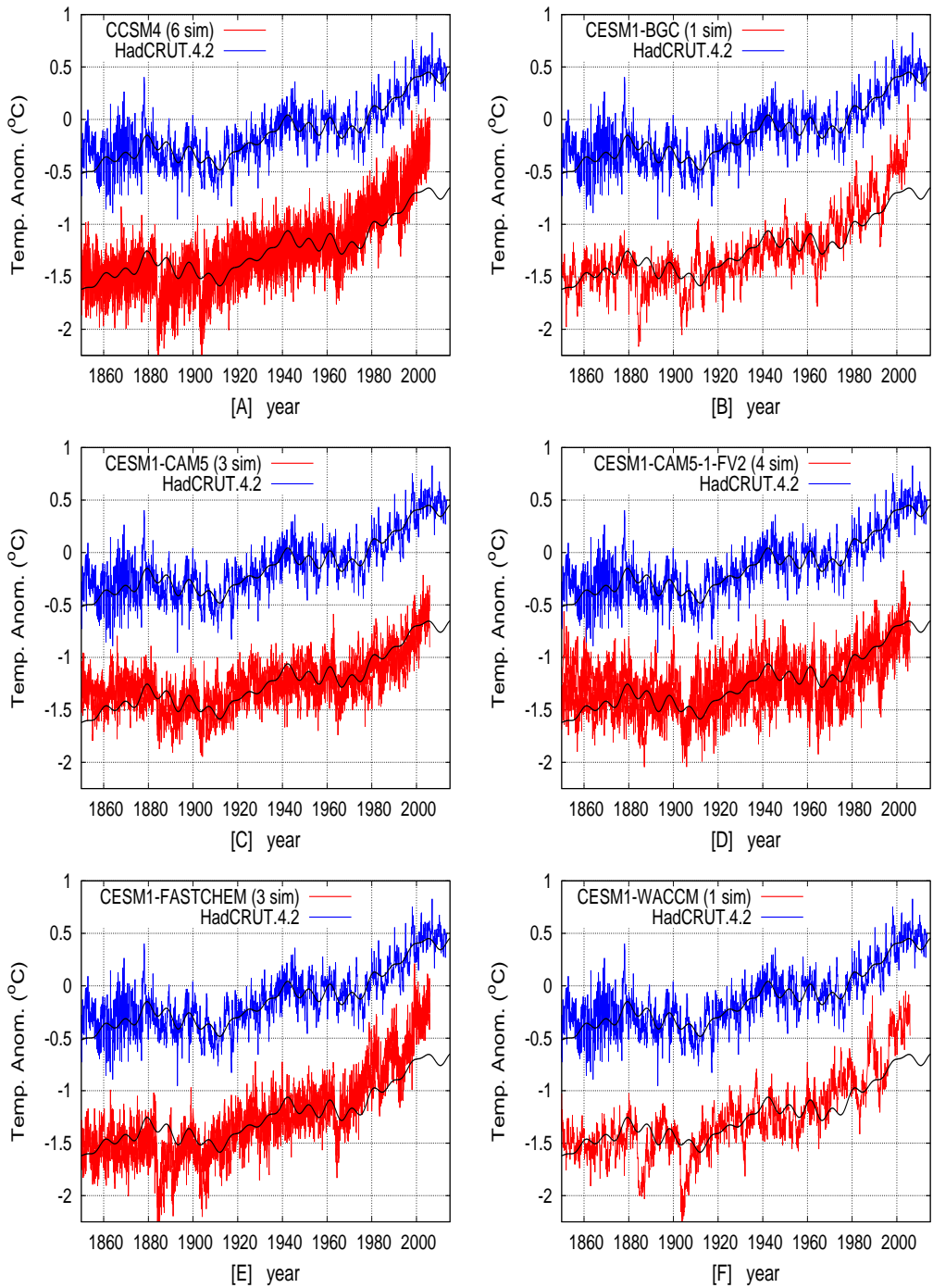}
\end{center}
 \caption{Continues:  HadCRUT4 GST (blue) vs. CMIP5 GCM simulations (red). Eq. \ref{eqf} (black). }
\end{figure*}

Table 1 compares the warming or cooling trends of the HadCRUT4 GST and of the  GCM ensemble mean simulations depicted in Figure 2 for the periods 1860-1880, 1880-1910, 1910-1940, 1940-1960, 1940-1970, 1970-2000 and 2000-2013.5: (1) from 1880 to 1910 the GST cooled while the GCMs predict a warming; (2) the 1910-1940 GST warming trend is almost twice than that predicted by the GCMs; (3) the GST cooled since the 1940s while the GCMs predict a warming from 1940 to  1960 which is interrupted by  volcano eruptions in the early 1960s; (4) from 1970 to 2000 there is an approximate agreement between the GST and the GCMs; (5) since 2000 a strong divergence between the modeled and observed temperatures is observed. Thus, only during the period 1970-2000 do the GCM simulations present  a mean warming trend somewhat compatible with that found in the GST. During the other intervals the GST trends  differ significantly from those predicted by the GCM ensemble mean simulations. In particular,  the 1910-1940 strong warming and the steady temperature since 2000 cannot be explained by anthropogenic emissions plus a small solar forcing effect, as assumed by the current CMIP3 and CMIP5 GCMs.

\begin{figure*}[!t]
\begin{center}
 \includegraphics[angle=0,width=0.8\textwidth]{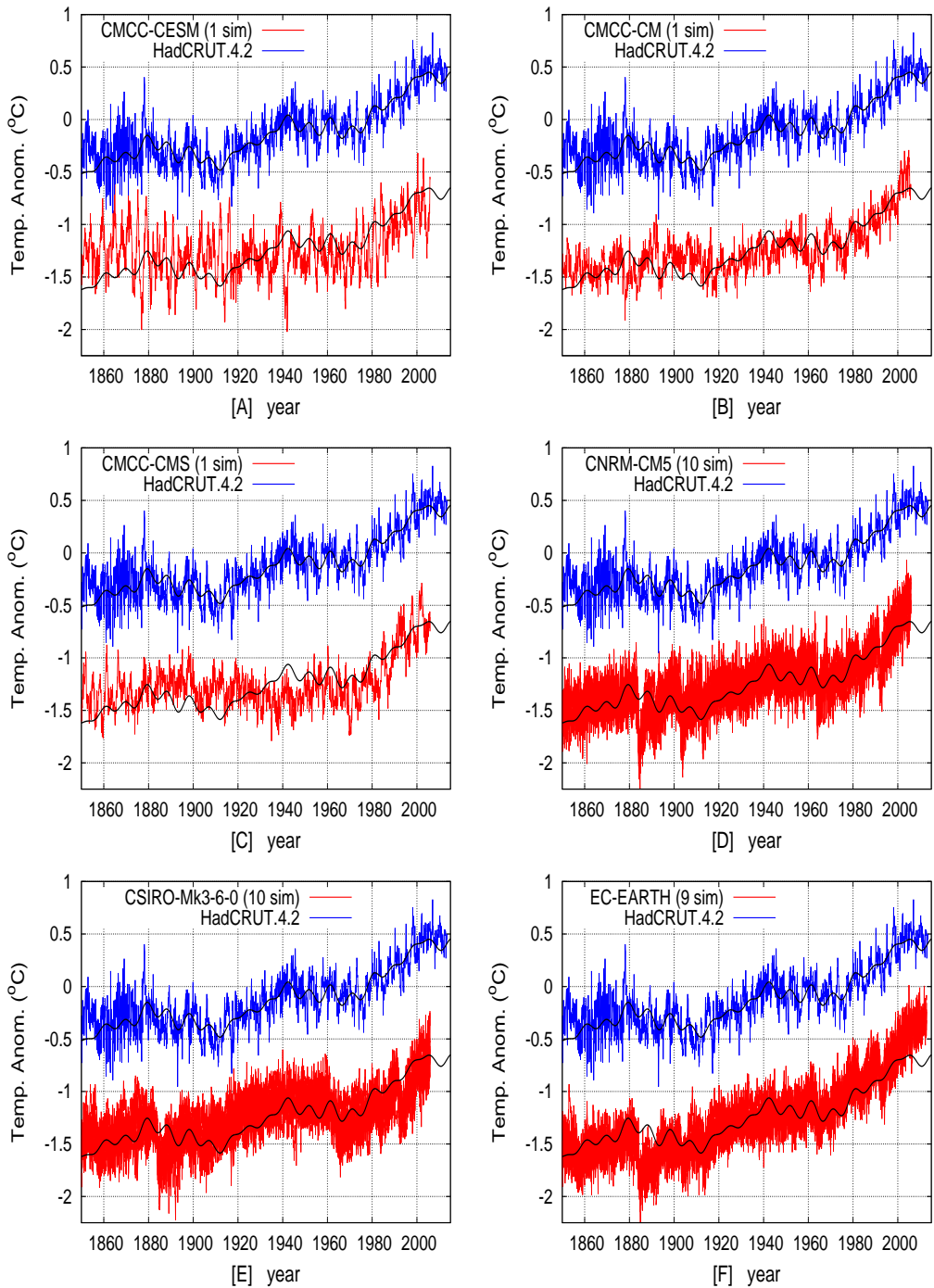}
\end{center}
 \caption{Continues:  HadCRUT4 GST (blue) vs. CMIP5 GCM simulations (red). Eq. \ref{eqf} (black).}
\end{figure*}

Also the GST volcano cooling spikes  are not as deep as those predicted by the synthetic record. For example, there is no observational evidence of the strong modeled volcano cooling associated with the eruption of the Krakatau in 1883. Also other volcano signatures appear to be overestimated by the GCMs and produce a spurious recurrence pattern at about 70-80 years \citep{Scafetta2012b}.

\begin{figure*}[!t]
\begin{center}
 \includegraphics[angle=0,width=0.8\textwidth]{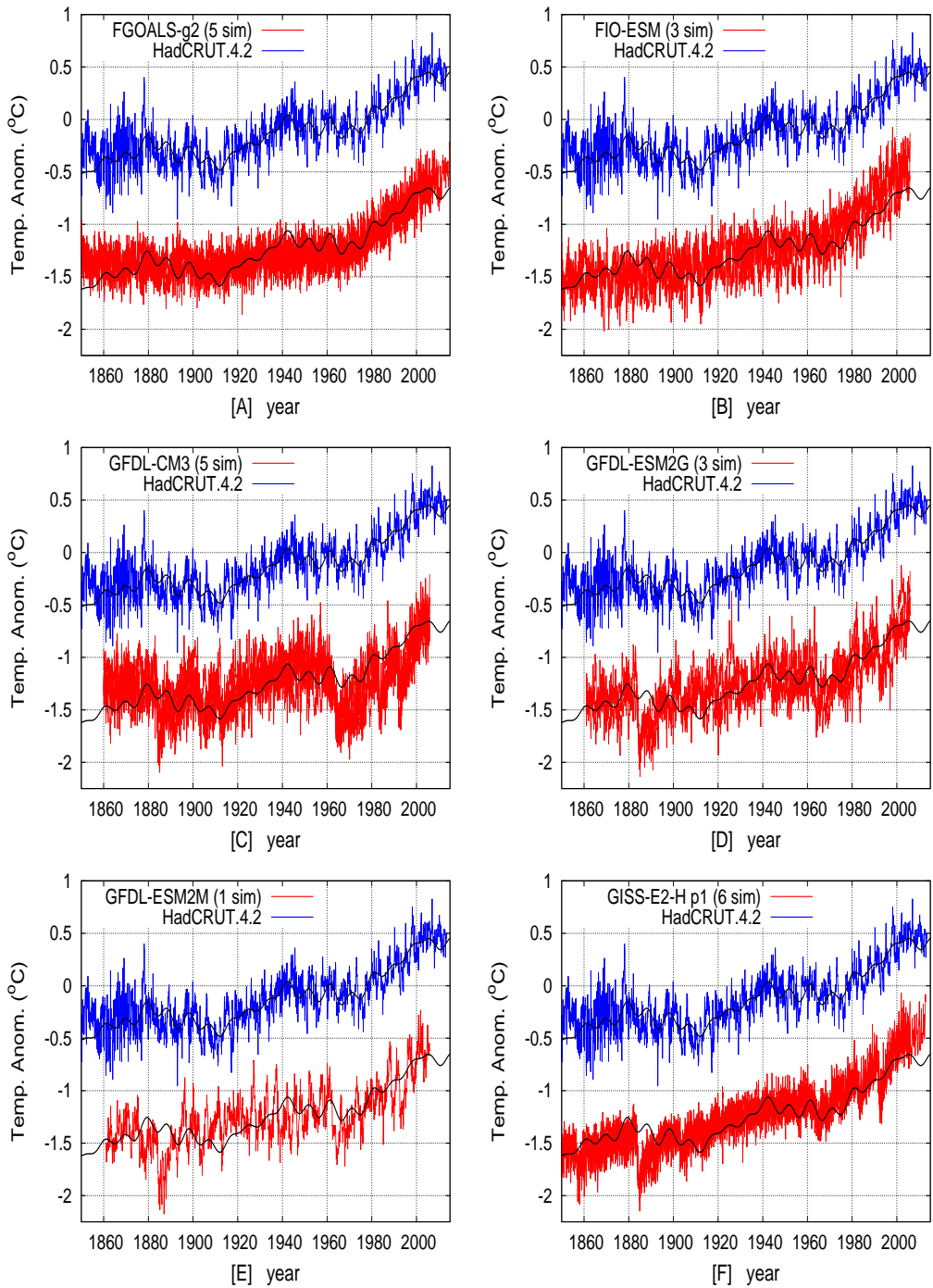}
\end{center}
 \caption{Continues:  HadCRUT4 GST (blue) vs. CMIP5 GCM simulations (red). Eq. \ref{eqf} (black).}
\end{figure*}

These results  suggest missing mechanisms and a significant  overestimation of the climate sensitivity to the adopted radiative forcings, which is particularly evident in the overestimation of the volcano signatures.  On a 163-year period since 1850 only the 1970–2000 warming trend (about 20\% of the total period) has been approximately recovered by using known forcings. Thus, the ability of the CMIP5 GCM ensemble mean simulations to project or predict climate change 30 years ahead with any reasonable accuracy is questionable. Indeed, it is possible that the  1970-2000 GCM-data matching  could be coincidental and simply due  to a fine-tuning calibration of the model parameters to reproduce this period.

\section{Scale-by-scale comparison between GST and the CMIP5 simulations}

In 48 panels (one for each GCM) Figures 4-11 depict all 162 CMIP5 GCM individual available simulations that are herein analyzed  against the HadCRUT4 GST record. The CMIP5 GCM simulations are shifted downward for visual convenience.

\begin{figure*}[!t]
\begin{center}
 \includegraphics[angle=0,width=0.8\textwidth]{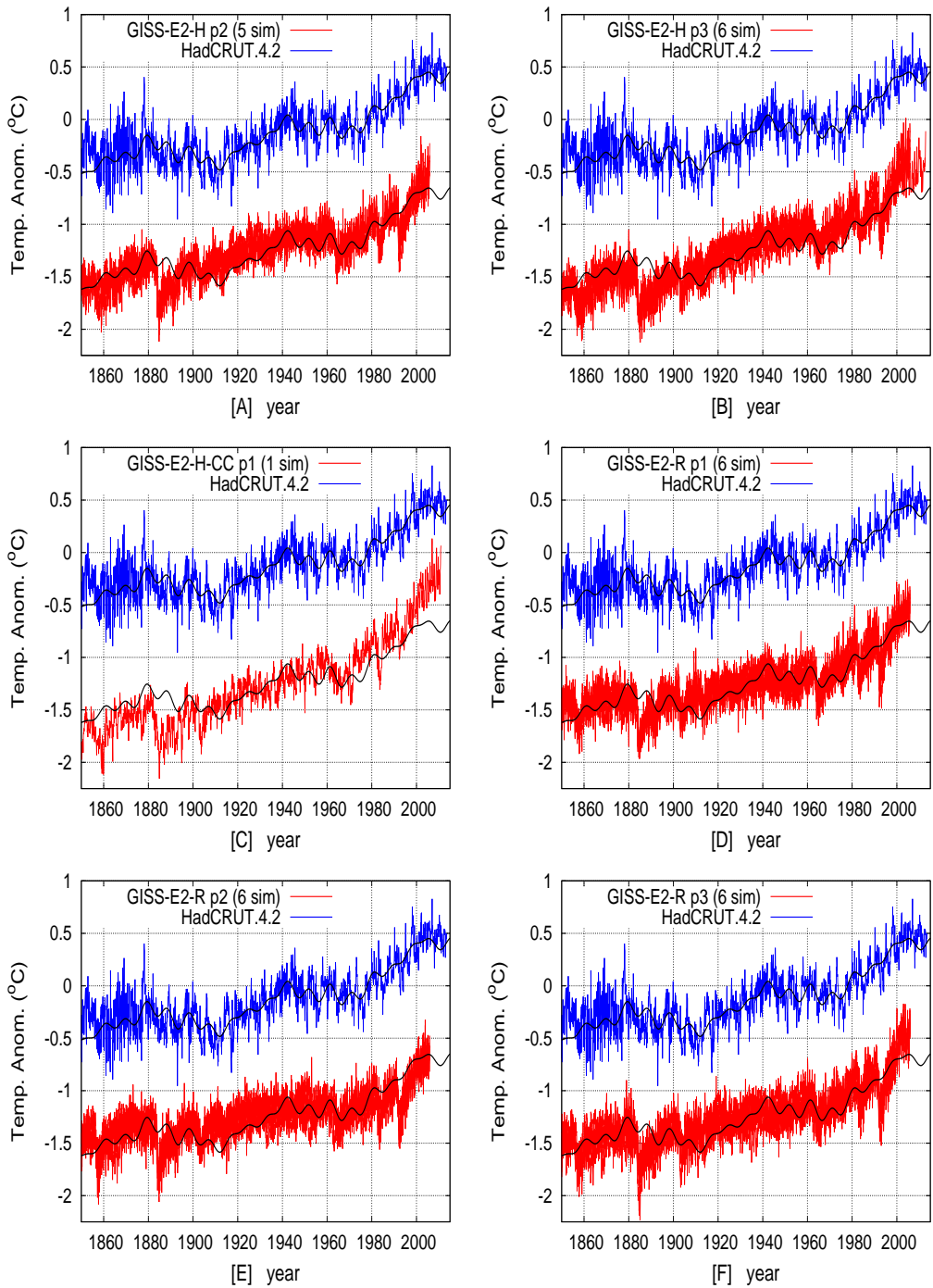}
\end{center}
 \caption{Continues:  HadCRUT4 GST (blue) vs. CMIP5 GCM simulations (red). Eq. \ref{eqf} (black). }
\end{figure*}

The GCM simulations vaguely reproduce a warming from 1860 to 2010.
However, significant discrepancies versus the GST record are observed at all time scales.
Often the volcano signatures appear significantly overestimated. Some model
simulations present a monotonic warming during the entire period, while others
show an almost flat temperature trend until 1970 and a rapid rise since then. In a number of cases the 1970-2000 GCM warming rate is visibly higher than the observed 1970-2000 GST warming rate.
By looking only at the decadal to the multidecadal scales, numerous mismatches
are observed as well. It is easy to notice periods as long as 10-30 years
showing divergent trends between the GST record and the GCM  simulations. The
fast fluctuations at a multi-annual scale are also not reproduced by the models.
Indeed, the GCMs  reproduce a  variability at all time scales, but it
appears to be uncorrelated with the GST observations. In general, all GCM
simulations significantly differ from each other.

\begin{figure*}[!t]
\begin{center}
 \includegraphics[angle=0,width=0.8\textwidth]{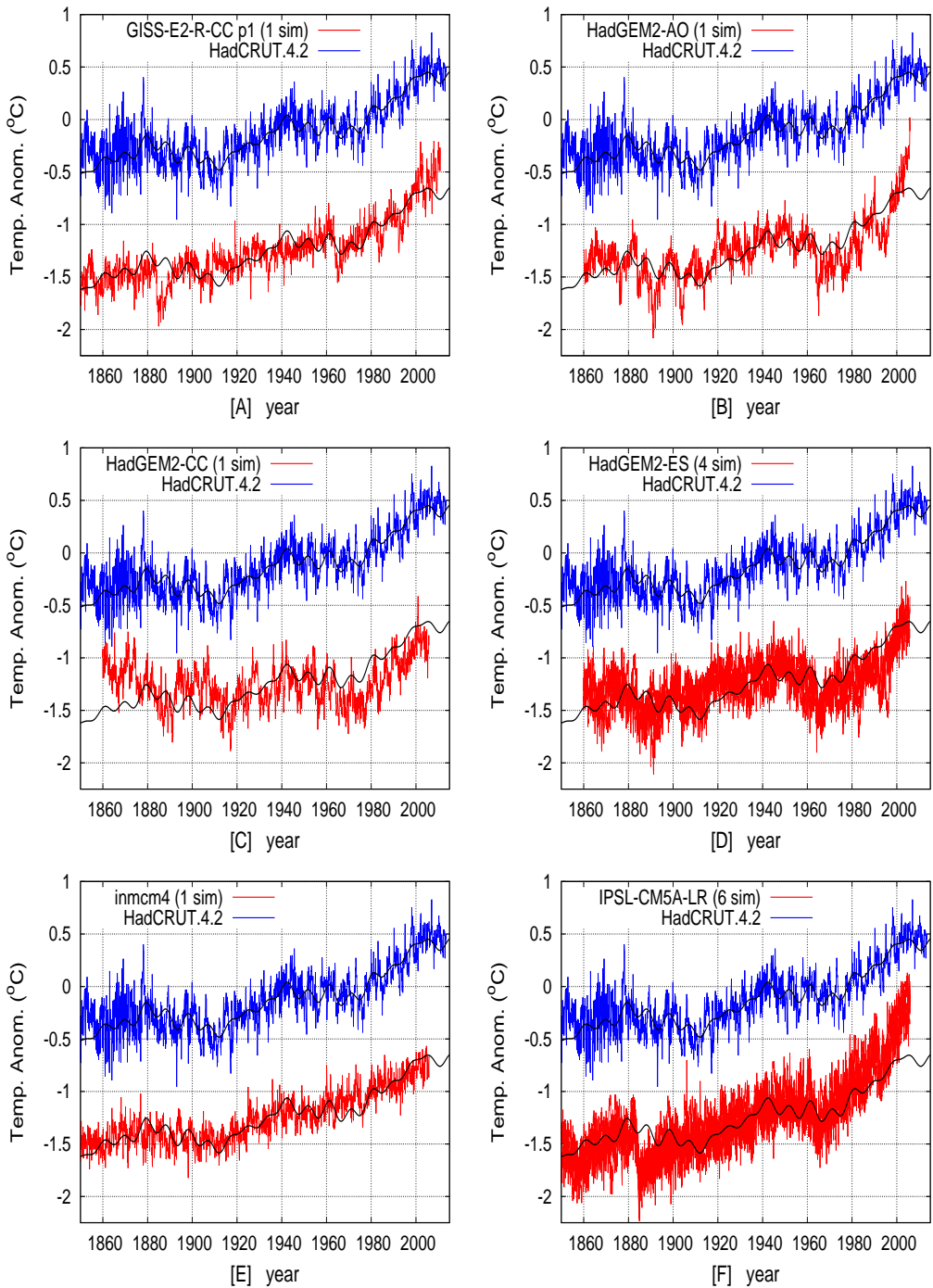}
\end{center}
 \caption{Continues:  HadCRUT4 GST (blue) vs. CMIP5 GCM simulations (red). Eq. \ref{eqf} (black).}
\end{figure*}

In the following subsections three alternative strategies are adopted to study how well the CMIP5 GCM individual simulations  reproduce the GST patterns at multiple scales. The following tests are discussed: (1)   the records are decomposed using a limited set of major harmonics detected by power spectrum analyses, as shown in Figure 1 and the ability of each GCM simulation to reproduce each GST component is tested; (2)  a Fourier band-pass filter decomposition methodology is adopted to decompose each sequence in four band-pass frequency components and the ability of each GCM simulation to reproduce each GST band-pass frequency component is tested; (3) power spectra in the period range from 7 years to 100 years are evaluated for all records and the ability of each GCM simulations to reproduce the GST power spectrum is tested. The scale-by-scale comparison uses  a technique similar to the multiresolution correlation analysis \citep{Scafetta2004}.  The  HadCRUT4  record and all model simulations are analyzed from Jan/1861 to Dec/2005, which is the common period.

\begin{figure*}[!t]
\begin{center}
 \includegraphics[angle=0,width=0.8\textwidth]{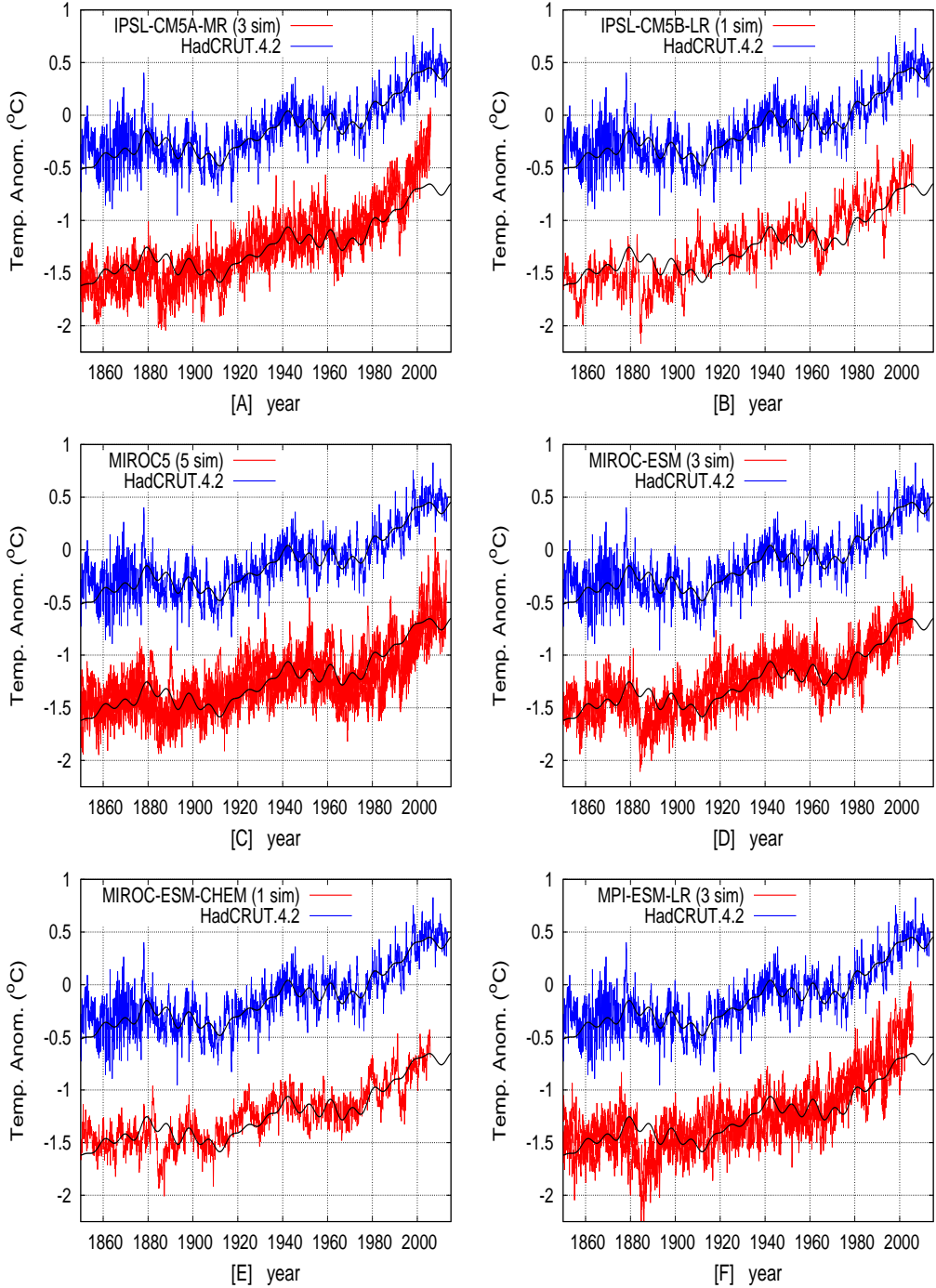}
\end{center}
 \caption{Continues:  HadCRUT4 GST (blue) vs. CMIP5 GCM simulations (red). Eq. \ref{eqf} (black).}
\end{figure*}

\subsection{Scale-by-scale harmonic decomposition comparison}

\cite{Scafetta2010,Scafetta2012a,Scafetta2012b} showed that to a first order approximation the HadCRUT3 GST record can be geometrically decomposed into four major harmonics found in astronomical records with periods of about 9.1, 10.4, 20 and 60 years plus a  second order polynomial trending component. Here  the calibration is repeated using the same harmonics and the  HadCRUT4 GST record from  Jan/1861 to Dec/2005 to obtain:

\begin{eqnarray}
  h_{60}(t) &=& 0.111\cos(2\pi (t-2001.29)/60) \\
  h_{20}(t) &=& 0.043 \cos(2\pi (t-2001.43)/20)
\end{eqnarray}

\begin{eqnarray}
  h_{10.4}(t) &=& 0.030 \cos(2\pi (t-2002.93)/10.4) \\
  h_{9.1}(t) &=&  0.044 \cos(2\pi (t-1997.82)/9.1)\\
  p(t) &=& \frac{3.39}{10^5}(t-1850)^2 -\frac{9.46}{10^4}(t-1850)-0.36~.
\end{eqnarray}

\begin{figure*}[!t]
\begin{center}
 \includegraphics[angle=0,width=0.8\textwidth]{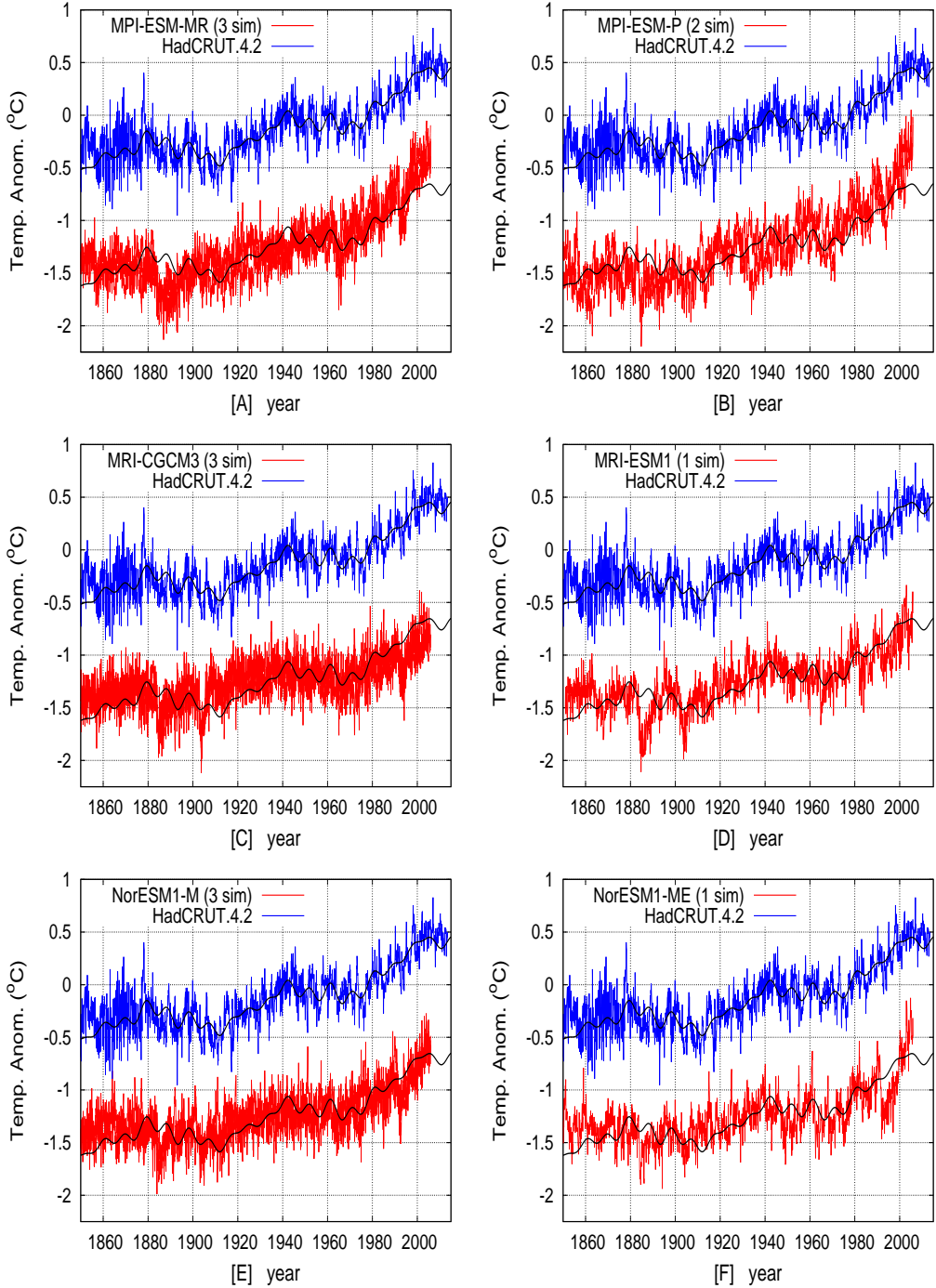}
\end{center}
 \caption{Continues:  HadCRUT4 GST (blue) vs. CMIP5 GCM simulations (red). Eq. \ref{eqf} (black). }
\end{figure*}

The above amplitudes present a statistical error of about 15\% and, together with the phases, they were estimated by linear regression on the GST.  This yields the following  function:

\begin{equation}\label{eqf}
    f(t)=h_{60}(t)+h_{20}(t)+h_{10.4}(t)+h_{9.1}(t)+p(t).
\end{equation}
Eq. \ref{eqf} is plotted in black in Figures 4-11 and reproduces the decadal and multidecadal GST patterns relatively well.  Figures 4-11 also plot Eq. \ref{eqf} against the GCM simulations on the common baseline. The latter comparison assists a visual check of  the ability of the GCM simulations to reproduce the decadal and multidecadal GST patterns, which is generally poor.


\begin{figure*}[!t]
\begin{center}
 \includegraphics[angle=0,width=0.7\textwidth]{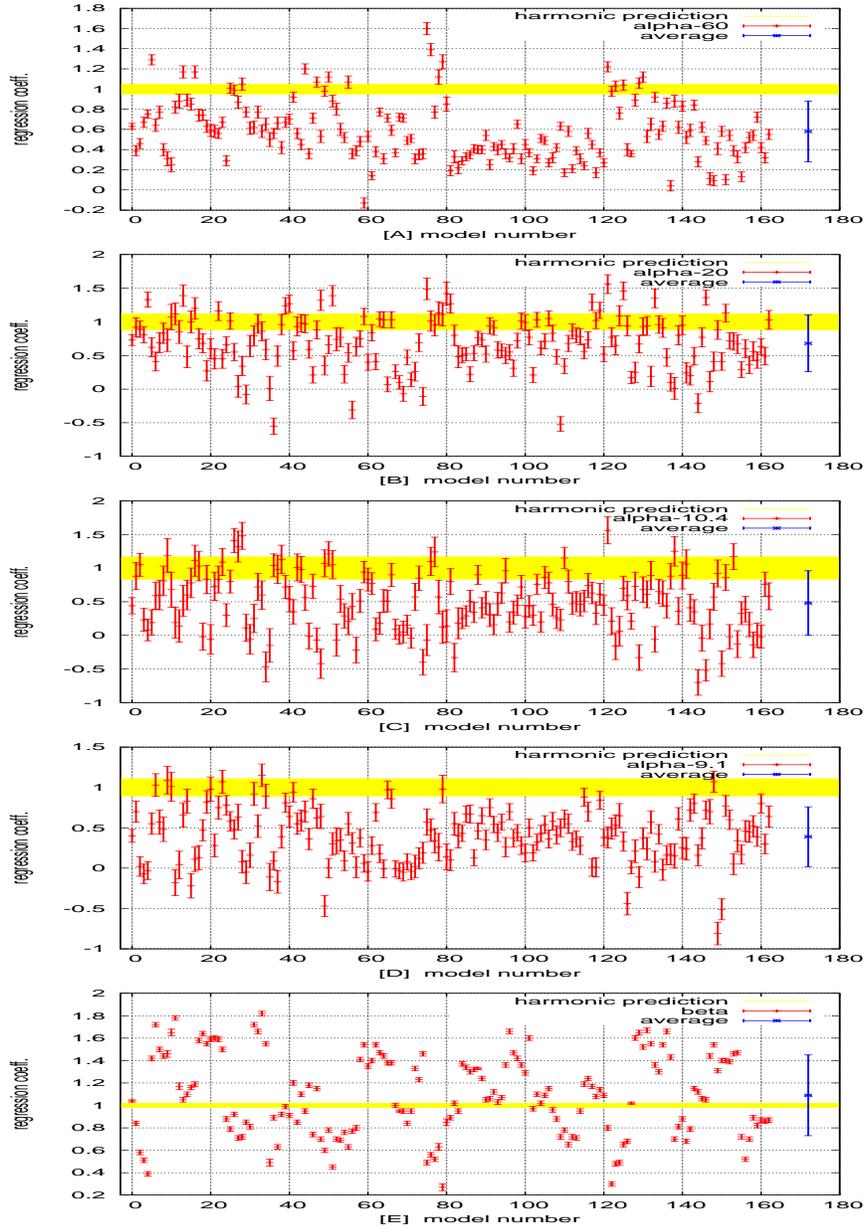}
\end{center}
 \caption{Regression coefficients calculated using Eq. \ref{eqg} referring to all GCM runs, as reported in Table 2. The GCM number -0- corresponds to the CMIP5 model mean simulation depicted in Figure 2. The yellow area around 1 corresponds to the harmonic model confidence region. The blue bars are the average values with relative error bars.}
\end{figure*}

Note that $p(t)$ is only a convenient second order polynomial geometrical description of the upward warming trend from Jan/1861 to Dec/2005. This function does not have any hindcasting or forecasting ability outside
the interval 1861–2005 because its derivation is  geometrical, not physical.
The justification of $p(t)$  is purely mathematical and derives directly from Taylor's theorem. Essentially, a second order polynomial is used because this function is required to capture the accelerating GST warming trend from 1861 to 2005. Simultaneously, it should be as orthogonal as possible to a 60-year cycle on a 145-year period: see also \citet[][figure 4]{Scafetta2013}.  In  section 5  the geometrical function $p(t)$ is substituted with a more physically-based function by taking into account secular and millennial cycles  \citep{Humlum,Scafetta2012c,Scafetta2013a} plus the contribution from volcano and anthropogenic forcings.

In contrast, the four chosen harmonics  are supposed to represent  real dynamical mechanisms related to astronomical cycles. Thus, they may have, within certain limits,  hindcast/forecast capabilities outside the interval 1850-2010, as it was explicitly tested in multiple ways \citep{Scafetta2010,Scafetta2012a,Scafetta2012b,Scafetta2012c}. However,  numerous other harmonics may exist, as it happens for the tidal system where up to 40 harmonics are used. Additional harmonics are ignored in this analysis.

The GCM simulations should simultaneously reproduce  harmonics and an upward trend statistically compatible with those found in the GST record. Thus, we use the same strategy implemented in \cite{Scafetta2012b}, and adopt the following regression model

\begin{equation}\label{eqg}
\begin{split}
    g(t)=\alpha_{60}h_{60}(t)+\alpha_{20}h_{20}(t)+\alpha_{10.4}h_{10.4}(t)\\+\alpha_{9.1}h_{9.1}(t)+\beta p(t) + \gamma
\end{split}
\end{equation}
where $\alpha$, $\beta$ and $\gamma$ are appropriate regression coefficients that are evaluated for each GCM simulation, $m(t)$, by using the usual minimization of the residual square function, $\sum_x[m(t)-g(t)]^2$, during the period Jan/1861 - Dec/2005.  The parameter $\gamma$ has only a baseline meaning.

If the evaluated regression coefficients $\alpha$ and  $\beta$ are close to 1, then the GCM simulation could well reproduce the corresponding temperature constituent pattern captured by Eq. \ref{eqf}. For example, if $\alpha_{60}\approx1$, then  the analyzed GCM simulation function, $m(t)$, could reconstruct the 60-year modulation found in the temperature record as captured by the constituent function $h_{60}(t)$. However, such a condition is \emph{necessary} but not \emph{sufficient} to guarantee the matching between the two records at the given frequency. On the contrary, if the evaluated regression coefficients are statistically different from 1, then the GCM simulation can not reconstruct the correspondent GST pattern.

Table 2 reports the six regression coefficients $\alpha$, $\beta$ and $\gamma$ for the ensemble mean and for each of the 162 GCM simulations. These results are also depicted in Figure 12.
 The  results are  quite unsatisfactory. All GCM simulations fail to  reproduce the four decadal and multidecadal GST oscillations sufficiently well.

\begin{figure*}[!t]
\begin{center}
 \includegraphics[angle=90,width=1\textwidth]{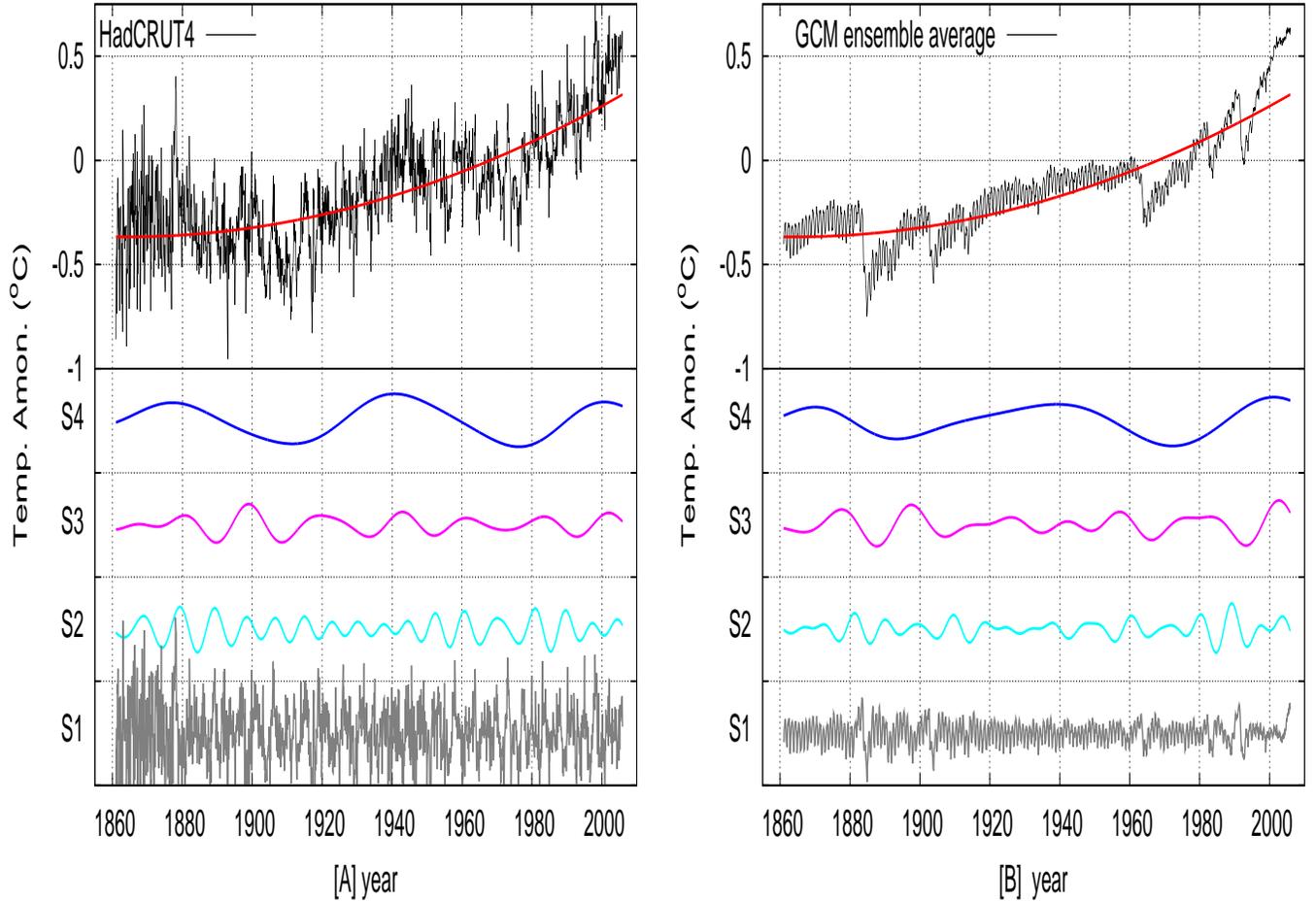}
\end{center}
 \caption{Scale-by-scale Fourier band-pass frequency decomposition: [A] HadCRUT4 GST; [B] GCM ensemble mean. Quadratic polynomial fit, Eq. 5 (red). S1 captures time-scales shorter than 7 years; S2 captures scales between 7 and 14 years; S3 captures scales between  14 and 28 years; S4 captures scales between 28 and 104 years. Note in [A] the clear quasi 60-year oscillation in the HadCRUT4 GST S4 record and the beat modulation in  the HadCRUT4 GST S2 curve caused by the 9.1 and 10.4 year oscillations. }
\end{figure*}

  The averages of the 162 results give values between 0.4 and 0.7 with a standard deviation  between 0.3 and 0.5. Thus, most regression coefficients fall outside the yellow area of the harmonic model confidence reported in the first line of Table 2. Even if occasionally a regression coefficient falls close to 1, as shown in Figure 13, it is likely a coincidence as the GCM simulations should simultaneously reproduce all decadal and multidecadal patterns. A comprehensive statistical test is provided by using the following reduced $\chi^2$ values that are reported in Table 2 and are defined as:

\begin{equation}
\label{eq233}
\begin{split}
 \chi^2= & \frac{1}{5} \left[
 \frac{(\alpha_{60m} - \alpha_{60T})^2}{\Delta\alpha_{60m}^2+ \Delta \alpha_{60T}^2} +
 \frac{(\alpha_{20m} - \alpha_{20T})^2}{\Delta\alpha_{20m}^2+ \Delta \alpha_{20T}^2} +  \right. \\
 & \frac{(\alpha_{10.4m} - \alpha_{10.4T})^2}{\Delta\alpha_{10.4m}^2+ \Delta \alpha_{10.4T}^2} +
 \frac{(\alpha_{9.1m} - \alpha_{9.1T})^2}{\Delta\alpha_{9.1m}^2+ \Delta \alpha_{9.1T}^2} + \\
& \left. \frac{(\beta_{m} - \beta_{T})^2}{\Delta\beta_{m}^2+ \Delta \beta_{T}^2} \right].
\end{split}
\end{equation}
 The suffix `m' indicates the correspondent regression coefficient for the GCM model; the suffix `T'   indicates the correspondent regression coefficient for the harmonic model reported in the first line of Table 2. Values of $\chi^2 \lessapprox 1$ would indicate that the model performs well in reproducing the decadal and multidecadal patterns shown in the data. However, as the table reports, $\chi^2$ varies between 3.5 and 161 (mean=$51$) among the 162 individual simulations.  $\chi^2$  is 14 for the ensemble mean model. Values $\chi^2\gg1$ indicate that these  GCMs do not reconstruct the GST decadal and multidecadal  scales.

Table 2 also reports the root mean square deviation \\ (RMSD) from Eq. \ref{eqf} for both the GST record and for all GCM simulations  shown in Figure 4-11:
 $RMSD(\xi,\theta)=$ \\ $\sqrt{\sum_{i=1}^N(\xi_i-\theta_i)^2/N}$. This function   measures the ability of a simulated sequence, $\{\xi_i\}$, to reconstruct the observed sequence, $\{\theta_i\}$. The RMSD is calculated after  both the GST record  and the GCM simulations are smoothed with a 49-month moving average algorithm to highlight the decadal modulation, as done in Figure 3B. For the GCMs, RMSD ranges between 0.08 $^oC$ and 0.22 $^oC$, with an average of 0.14 $^oC$. On the contrary, the harmonic model (Eq. 6) presents a RMSD of 0.05 $^oC$, indicating that the latter is statistically more accurate in representing the GST records by a factor between 2 and 4.4 compared with  the  CMIP5 GCMs, as  also found for  the CMIP3 GCMs \citep{Scafetta2012b}.

  \subsection{Scale-by-scale band-pass filter decomposition comparison}

A Fourier band-pass filter decomposition  captures all dynamical details of a time sequence at complementary time-scales. Figures 13A and 13B show the decomposition of the HadCRUT4 GST and of the GCM ensemble mean  simulation. S1 corresponds to time-scales larger than 6 months and shorter than 7 years, and captures most of the fast variability of the signal such as ENSO oscillations and volcano eruptions; S2 corresponds to scales between 7 and 14 years, and captures the decadal scale; S3 corresponds to scales between  14 and 28 years, and captures the bi-decadal oscillation; S4 corresponds to scales between 28 and 104 years, and captures the multidecadal scale such as the quasi 60-year oscillation. A second order polynomial, given by Eq. 5, is detrended from all sequences to make them stationary to a first order  approximation.

If $T_c(t)$ is a measured band-pass temperature component function, and  $M_c(t)$ is the correspondent  GCM temperature band-pass component function, the regression coefficient $\Phi$ is given by the formula

\begin{equation}\label{phi}
\Phi=\frac{\sum M_c(t)T_c(t)}{\sum T_c(t)T_c(t)}~.
\end{equation}
If $\Phi$ is close to 1, then the functions $M_c(t)$ and $T_c(t)$ are statistically compatible according to this measure. Again,  such a condition is \emph{necessary} but not \emph{sufficient} to guarantee the matching between the two records at the given frequency band.

\begin{figure*}[!t]
\begin{center}
 \includegraphics[angle=0,width=0.7\textwidth]{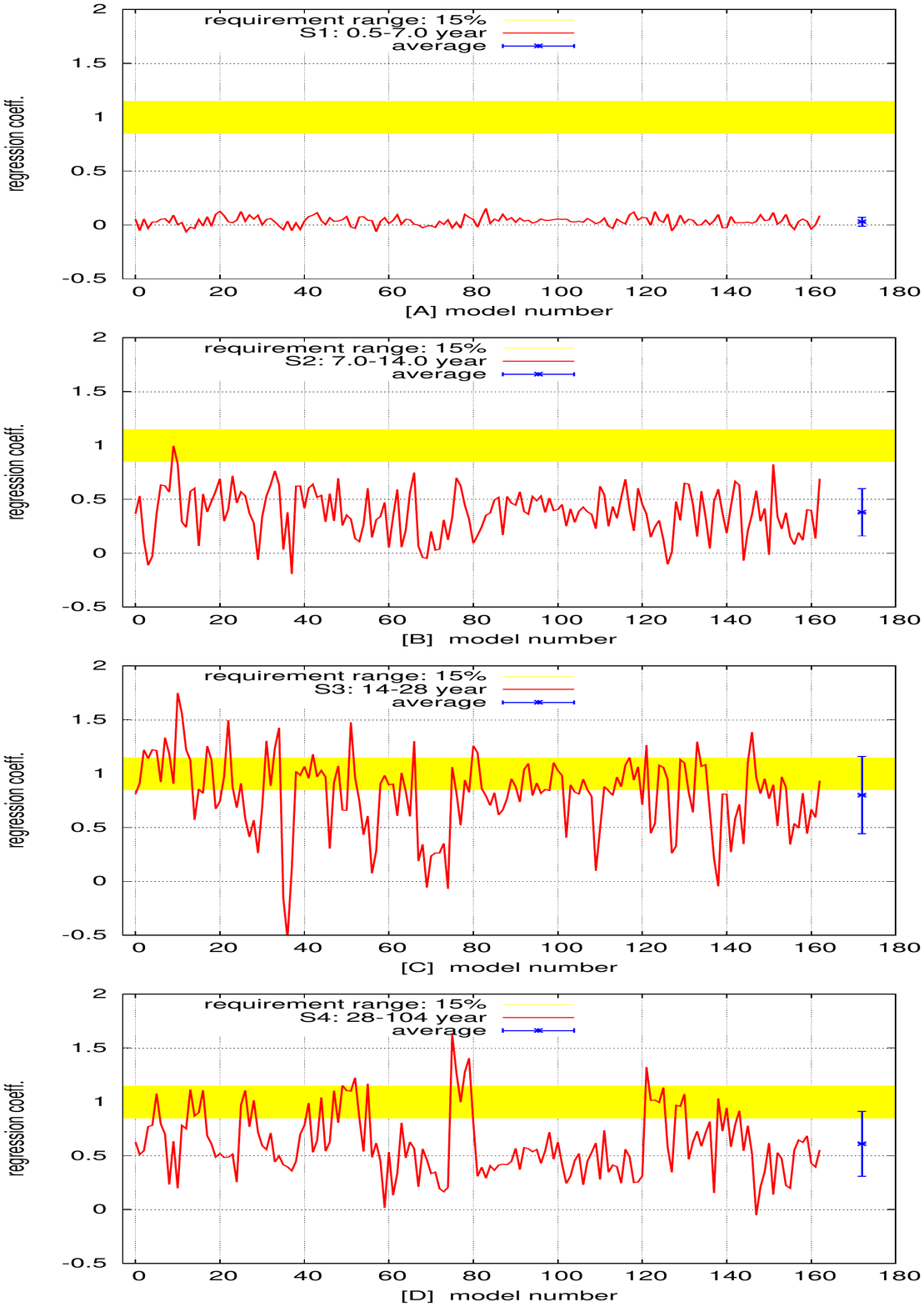}
\end{center}
 \caption{Regression coefficients $\phi$ calculated using Eq. \ref{phi} referring to all GCM runs, as reported in Table 3: [A]  $\phi_{S1}$, band-pass range 0.5-7.0 year; [B] $\phi_{S2}$, band-pass range 7.0-14.0 year; [C] $\phi_{S3}$, band-pass range 14-28 year; [D] $\phi_{S4}$, band-pass range 28-104 year. The GCM number -0- corresponds to the CMIP5 model mean simulation depicted in Figure 1. The yellow area around 1 corresponds to the chosen confidence region of 15\%. The blue bars are the average values with relative error bars.}
\end{figure*}

Table 3  reports the four scale-by-scale  regression coefficients calculated for the GCM ensemble mean and the 162 GCM simulations versus  the correspondent GST frequency band-pass components. The same coefficients are depicted in Figure 14. All GCM simulations, including the GCM ensemble mean (number -0- in the figure), perform much less well  in reconstructing the temperature components although occasionally a single regression coefficient may fall close to 1. The average values relative to the four band-pass components are statistically different from $\langle \Phi \rangle =1\pm0.15$ (a reasonable 15\% error from the ideal $\Phi=1$ is assumed for all values as approximately found in Table 2 for the harmonic model components): for S1, $\langle \Phi \rangle =0.03\pm0.04$; for S2, $\langle \Phi \rangle =0.38\pm0.22$; for S3, $\langle \Phi \rangle =0.80\pm0.36$;  for S4, $\langle \Phi \rangle =0.61\pm0.30$.

Table 3  also reports a reduced $\chi^2$ test using an equation similar to Eq. \ref{eqg33}. However  only the three components related to the S2-scale, S3-scale and S4-scale are used in the test as

\begin{equation}
\label{eq23355}
 \chi^2=  \frac{1}{3} \left[
 \frac{(\Phi_{S4} - 1)^2}{0.15^2} +
 \frac{(\Phi_{S3} - 1)^2}{0.15^2} +
 \frac{(\Phi_{S2} - 1)^2}{0.15^2}  \right] .
 \end{equation}
The S1-scale is excluded because the GCM models evidently do not  reconstruct the fast fluctuations at scales shorter than 7 years. Again, $\chi^2\gg1$ for all models. Thus, as in the previous subsection, also these results suggest that the CMIP5 GCMs do not properly reconstruct the observed GST dynamics at  multiple time scales, although a significant variation among the individual  simulations is observed.

\begin{figure*}[!t]
\begin{center}
 \includegraphics[angle=0,width=0.7\textwidth]{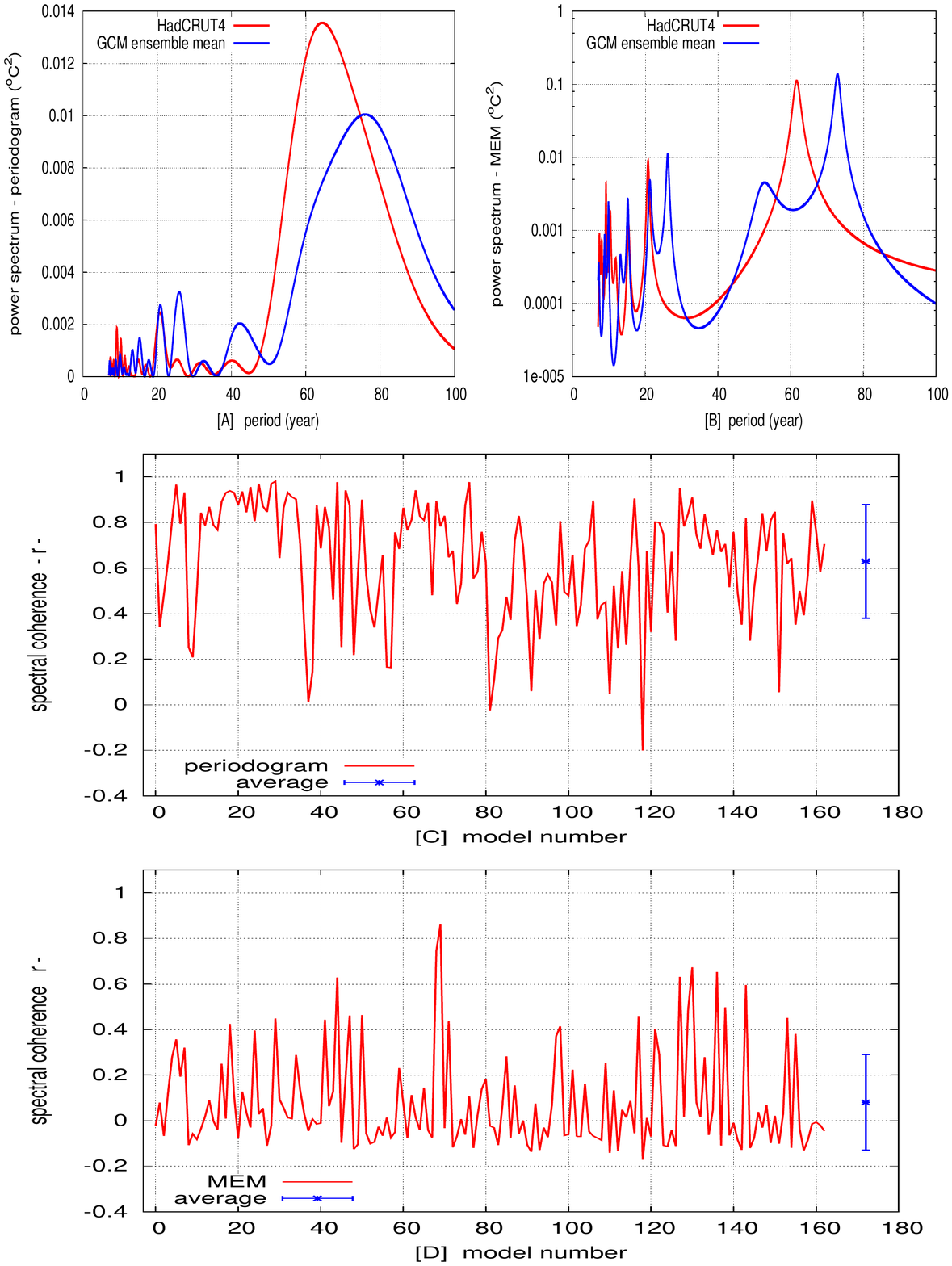}
\end{center}
 \caption{[A] Periodograms of the GST record (red) and of the GCM ensemble mean simulation (blue); [B] Maximum entropy method (MEM) power spectrum  of the GST record (red) and of the GCM ensemble mean simulation (blue); [C] Spectral correlation coefficients between the GST recorcd and all GCM simulations using the periodograms; [D] Spectral correlation coefficients between the GST record and all GCM simulations using the MEM power spectra. See Table 3, last two columns.}
\end{figure*}

\subsection{Direct power spectrum comparison}

Figure 15 depicts a final statistical test that estimates the ability of the GCMs in reconstructing the power spectrum  of the  GST within the period range from 7 to 100 years.

Figure 15A shows in red the periodogram of the HadCRUT4 GST and of the GCM ensemble mean; Figure 15B shows in red the maximum entropy method (MEM) power spectrum  of the HadCRUT4 GST and of the GCM ensemble mean. The periodogram and MEM algorithms were taken from \citet{Press}  and calculated after detrending from all records Eq. 5.  The figures highlight, for example, that the  GCM ensemble mean macroscopically fails to reproduce the quasi 60-year oscillation by presenting a spectral peak at a period of $\sim75$ years instead of the observed $\sim61$ years: see also Figure 3A.

As alternatively demonstrated in \citet[][figure 2]{Scafetta2012b} the autocorrelation peak at a time-lag of 70-80 years found in the GCM simulations is mostly driven by the strong GCM volcano eruption signatures. In fact, there is a quasi 80-year lag between the two GCM large volcano eruption signatures of Krakatoa (1883) and Agung (1963-1964), and between the volcano signatures of Santa Maria (1902) and El Chichon (1982).
 Note that the quasi 80-year recurrent pattern in the volcano signature could have been responsible for the slight shift of the peak of the GST periodogram at a value slightly larger than 60 years, as mostly observed in Figure 15A.

A coherence test is made by simply calculating the cross-correlation coefficient $r$ between the GST power spectrum depicted in the figure and each of the GCM power spectra. Values of $r$ close to 1 indicate a good spectral coherence between the GST record and the GCM simulations. Note, however, that this is a less stringent test than the  cases discussed in subsections 3.1 and 3.2 because a mere power spectrum correlation test  ignores the phase positions of the harmonics which is a necessary component that the models should reconstruct as well.

The comparison between the GST record and the GCM ensamble mean record gives  $r=0.79$ using the periodograms, and $r=-0.02$ using the MEM power spectra, which present sharper peaks.
The test is repeated for all 162 GCM simulations. The results are depicted in Figues 15C and 15D and reported in the last two columns of Table 3. The average correlation coefficient is $\langle r\rangle =0.63 \pm 0.25$ using the periodograms and $\langle r\rangle =0.08 \pm 0.21$ using the MEM power spectra.

\begin{figure*}[!t]
\begin{center}
 \includegraphics[angle=-90,width=1\textwidth]{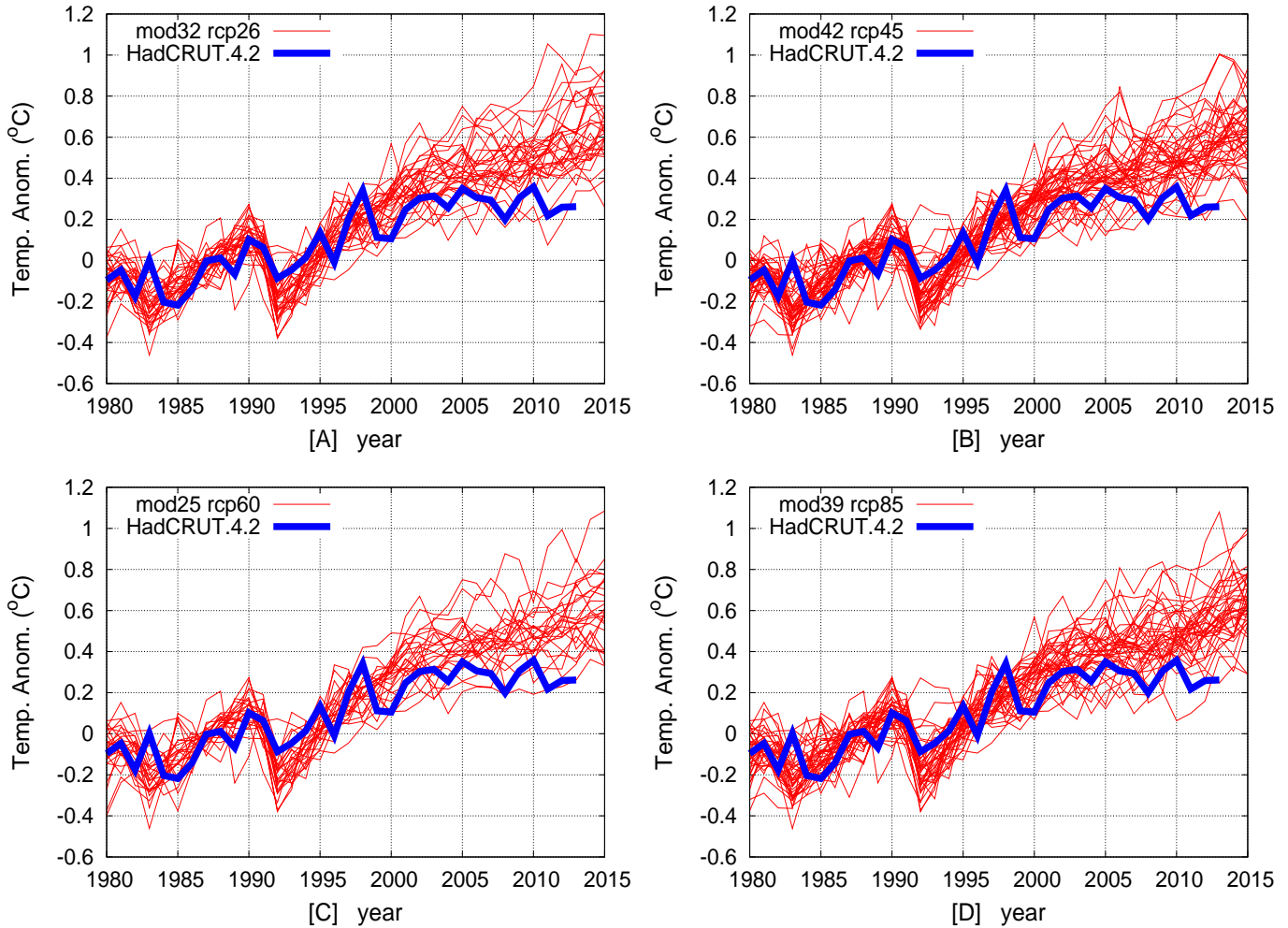}
\end{center}
 \caption{HadCRUT4 record (blue) versus the CMIP5 GCM projections: [A] the rcp26 simulations made of 32 models; [B] the  rcp45 simulations made of 42 models; [C] the rcp60 simulations made of 25 models; [D] the  rcp85 simulations made of 39 models. All records are annually resolved and baselined in the 1980-2000 period. Note that in 2013 the temperature, which is approximately steady since 1997, runs cooler than all CMIP5 GCM simulations.}
\end{figure*}

Thus,  the GCMs do not reproduce the natural harmonics of the climate system even in the less stringent sense of a mere power spectrum correlation test that ignores the phase positions of the harmonics. These results indicate a relatively poor spectral coherence at the dacadal and multidecadal scale between the GST and the GCM simulations. The performance of the individual GCM simulations varies greatly.

\section{Visual examples of CMIP5 GCM deficiencies}

Typical major deficiencies found in the GCM simulations are briefly discussed in the following two subsections.
A simple visual analysis of the records suffices to highlight severe mismatches between the GCM simulations and the GST record.

\begin{figure*}[!t]
\begin{center}
 \includegraphics[angle=0,width=1\textwidth]{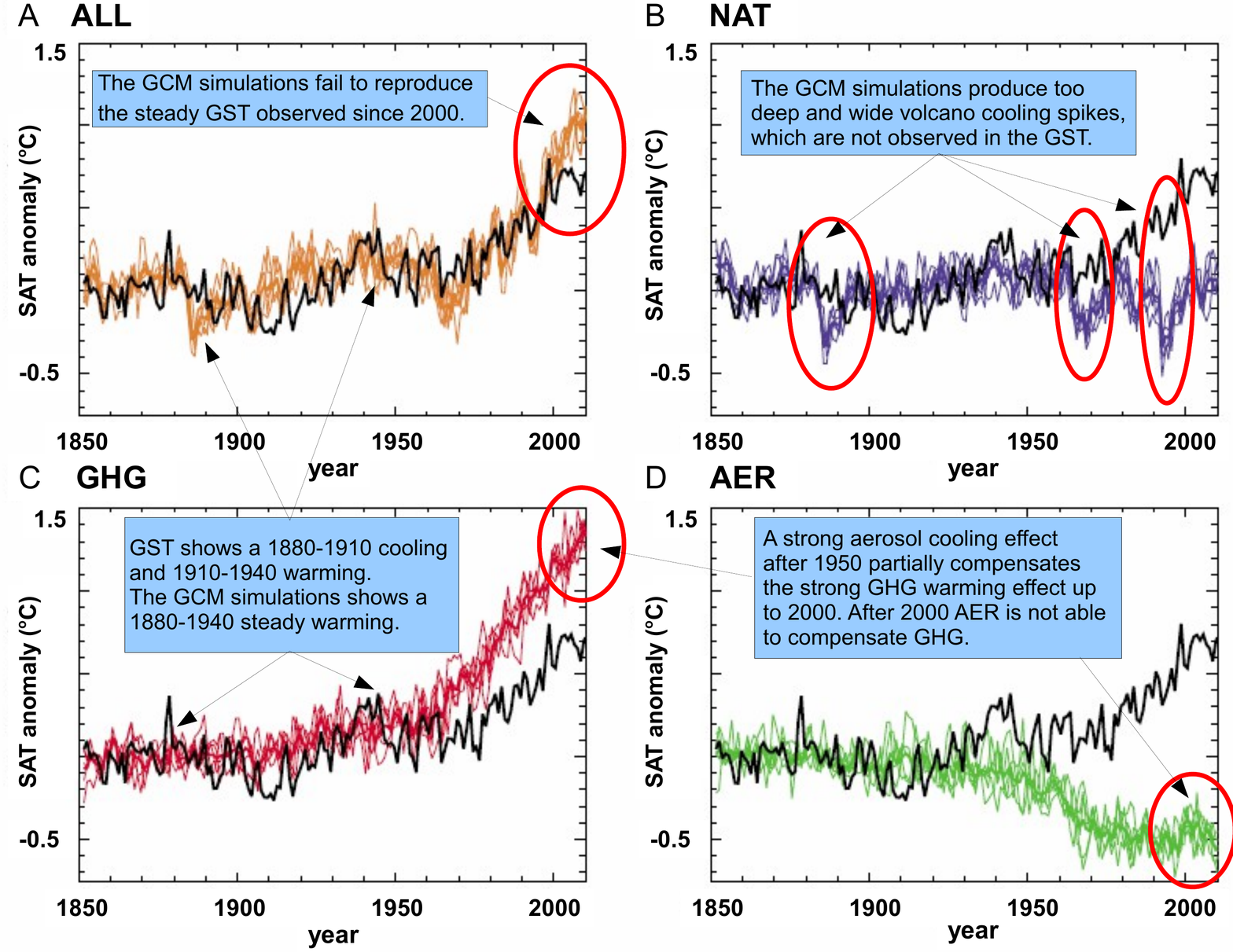}
\end{center}
 \caption{A reproduction of figure 1 in \cite{Gillett2012} with additional comments that highlight major mismatches between the GST record (black)   and a set of simulations made with CanESM2. The figure highlights problems common to all CMIP5  GCMs. }
\end{figure*}

\subsection{Example 1: the GCM simulations significantly diverge from the GST record since 2000}

Most CMIP5 GCM simulations using the historical forcings end in Dec/2005. Since 2006 the models are run using projected forcings for the 21st century. Four scenarios have been proposed and labeled  as: rcp26, rcp45, rcp60, rcp85, as also shown in Figure 2.

Figure 16 shows the annually resolved available simulations in the four cases versus the annually resolved HadCRUT4 record (the 2013 value is based on  data from January to June). The records are baselined during the period 1980-2000.  The rcp26 simulations are made of 32 models, rcp45 of 42 models, rcp60 of 25 models and  rcp85 of 39 models.

Figure 16 clearly shows that the models have significantly overestimated the global warming rate since 2000: see also the detailed analysis proposed by  \citet{Fyfe}. For the year 2013 all models have predicted a global mean surface temperature that is found to be 0.0-0.5 $^oC$ warmer   than the GST record.  The linear rate of the HARCRUT4 record since 2000 is $0.3\pm0.4$ $^oC/century$, which indicates that no warming has been observed in the GST record since 2000. In contrast, the CMIP5 simulations have predicted a strong warming rate of: [A] rcp26, $2.2\pm0.2$ $^oC/century$; [B] rcp45, $2.1\pm0.2$ $^oC/century$; [C] rcp60, $1.9\pm0.2$ $^oC/century$; [D] rcp85, $2.1\pm0.2$ $^oC/century$.

\subsection{Example 2: a detailed qualitative visual study of the CanESM2 GCM simulations}

Figure 17  reinterprets figure 1 of  \cite{Gillett2012} that  compares the GST record and a set of simulations of CanESM2 GCM, which  produces some of the best results among all GCMs. For example its simulations  produce the best multidecadal result with a $3< \chi^2 <15$ (see  Tables 2-3 and the correspondent figures). However, even the CanESM2 simulations  macroscopically fail to reproduce the observed steady GST pattern since 2000. CanESM2  appears to be finely tweaked with the aerosol uncertain forcing, but still fails to reconstruct important temperature patterns, which are far better reconstructed by the harmonic model discussed in Section 6. In fact, for the CanESM2 simulations RMSD ranges from 0.13 to 0.15 $^oC$ while Eq. \ref{eqg33} has a RMSD of about 0.04  $^oC$.

 In the simulations depicted in Figure 17, CanESM2 GCM is forced with: (a) anthropogenic and natural forcings (ALL), (b) natural forcings only (NAT), (c) greenhouse gases only (GHG), and (d) aerosols only (AER).
 The comments added to the figure highlight typical common problems found in all CMIP5 GCMs when compared with the  GST record. (1) The GCM does not reproduce the 2000-2012 steady GST trend. (2) The volcano cooling spikes are too large compared to the signature that can be qualitatively deduced from the GST record. (3) GST shows a clear 60-year modulation made of a 1880-1910 cooling plus a 1910-1940 warming, while the GCM shows a 1880-1940 steady warming. (4) The GCM needs a strong aerosol cooling effect after 1950 to partially compensate the strong GHG warming effect up to 2000, but after 2000 the aerosol cooling effect is not able to compensate the strong GHG warming effect, and the simulations strongly diverge from the observations.

\section{Discussion}

The general failure of the CMIP5 GCMs to accurately reconstruct the dacadal and multidecadal GST scales, including that the  GST has not warmed during the last 17 years (essentially since about 1997), brings into question the reliability of these models. In fact,   \cite{Knight} observed that: \emph{``Near-zero and even negative trends are common for intervals of a decade or less in the simulations, due to the model's internal climate variability. The simulations rule out (at the 95\% level) zero trends for intervals of 15 year or more, suggesting that an observed absence of warming of this duration is needed to create a discrepancy with the expected present-day warming rate."} The results of this analysis suggest that major physical flaws exist in the CMIP3 and CMIP5  GCMs, which may cast doubts on  their 21st century projections as well.

\begin{figure*}[!t]
\begin{center}
 \includegraphics[angle=0,width=1\textwidth]{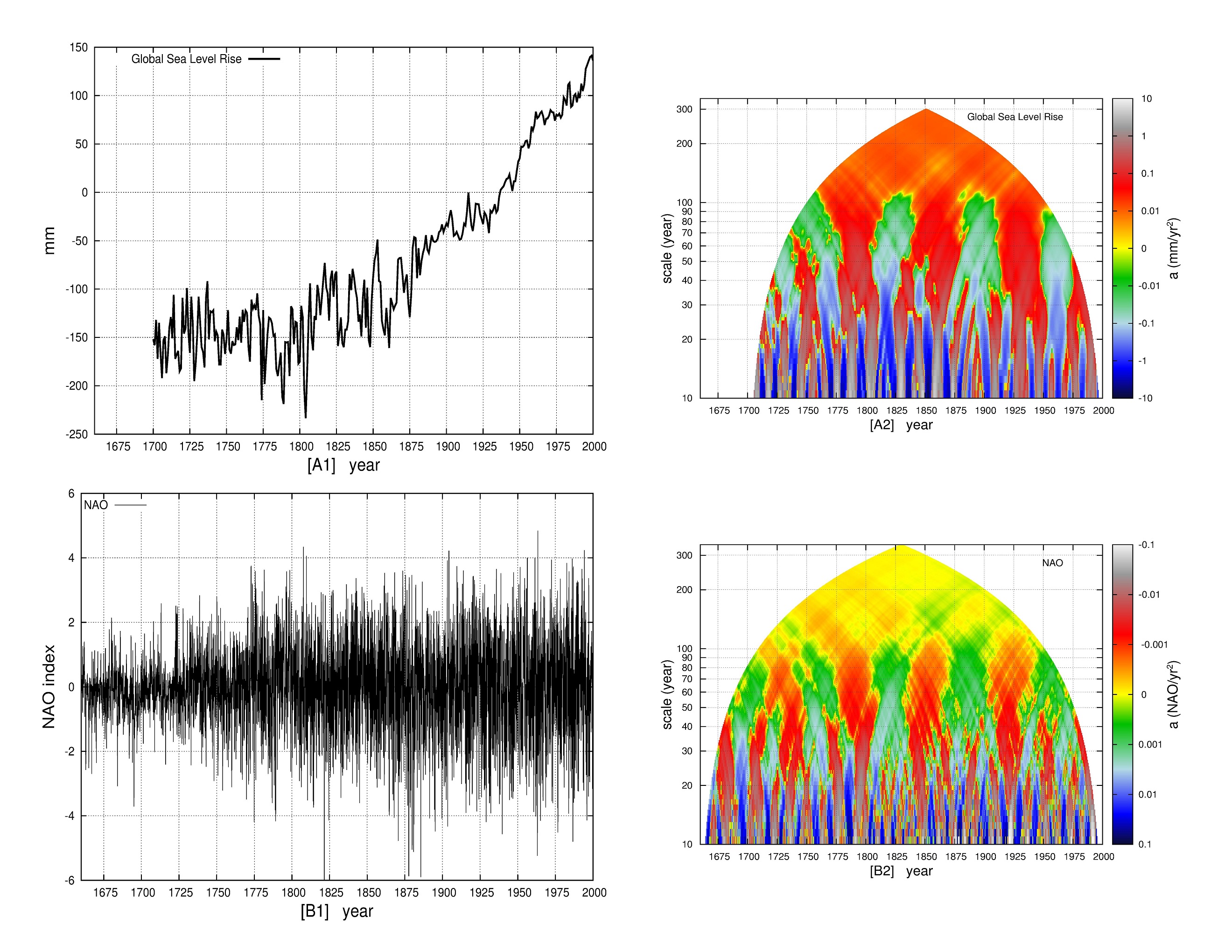}
\end{center}
 \caption{[A] Global sea level record \citep{Jevrejeva} (left) and its
MSAA colored diagram (right). [B] North Atlantic Oscillation (NAO)
\citep{Luterbacher1999,Luterbacher2002} (left) and its MSAA colored diagram (right).
In [B] the colors are inverted. Note the common quasi 60 year oscillation
since 1700 indicated by the alternating green and red regions within
the 30-100 year scales. From \cite{Scafetta2013c}.}
\end{figure*}

The inability of the GCMs to model the observed oscillations and, in particular, the post-2000 temperature plateau  has been justified in various ways. In general, it has been  speculated that the models and/or their forcings need to be slightly corrected. Effects of missing volcano forcing, aerosol forcing and/or  some internal unforced dynamics of the climate system have typically been speculated, but problems arise with these interpretations.

For example, \cite{Booth} speculated that the cooling from 1940 to 1970 was due to poorly modeled aerosol forcing. However, the temperature also presents an equivalent cooling period from 1880 to 1910 that was not reconstructed by their model. \cite{Kaufmann} speculated that the steady GST between 1998-2008 was caused by an increase of atmospheric sulfate production (primarily from China) that  countered the greenhouse gas warming. However, \cite{Remer} (see their figure 5) showed no change in global aerosol optical depth during the period 2000-2007. \cite{Meehl} speculated that GST \emph{hiatus} periods could be caused by occasional deep-ocean heat uptake, and showed that GCM simulations may occasionally present, at random times, an up-to-a-decade of steady temperature despite an increasing anthropogenic forcing. However, the CCSM4 GCM used in \citet{Meehl} (which is one of the models analyzed above) does not reproduce the steady temperature observed from 2000 to 2013. The  CCSM4 GCMs only produce \textit{hiatus} periods occurring in 2040-2050 and 2070-2080, which appear as random red-noise fluctuations of the model (see their figure 1a). The latter variability is commonly referred to as internal unforced dynamics of the climate system and it is claimed to be unpredictable.

However,  because the lack of warming since 1997–1998 is just an aspect of the problem, the above speculations appear physically unsatisfactory. A comprehensive and consistent theory of climate change must simultaneously   interpret  the entire GST dynamics observed since 1850 at least from the decadal scale  up. Although aerosols and internal dynamics certainly may have some effects, the GST also appears to present quasi regular oscillations. The findings of the previous sections  indicate that the CMIP5 GCMs fail  to simultaneously capture  the decadal and multidecadal GST dynamical patterns observed since 1850 such as the four identified major oscillations with approximate periods of 9.1, 10-11, 20 and 60 years \citep{Scafetta2010}. These oscillations generate a network of dynamical synchronization within the climate system also not reproduced by the models  \citep{Wyatt}. The tables 2 and 3 show that the GCM performance varies greatly both among the models and among the individual model runs produced by the same model.

 Quasi-decadal, bidecadal and 60-year oscillations and other longer oscillations have been detected in numerous records covering centuries and millennia.   For example, \cite{Jevrejeva} and \cite{Chambers} showed a quasi 60-year cycle in the sea level rise rate since 1700; \cite{Klyashtorin} showed that  numerous climate indexes present a long-term 50-70 year oscillations during the last 1500 years; \cite{Knudsen} showed a persistent quasi 60-year cycle in the Atlantic Multidecadal Oscillation throughout the last 8,000 years; a quasi 20-year and 60-year oscillations also appear for centuries and millennia  in some Greenland temperature records \citep{Davis,Chylek2012}. The Introduction section contains additional suggested references showing these oscillations.

 For example, Figure 18 reproduces figure 10 in \cite{Scafetta2013c} that shows two relatively global climatic indexes since 1700: the global sea level record \citep{Jevrejeva} and the North Atlantic Oscillation (NAO) reconstruction \citep{Luterbacher1999,Luterbacher2002}. The right panels show the multi-scale acceleration analysis (MSAA) of these two records and highlight the presence  of a common major quasi 60-year oscillation since 1700. This oscillation is revealed  by the alternating green and red colors indicating that the local acceleration of the records varies from negative to positive values, that is, there is an oscillation.

 \begin{figure*}[!t]
\begin{center}
 \includegraphics[angle=-90,width=0.8\textwidth]{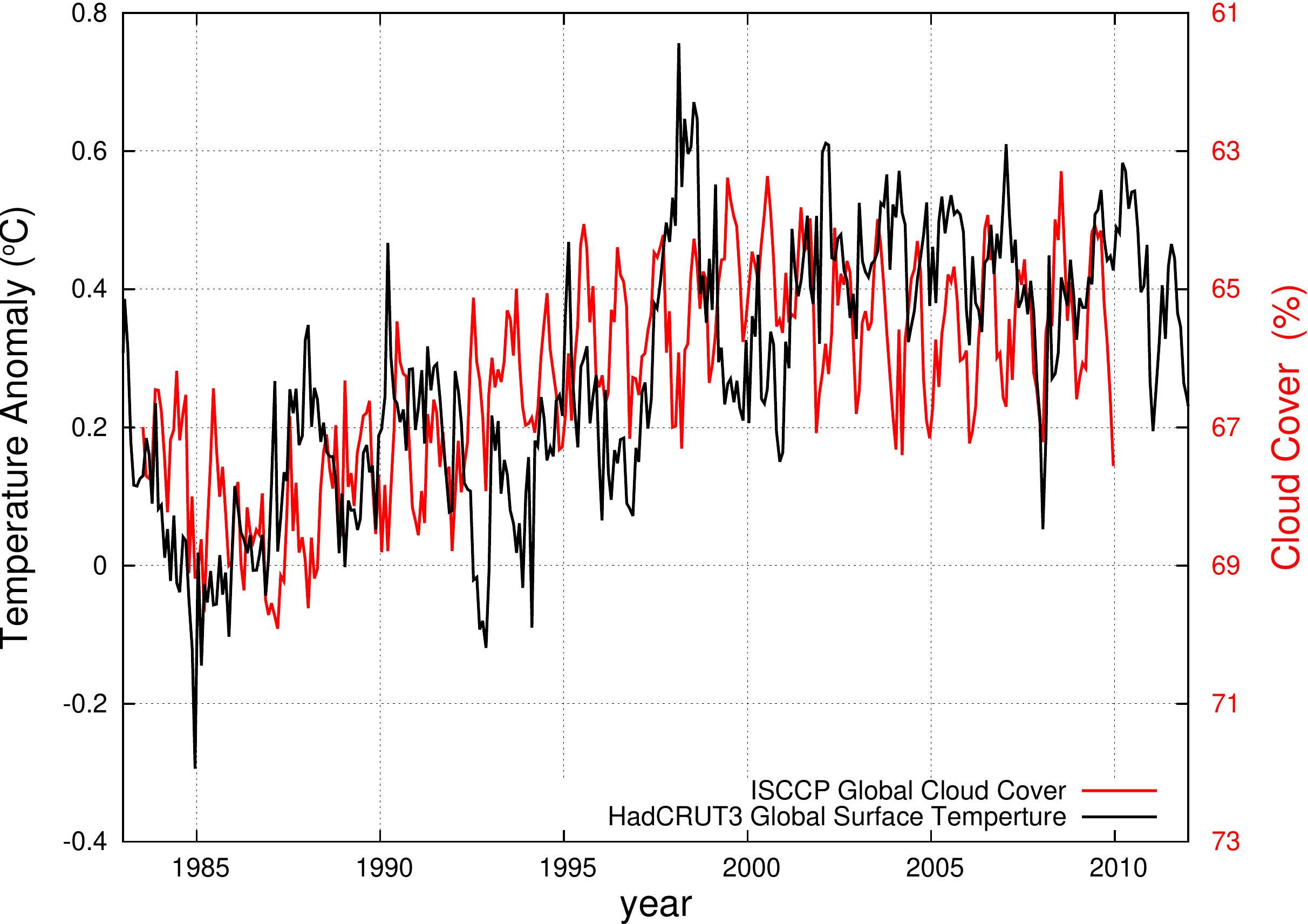}
\end{center}
 \caption{ Global surface temperature (black) against monthly variations in total global cloud cover since July 1983 (red). Correlation coefficient: $r_o=-0.52$, for 318 points $P(|r| \geq |r_o|)<0.0005$.  The cloud data are from the International Satellite Cloud Climatology Project (ISCCP). Cloud data from  {http://isccp.giss.nasa.gov/pub/data/D2BASICS/B8glbp.dat}}
\end{figure*}

 The existence of large quasi-60 year oscillations lasting centuries (for example, \cite{Scafetta2013al} found a quasi-60 year oscillation in the ice core GISP2 temperature record since 1350)   questions  interpretations such as  that proposed by \cite{Imbers} that the quasi 60-year modulation observed in the GST record since 1850 could be due to some kind of red noise produced by short memory processes exemplified by  AR(1) models or by long memory processes, which were claimed to modulate the internal variability of the global mean surface temperature. In fact, a generic stochastic model would not be able to reproduce a quasi harmonic signal lasting centuries in an energetically dissipative system such as the climate without the help of a specific harmonic forcing.

 Indeed, the climate system appears to be chaotically oscillating around a dominant complex harmonic component made of multiple specific frequencies plus some additional contribution from volcano and anthropogenic forcings. The internal variability more likely  chaotically perturbs  the oscillations, but does not produce them. This harmonic component looks complex but also predictable. It appears more similar in principle to the tidal oscillations of the ocean, which are forced by numerous astronomical harmonics, than to a hypothetical random internal unforced dynamics of the climate system.

Evidence has been presented   that the observed decadal and multidecadal oscillations might have an astronomical origin \citep{Scafetta2010,Scafetta2012a,Scafetta2012b,Scafetta2012c,Scafetta2012d,Scafetta2013a}.

For example, the quasi 9.1-year oscillation appears to be related to long solar/lunar tidal oscillations: see also \cite{Keelinga} and \cite{Wang}. The rationale is the following. The lunar nodes complete a revolution in 18.6 years, and the Saros soli-lunar eclipse cycle completes
 a revolution in 18 years and 11 days. These two cycles induce 9.3 year and 9.015 year tidal oscillations
 corresponding respectively to Sun-Earth-Moon and Sun-Moon-Earth tidal configurations. Moreover,
 the lunar apsidal precession completes one rotation in 8.85 years causing a corresponding lunar tidal cycle.
 Thus,  three interfering major soli-lunar tidal cycles clustered between 8.85 year and 9.3 year periods are expected, which should generate a major varying oscillation with an average period around 9.06 years. This soli-lunar tidal induced cycle  could peak, for example, in 1997-1998 when the solar and lunar eclipses occurred close to the equinoxes (this happens every $\sim 9$  years) when the soli/lunar tidal torque is reasonably  at the equator. In general, it is evident that the GST can be influenced by oceanic oscillations induced by soli-lunar gravitational tides, which produce a very complex set of harmonics at multiple time scales \citep{Keelinga,Keelingb,Wang}.

 The other decadal and multidecadal oscillations shown in Figure 1 appear to be mostly related to solar/planetary oscillations induced by Jupiter and Saturn and are seen  in the solar wobbling and in the solar activity.  These astronomical oscillations are  indicated by the black curves  depicted in Figure 2C and 2D which refer to the speed of the sun relative to the barycenter of the solar system.  Scafetta   analysed  several other multisecular proxy temperature models and also showed that it is possible to hindcast the 1950-2010 GST oscillations using a harmonic model calibrated on the period 1850-1950, and vice versa.

There is considerable empirical evidence  showing a strong correlation between climate and solar records at multiple scales  \citep{Hoyt,Bond,Kerr,Kirkby,Svensmark,Svensmark1,Steinhilber,Kokfelt}. Numerous authors  \citep[e.g.:][etc]{Scafetta2007,Scafetta2009,Kirkby}  have argued that to interpret recent paleoclimate temperature reconstructions and their patterns since the Maunder solar minimum (1640-1715), it is necessary to postulate a climatic response to solar variations  significantly larger than that predicted by the current GCMs. GCMs assume the existence of  a total solar irradiance (TSI) forcing  although this is a very small contribution to climate change \citep{IPCC}. However, a 1-3\%  astronomically-induced modulation of the Earth's albedo can easily provide the strong needed climatic amplification effect to solar variation up to a  factor of 10. For example \cite{Scafetta2012a} calculated   that  differentiating directly  the Stefan–Boltzmann’s black-body equation a climate sensitivity of $k_S=0.053~K/Wm^{-2}$ is found (this value uses the metric adopted in \citet{Scafetta2012a} which differs from the common metric). However, if  a solar activity variation  of 1 $W/m^2$ induces also  a 1\% variation of the albedo, a climate sensitivity of $k_S=0.36~K/Wm^{-2}$ would be found, which is about an order of magnitude larger than the previous value.

\begin{figure*}[!t]
\begin{center}
\includegraphics[angle=0,width=0.8\textwidth]{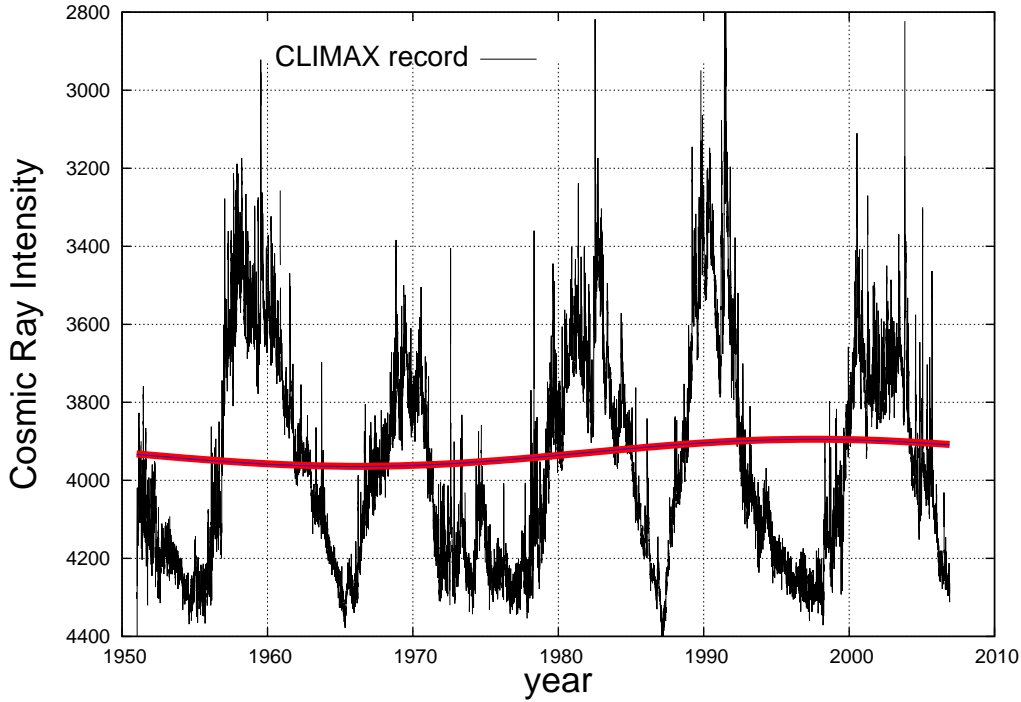}
\end{center}
\caption{CLIMAX cosmic ray count. The scale is inverted because cosmic ray flux is negatively correlated with solar activity.
The figure suggests that  solar activity might have increased on average from 1970 to 2000 (similarly to ACRIM TSI satellite composite \citep{Willson2003,Willson}) causing a global warming by inducing a decrease in the cloud cover (observed in the data, Figure 19) via a reduction of cosmic ray flux.  The red curve is a simple regression model made of a linear trend plus a 60-year oscillation used to highlight the multidecadal modulating pattern.
Data from {http://ulysses.sr.unh.edu/NeutronMonitor/Misc/neutron2.html} .}
\end{figure*}

 \cite{ScafettaW2013a} found evidence for a quasi 60-year oscillation (and others) in aurora records since 1530 linked to astronomical oscillations.
If  electromagnetic space weather mechanisms  induce small  oscillations in the upper strata of the atmosphere that drive coherent oscillations  (approximately 1-3\%) in the albedo by regulating the cloud cover system, this may suffice to produce GST oscillations synchronized with astronomical oscillations by means of an albedo-related modulation of the amount of solar radiation reaching and warming the surface \citep{Svensmark,Tinsley,Scafetta2012a}. This may happen because solar activity can modulate the incoming flux of galactic cosmic ray or other electromagnetic mechanisms related to the physical properties of the Parker spiral of the Sun's magnetic field as it extends through the solar system.  Despite the fact that preliminary  attempts to include some physical connections such as those between cosmic rays, ions, nucleation, and cloud drops have showed up to now a relatively weak model response \citep{Pierce}  there still exists much debate \citep{Svensmark2012} and  more advanced models and alternative space weather mechanisms may be better understood in the future. For example, \cite{Svensmark2013}  have recently discovered physical processes not included yet in current theoretical models. Moreover, solar UV radiation can also influence the stratospheric ozone variability \citep{Lu2009}. Indeed,  UV varies in percentage significantly more than total solar irradiance (TSI).  \cite{Reichler} have also proposed a  stratospheric direct driving of the  oceanic climate variability.

In support of the above theory,  Figure 19 shows the global surface temperature  plotted against the monthly variations in the total global cloud cover (TGCC) available since July 1983, obtained from the International Satellite Cloud Climatology Project (ISCCP) (\url{http://isccp.giss.nasa.gov/index.html}). The TGCC record is flipped upside-down for visual convenience. The temperature record is well correlated (negatively) with TGCC (correlation coefficient: $r_o=-0.52$, for 318 points $P(|r| \geq |r_o|)<0.0005$). In fact, TGCC decreases from  69\% to 64.5\% during the 1983-2000 warming period, and increased slightly from  64.5\% to 65.5\% during the 2000-2010 quasi-plateau temperature period. A similar pattern is observed in the record of total precipitable water (TPW) since 1988 \citep{Vonder}. A variation of a few percent in global cloud cover can easily cause a variation of a fraction of Celsius degree on the surface global temperature \citep{Scafetta2012b}. Moreover, \cite{Soon2011} showed a good correlation between a solar activity proxy model, the surface temperature of China (which also shows a clear cooling from 1940 to 1970) and a record of sunshine duration over Japan, which is  related to cloud cover variation since 1890 \citep{Stanhill}. These records also present a clear 60-year cyclical modulation synchronous to the temperature record.  The cooling from 1940 to 1970  was a global phenomenon \cite{LeMouel} also highlighted in the newspapers of the time \cite{Gwynee}. \cite{Xia} showed that cloud cover decreased slightly in China  during 1954-2005, although most decrease occurred from 1970 to 2000 when GST increased. Indeed, from 1970 to 2000 the cosmic ray count decreased slightly on average, as shown in Figure 20. According Svensmark's theory this would imply a decreasing cloud cover (as the data in Figure 19 show) and cause a global warming.

On the contrary, $CO_2$ atmospheric concentration and, in general, the net anthropogenic forcing monotonically increased since the 1980s \citep[e.g.][]{Hansen2011} and, after 2000, do not correlate with the observed temperature plateau.   The above results indicate that the cloud cover and the temperature are  responding to some other physical mechanism, possibly driven by  solar and lunar forcings \citep{Scafetta2010, Scafetta2012b},  rather than to the forcings currently included in the GCMs. Since 1980, the latter  are dominated by anthropogenic forcing which was still increasing from 2000 to 2013. In general, the net radiative forcings as used in the CMIP5 GCMs has increased since 2000, as implicitly demonstrated in the GCM simulations shown in Figures 16 and 17. Indeed, despite the importance of the cloud cover system in shaping the GST records, the CMIP5 GCMs are found to poorly reconstruct the cloud system \citep{Nam}.

\begin{figure*}[!t]
\begin{center}
\includegraphics[angle=0,width=1\textwidth]{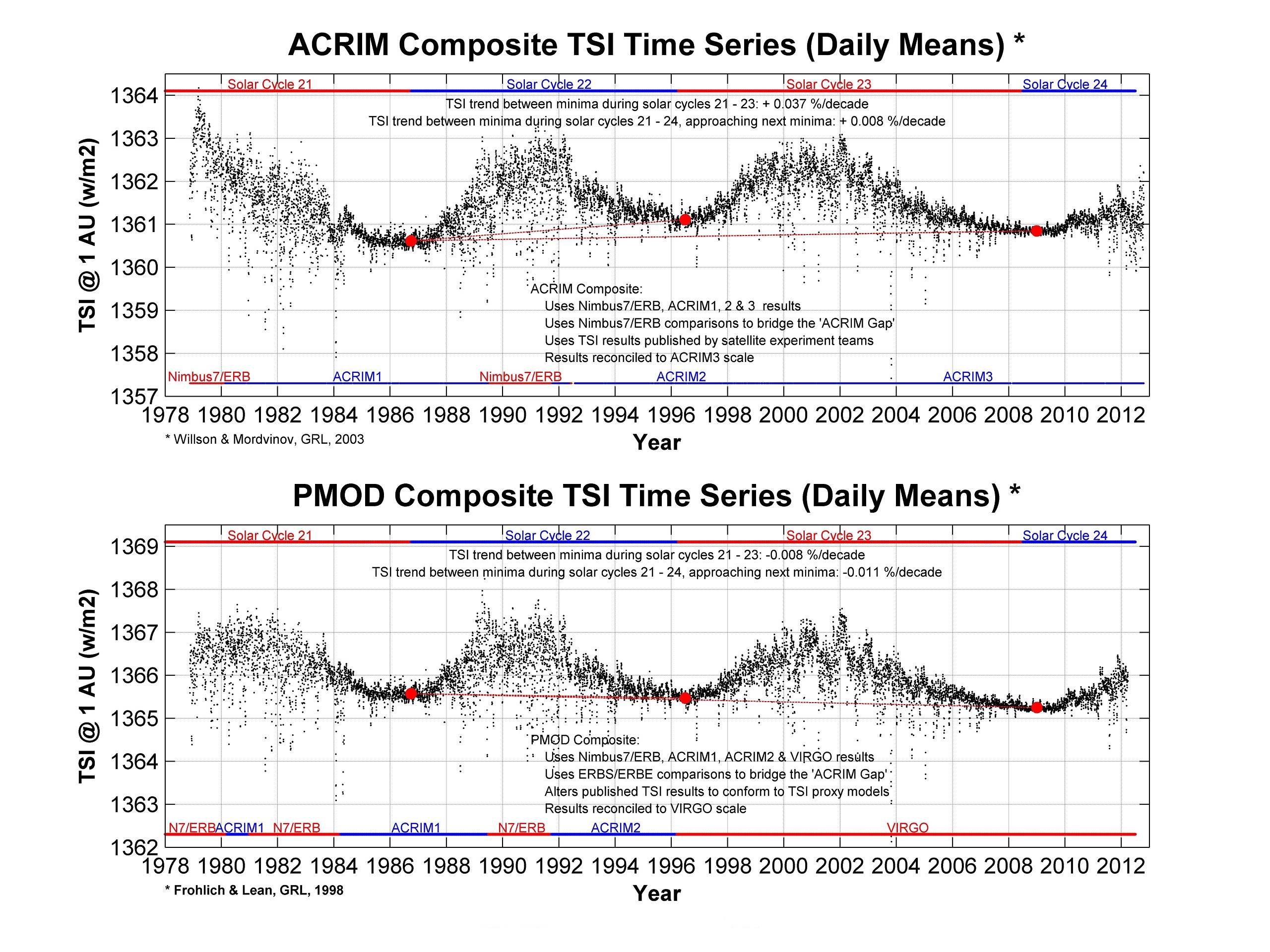}
\end{center}
\caption{ACRIM \citep{Willson2003} and PMOD \cite{Frohlich} total solar irradiance satellite composites.}
\end{figure*}

A serious source of uncertainty refers to   the  solar forcing functions that have to be used in the GCMs:  see also \cite{Gray} for a general discussion on the topic.  For example, CMIP5 GCMs used only a TSI forcing function deduced from \cite{Lean}, but TSI functions are currently extremely uncertain \citep[e.g.:][]{Hoyt1993,Lockwood2011,Shapiro}. Using correct  solar forcing functions  in the GCMs is fundamental if the climate system is very sensitive to solar variations as the above studies suggest.
Direct TSI satellite measurements  started in 1978. However,  an upward TSI trend from 1980 to 2000 followed by a decrease since 2000 is implied by the ACRIM TSI satellite composite \citep{Willson2003,Willson}, which uses the TSI experimental data as published by the original science teams. An alternative TSI satellite composite, the PMOD, based on altered TSI data \citep{Frohlich2006,Frohlich}, shows a gradual TSI decrease from 1980 to 2010. ACRIM and PMOD TSI satellite composites are compared in Figure 21.
Before 1980 only highly controversial solar proxy reconstructions exist.

The CMIP5 GCMs use the recommended TSI proxy model prepared by Lean and collaborators \citep{Lean,Kopp}, which does not show a TSI increase from 1980 to 2000, presents a peak around 1960, as also shown by the sunspot number record, and presents a relatively small secular variability since the Maunder solar minimum of the 17th century. On the contrary, some alternative TSI reconstructions present a larger secular variability  \citep{Hoyt1993,Hoyt}, peak in the 1940s and in the 2000, and correlate with the GST records far better than Lean's TSI proxy models \citep{Loehle,Soon2005,Soon2009,Soon2011}. Also solar cycle length models \citep{Thejll} peak in the 1940s instead of $\sim$1960. \cite{Schrijver},  \cite{Shapiro} and \cite{Vieira} have recently proposed quite contrasting TSI reconstructions with a range of secular variability from  very small to  very large secular variability and significantly differ from Lean's TSI models.

Figure 22A compares the latest Lean model \citep{Kopp} and  the model proposed by \cite{Hoyt1993}. Figure 22B shows that the solar model proposed by \cite{Hoyt1993,Hoyt} well correlates with the central England temperature (CET) reconstruction \citep{Parker} since 1700, suggesting a strong climate sensitivity to solar changes. Figure 22B suggests that the sun could have contributed about half of the 20th century warming in England: see \cite{Scafetta2013a,Scafetta2013} for additional details.  In any case, even if the proposed TSI reconstructions differ from each other in important details, there exists a general agreement that solar activity during the second half of the 20th century was higher than the previous centuries suggesting that the observed global warming since 1900 could have been partially caused by the increased solar activity \citep{Scafetta2007,Scafetta2009}.

\begin{figure*}[!t]
\begin{center}
\includegraphics[angle=0,width=0.7\textwidth]{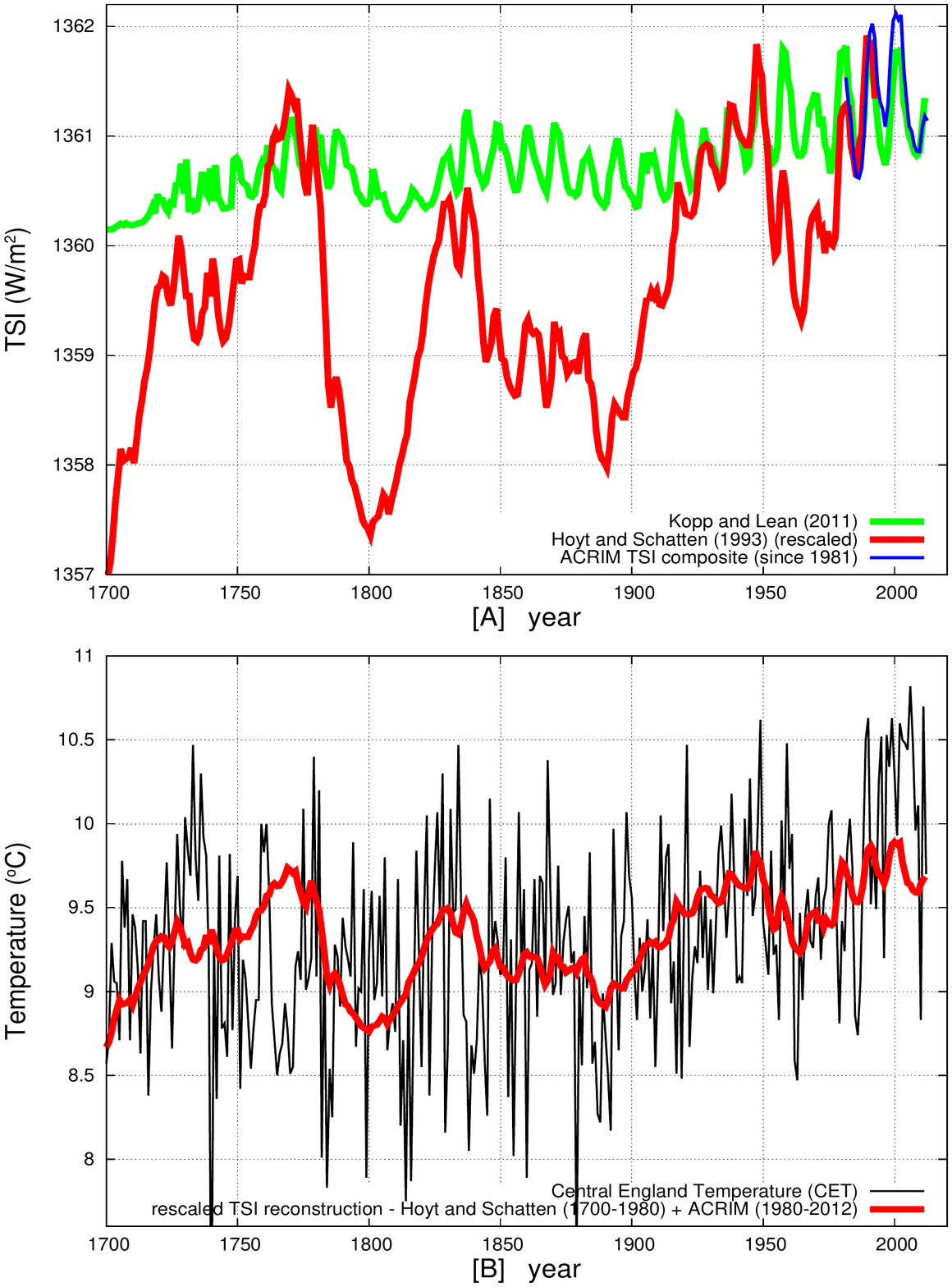}
\end{center}
\caption{[A] Total solar irradiance (TSI) reconstruction by \citet{Hoyt1993}
(red) rescaled on the updated ACRIM3 record \citep{Willson2003} (since 1981)
(blue) vs. the updated Lean model \citep{Wang,Kopp} (green). [B]
Comparison between the Central England Temperature (CET) record (black)
\citet{Parker} and the TSI model by Hoyt and Schatten merged with the ACRIM
TSI record since 1981. Good correlation is observed at least since 1772.}
\end{figure*}

Because the CMIP5 GCMs use Lean's TSI model, it is also  important to point out that despite serious controversy over the TSI dynamical behavior before 1992, the experimental TSI satellite groups  (ACRIM and PMOD) agree that the TSI minimum in 1996   was higher than the TSI minimum in 2008 by at least 0.2-0.3 $W/m^2$: see ACRIM and PMOD TSI satellite composites at \url{http://acrim.com/TSI%20Monitoring.htm}).
The open solar magnetic flux,  the galactic
cosmic ray (GCR) flux and other solar indexes also suggest that solar activity was higher in 1996 than in 2008 \citep{Lockwood,Schrijver,Shapiro,Vieira}.  However, the updated Lean's TSI proxy model \citep{Kopp} fails to reproduce this pattern by predicting  a 1996 TSI minimum  (1360.7370 $W/m^2$) lower than the 2008 TSI  minimum (1360.8217 $W/m^2$) \\
(\url{http://lasp.colorado.edu/data/sorce/tsi_data/TSI_TIM_Reconstruction.txt}).

Recently, \citet{Liu} used
the ECHO-G model and showed that to reproduce the $\sim0.7$ $^{o}C$
global cooling observed from the Medieval Warm Period (MWP: 900-1300)
to the Little Ice Age (LIA: 1400-1800) according to  recent paleoclimatic
temperature reconstructions \citep[e.g.: ][]{Christiansen,Ljungqvist2010,Mann3,Moberg},
a TSI model with a secular variability $\sim3.5$ times larger than
that shown by Lean's TSI model would be required.

Thus, there is a realistic  possibility that for their climatic simulations the current GCMs are not using sufficient solar-climate  physical mechanisms and, by adopting Lean's TSI model, are  not even using a sufficiently accurate solar radiative forcing record. In the next section,  an alternative solar model based on astronomical harmonic constituents will be used. This model  is constructed adopting a very different methodology than those used to construct the above TSI proxy models. A strength of the proposed harmonic solar model is that it has been shown to hindcast quite well major solar and climate patterns during the Holocene \citep{Scafetta2012c}.

The author notes that \cite{Benestad},  using theoretical results derived from the GISS GCM simulations,  criticized some of Scafetta's preliminary studies \citep{Scafetta2005,Scafetta2006} that demonstrated a significant solar contribution (up to 40-70\%) to the 1850-2000 warming: they claimed that the sun contributed about 7\% of the 20th century warming.   However,  GISS models do not reproduce the observed oscillations at multiple time scales \citep{Scafetta2010,Scafetta2012b} and cannot be used to validate or contradict studies based on data analysis.  \cite{Scafetta2009} confirmed his previous results with hindcast based models. Moreover,  \cite{Benestad}'s work also contains flawed models in  particular with regard to use of linear regression algorithms  and  the wavelet decomposition algorithm \citep{Scafetta2009b,Scafetta2013}. Linear regression algorithms are inefficient when the constructors are collinear, and during the 20th century the solar forcing is collinear with the anthropogenic forcing:  both trend upward. Using linear regression algorithms in  non-collinear situations  \cite{Scafetta2013} showed: (1)  GISS ModelE significantly underestimate the solar signature by a factor from 3 to 8; and (2)  modern paleoclimatic temperature reconstructions imply that the sun contributed significantly to the 20th century warming.  Moreover,  \cite{Benestad} used the periodic padding in the wavelet algorithm, which is highly inappropriate for decomposing trending sequences, instead of the reflection one, which minimizes Gibbs boundary artifacts. Figures 7 and 8 in \cite{Benestad} demonstrate the mathematical error: the increased solar activity from the solar minimum in 1995 to the solar maximum in 2002 is claimed to have induced a significant \textit{cooling} in the climate system, which would imply a nonphysical negative climate sensitivity to radiative forcing from 1995 to 2002. This is just a boundary artifact due to \cite{Benestad}'s erroneous implementation of the wavelet algorithm: see \cite{Scafetta2013,Scafetta2013d} for details.

In conclusion,  the GCM implemented analytical approach cannot take into account unknown physical mechanisms and uncertain forcings. On the contrary,  empirical modeling may reconstruct geometrical dynamical patterns, such as cycles, independently of their microscopic physical cause. It is important not to draw a logically flawed conclusion  that a strong climatic response to solar/astronomical inputs does not exist simply because current GCMs are not able to reproduce it   \citep{Lockwood}. Because of the existence of numerous physical uncertainties, it may be useful to investigate  the possibility of an empirical methodology alternative to the analytical GCM one.

\section{The astronomically-based empirical harmonic \\ model}

In the following subsections I summarize a number of pieces of empirical evidence suggesting a significant climate sensitivity  to astronomical/solar forcings. A semi-empirical  harmonic constituent  climate model is proposed. It is  made of specific astronomical oscillations that can be used to simulate natural climatic variability. The proposed model outperforms all CMIP5 GCMs.

\subsection{Millennial solar cycle, paleoclimatic temperature reconstructions, and their interpretation.}

Understanding the natural variability of the  climate of the past is necessary to properly interpret the climate changes occurred  since 1850. If   pre-industrial  climate changes were  similar to  those observed since the industrialization period, natural variability might have been the major determinant of the present climatic changes. On the contrary, if  the climatic changes that occurred since 1850 were anomalous relative to the preceding climate, this would support anthropogenic forcing as the major determinant of  the 20th century global warming.

 From 1998 to 2004 some preliminary studies claimed that the preindustrial GST since  Medieval times varied very little, by about 0.2 $^oC$, while GST anomalously increased since 1900   \citep{Mann1,Mann2}: the shape of these proxy temperature reconstructions resembled that of an \emph{hockey stick} with a MWP as warm as the 1900-1920 period. A number of climate model studies  concluded that the solar radiative forcing  plus  volcanic  and  anthropogenic forcings were sufficient to explain those paleoclimatic GST records for the last millennium. This type of analysis led to the conclusion  that the warming observed since 1900 could be due only to anthropogenic forcing \citep{Crowley,Shindell,Hegerl,Foukal}.

 \cite{Crowley} explicitly stated: \textit{``The very good agreement between models and data in the preanthropogenic interval also enhances confidence in the overall ability of climate models to simulate temperature variability on the largest scales''}, which suggests that in 2000 some climate scientists thought that the available climate models supporting the anthropogenic global warming theory for the 20th century were already sufficiently accurate (that is the science was considered sufficiently \emph{settled}) because of their ability to hindcast the \emph{hockey stick}  GST proxy reconstructions. This interpretation was strongly advocated and promoted by the IPCC in 2001 and 2007 and greatly contributed to support the anthropogenic global warming theory and the GCMs that predicted it.

\begin{figure*}[!t]
\begin{center}
 \includegraphics[angle=0,width=0.8\textwidth]{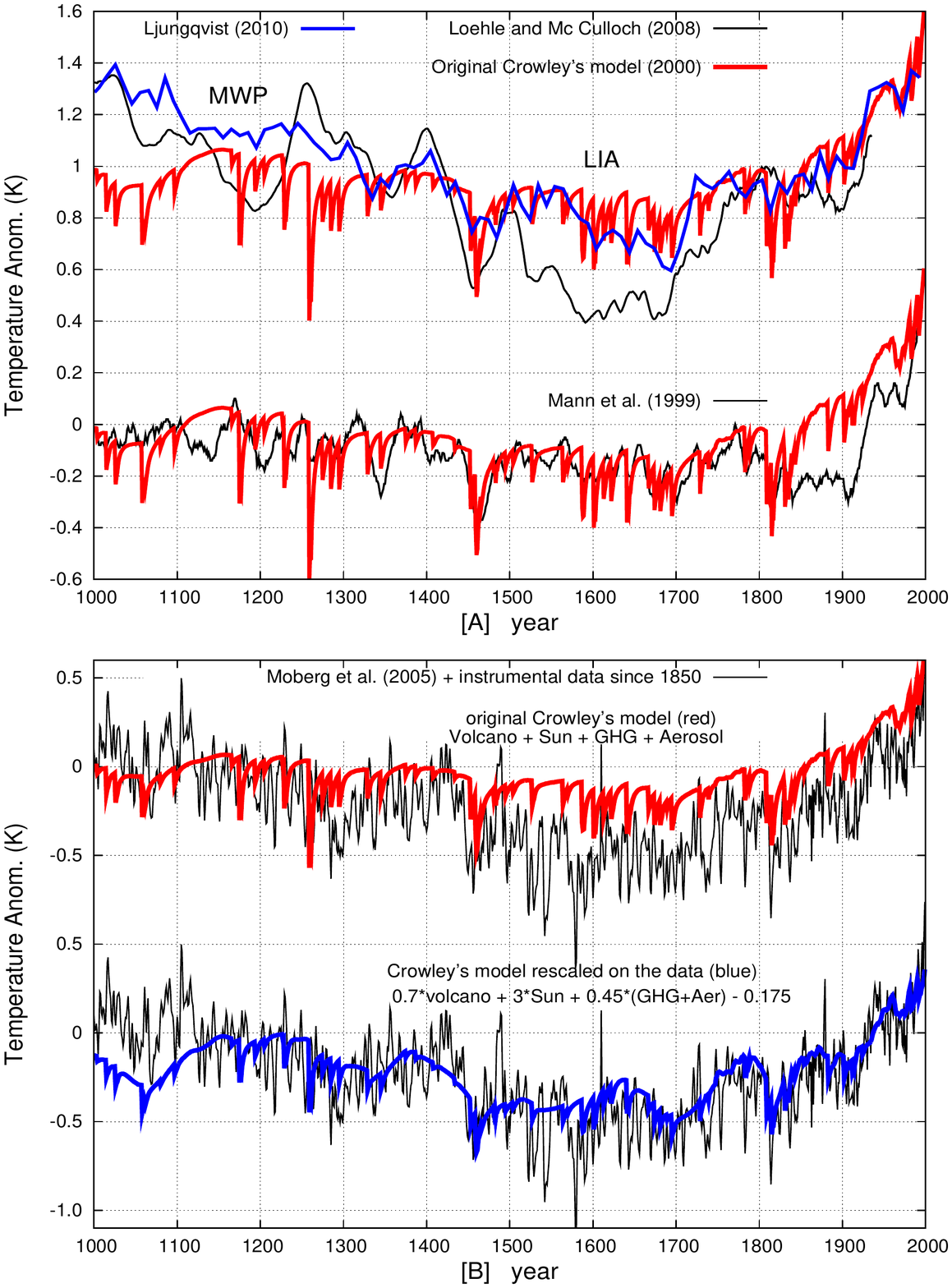}
\end{center}
 \caption{ [A] Comparison between the original energy balance model prediction by \cite{Crowley} versus the \emph{hockey stick} temperature graph by \cite{Mann1} implying a MWP as warm as the 1900-1920 period, and two non-\emph{hockey stick} recent paleoclimate GST reconstructions \citep{Loehle1,Ljungqvist2010} showing a far larger preindustrial variability and a MWP as warm as the 1940-2000 period. [B](Bottom) the volcano, solar and GHG+Aerosol temperature signature components produced by  \cite{Crowley} model are scaled to fit  \citep{Moberg} paleoclimate GST reconstruction corrected since 1850 by HadCRUT4, which also shows a MWP as warm as the 1940-1970 period. See \cite{Scafetta2013a,Scafetta2013} for more details.}
\end{figure*}

However, since 2005 a number of studies have demonstrated a larger global pre-industrial temperature variability. For example, it was found a cooling  of about 0.4-1.0 $^oC$ from the MWP to the LIA and a MWP as warm as the 1950-2000 period \citep{Moberg,Mann3,Loehle1,Kobashi,Ljungqvist2010,McShane2011,Christiansen}.  The new emerging millennial GST pattern  stresses  the existence of a large millennial climatic oscillation. A quasi millennial climatic oscillation  is also found to correlate well with the millennial solar oscillation observed throughout the Holocene  \citep{Bond,Kerr,Ogurtsov,Kirkby,Steinhilber,Scafetta2012c}.

The climate models   that predicted a very small natural variability and    that were used to  fit  the \textit{hockey stick} temperature records can not fit the recent proxy GST reconstructions casting doubts on their accuracy. Still recent millennium simulation  studies using modern solar models  \citep[that is,][]{Lean}  are able to predict only hockey-stick temperature graph showing  an average cooling from  the 900-1300 MWP to the 1300-1800 LIA up to $\sim$0.3 $^oC$, and just half of the empirically measured 11-year solar signature on the climate (see \cite{Feulner} and \cite{IPCC}  figure 6.14: \url{http:/www.ipcc.ch/publications_and_data/ar4/wg1/en/figure-6-14.html}). These  results are unsatisfactory, as also \cite{Liu} noted by demonstrating the need of assuming  a far stronger solar effect to properly interpret the modern paleoclimatic temperature reconstructions.

 For example, \cite{Ljungqvist2010} and \cite{Christiansen} estimations of the last 2000 years of extra-tropical Northern Hemisphere (30-90 $^oN$) decadal mean temperature variations
present two large millennial cycles. These GST reconstructions claim  that  the Roman Maximum and the MWP were as warm as today's temperatures. Indeed, these reconstructions might be quite plausible because they   agree with inferences deduced from numerous historical documents \citep{Guidoboni}. For example, the medieval Vikings'  villages in Greenland clearly indicate a MWP warmer than today's temperature at least in the North Atlantic \citep{Esper,Surge}.  However, a medieval warm period does not appear to be limited to the Northern Atlantic region. Similar evidence exists also for China \citep{Ge},  South America \citep{Neukom}, South Africa \citep{Tyson}   the Indo-Pacific region \citep{Oppo} and other locations \citep{Soon2003}. Thus, the MWP phenomenon was likely  more global than  was believed in 2000.

Figure 23  illustrates the paradigm-shift  issue related to the current understanding of the historical climate. Figure 23A depicts the original climate model by \cite{Crowley} against the temperature reconstruction by \cite{Mann1} (note the good pre-1900 fit) and against two non \emph{hockey-stick} temperature reconstructions \citep{Loehle1,Ljungqvist2010} (note the poor fit). Figure 23B depicts the temperature model by \cite{Moberg} (1000-1850) merged with the historical GST measurements since 1850  against the original climate model by \cite{Crowley}  (note the poor fit) and against an empirical model made by simply rescaling via linear regression  the same climatic (solar, volcano and GHG+Aerosol) components  predicted by Crowley's model  in such a way as to best fit the depicted temperature record (note the recovered overall good fit). The mathematical formula used in the regression model is reported in the figure: see \cite{Scafetta2013a,Scafetta2013} for an extended discussion on this exercise.

The rescaled climate model indicates that for reproducing recent paleoclimate temperature reconstructions with their larger millennial GST cycle,  the solar impact on the climate needs to be increased by at least a factor of three relative to the Crowley's original estimate, which was already twice that predicted by the current CMIP3 GCM models: see  \cite{Scafetta2013} for additional details. This  also means that a significant fraction of the warming observed since 1900 (up to around 50\% using different solar models) can be ascribed to the sun, as was calculated in \cite{Scafetta2007} and \cite{Scafetta2009} using alternative methods. The volcano effect needs to be reduced by  30\%, and the anthropogenic forcing effect (GHG plus Aerosol forcing) needs to be reduced by about 50\%. The latter result well agrees  with the correction implemented in \cite{Scafetta2012b} that used  an alternative reasoning based on the existence of a 60-year natural oscillation from 1970 to 2000 not modeled by the GCMs.  The above finding also quantitatively  confirms \cite{Eichler} and \cite{Zhou}, and  contradicts the \cite{IPCC}, \cite{Benestad} and \cite{Lean2008}, which claimed that 90\% or more of the 20th century warming had to be  caused by anthropogenic activity.

In conclusion,  around 2000 \emph{hockey-stick} shaped GST graphs implied a very small natural climatic variability (and a small solar effect) and a strong anthropogenic effect on climate. That evidence was consistent with the outputs of preliminary energy balance models, and is still consistent with the predictions of the CMIP3 and CMIP5 GCMs. However, recent paleoclimatic GST graphs have demonstrated a far larger preindustrial natural climate variability. The new evidence shifts the scientific paradigm.  The climate should be highly sensitive to solar/astronomical related forcings because the novel GST reconstructions show a  large millennial cycle that well correlates with  solar/astronomical records  \citep{Bond,Kerr,Kirkby,Ogurtsov,Steinhilber}.  Consequently, the current GCMs should overestimate the anthropogenic effect on climate.

As also commented in Scafetta (2013a), \cite{Crowley} and others  would have had a significantly lower confidence in the overall ability of climate models to simulate temperature variability if in 2000 the current paleoclimatic temperature reconstructions had been available. The scientific community would have more likely concluded that important astronomically-related climate change mechanisms were still unknown, and needed to be investigated before they could be implemented to make reliable analytical GCMs.

\begin{figure*}[!t]
\begin{center}
 \includegraphics[angle=0,width=0.8\textwidth]{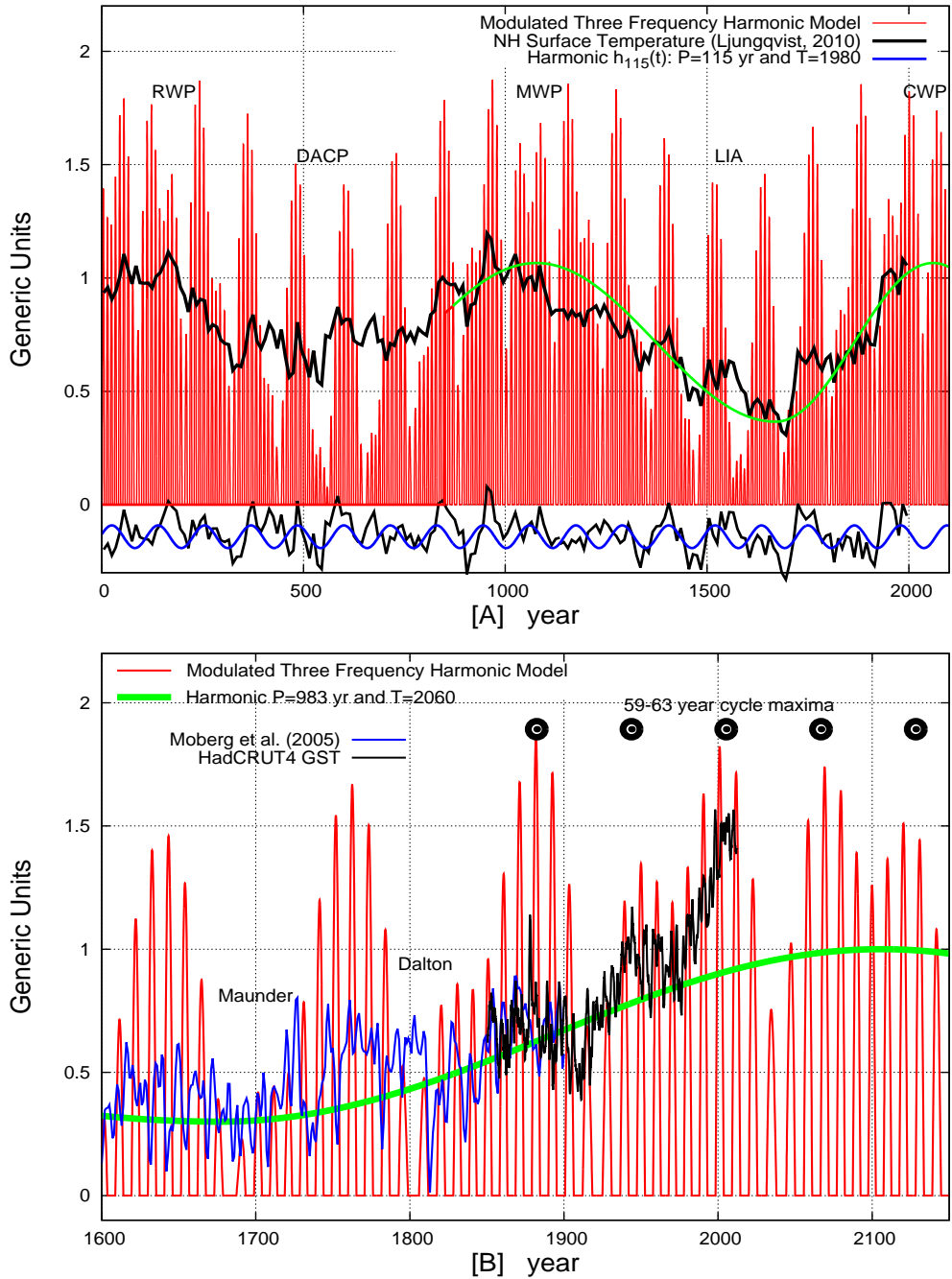}
\end{center}
 \caption{\cite{Scafetta2012c} three-frequency solar model (red). [A] Against the Norther Hemisphere temperature reconstruction by \cite{Ljungqvist2010} (black). The bottom depicts a filtering of the temperature reconstruction (black) that highlights the 115-year oscillation, $h_{115}(t)$, (blue). [B] The same solar model (red) is plotted against the HadCRUT4 GST (black) merged in 1850-1900 with the proxy temperature model by \cite{Moberg} (blue). The green curves highlight the quasi millennial oscillation, $h_{983}(t)$, with its skewness that approximately reproduces the millennial temperature oscillation. Note the hindcast of the Maunder and Dalton solar mimima and relative cool periods, and the projected quasi 61-year oscillation from 1850 to 2150. Adapted from \cite{Scafetta2013a}.}
\end{figure*}

\subsection{Construction of the astronomical/solar harmonics}

 \cite{Scafetta2010,Scafetta2012a,Scafetta2012b}   proposed that  the quasi 60-year  GST oscillation observed during  1850-2012, which has an amplitude of about 0.3 $^oC$ (see also Figure 2A), could  indicate that about  50\% of the 0.5 $^oC$ warming observed from 1970 to 2000 could have been  due to this natural oscillation during its warming phase. \cite{Zhou} reached a similar result using the Atlantic Multidecadal Oscillation, which also presents a clear quasi 60-year oscillation \citep[e.g.:][]{Morner1989,Morner1990,Manzi}, as a constituent regression model component to reconstruct the GST record.

 The existence of a large natural oscillation causing about  50\% of the 1970-2000 warming would mean that the net anthropogenic effect on the climate has been overestimated by the GCMs by at least the same percentage, and needs to be reduced on average by about a 0.5 factor, as alternatively demonstrated above (Section 6.1)  using an approach based on recent paleoclimate temperature records. Consequently,  also a significant fraction of the 1850-2013 warming (about 0.40-0.45 $^oC$) could not be reconstructed by the same GCMs.  Note that  part of the residual warming could also be due to poorly corrected urban heat island (UHI) and land use change (LUC) effects  \citep{McKitrick2007,McKitrick2010,Loehle}. Thus, a reduction to a 0.5 factor of the output of the GCMs may  be considered an upper limit. However, herein    such a hypothesis is not taken into consideration and the GST records are assumed to show  true climatic changes.

Figures 1 and 2 and Eq. 1-4 proposed a possible astronomical origin of the decadal and multidecadal GST oscillations. Interestingly, natural climatic oscillations linked to astronomical cycles with multidecadal periods of about 20 and 60 years, and longer secular and quasi-millennial cycles appear to have been  well-known in  ancient times, and in the Middle Ages through the  Renascence. These astronomical oscillations  could be easily deduced from the conjunction periods of Jupiter and Saturn and from their dynamical rotation along the Zodiac. These oscillations were included, for example, in Chinese and Indian calendars and  constituted the basis for a kind of astrological climatology. Indeed,  these oscillations could be observed in climate records, e.g. in the monsoon oscillations, and inferred from historical chronologies describing the fall and  rise of human civilizations \citep{Ptolemy,Masar,Kepler,Iyengar}: see more details about the ancient understanding of climate changes in \cite{Scafetta2013a}.

A link between planetary oscillations and climatic cycles  could be  indirect. Planetary oscillations may modulate solar changes that then induce climatic changes. Indeed,
\cite{Scafetta2012c} analyzed in details the sunspot number record and  noted that the 11-year Schwabe sunspot cycle is made of at least three harmonics interfering together at about 9.93 years, 10.87 years, and 11.86 years. The harmonic at 9.93 years corresponds to the Jupiter/Saturn spring tidal period and the 11.86 year corresponds to the Jupiter orbital period. The central period at 10.87 years could be generated by the solar dynamo itself by means of a dynamical synchronization process with the other two cycles or by a combination of the recurrent tidal cycles produced by Venus, Earth and Jupiter that present peaks at about 10.4 and 11.1 years. This finding suggests a planetary modulation of solar activity, as first proposed by \cite{Wolf}, whose physical mechanisms and additional empirical evidences are extensively discussed in a number of publications  \citep{Abreu,Brown,Fairbridge,Hung,Landscheidt,Wolff,Scafetta2010,Scafetta2012a,Scafetta2012c,Scafetta2012d,ScafettaW2013a,ScafettaW2013b}.

In particular \cite{Scafetta2012d} argued that the Sun might be working as a huge amplifier of planetary gravitational oscillations because the planetary tidal work released to the sun, although quite small,  could nevertheless be greatly amplified up to a 4 million factor by triggering  a modulation of the core nuclear fusion rate. Electromagnetic planet-sun interactions can also be hypothesized \citep{ScafettaW2013a,ScafettaW2013b}. Preliminary calculations suggests that Scafetta's model could produce luminosity oscillations up to one order of magnitude compatible with  the observed TSI oscillation. Such signal could be sufficiently powerful to modulate the solar dynamo mechanisms and produce a final TSI output  approximately synchronized to planetary harmonics.

The  three harmonics at  9.93 years, 10.87 years, and 11.86 years beat together forming a complex dynamics as shown in Figure 24 (red curve). Four major additional multidecadal, secular and millennial solar/astronomical oscillations emerge: a quasi 61-year oscillation (maximum around 2002), a 115-year oscillation (maximum around 1980) and a minor 130-year oscillation (maximum around 2035),  and a large quasi 983-year oscillation (maximum around 2060). The beats of \cite{Scafetta2012c} three-frequency model well correlate with the observed major solar and climatic variations for millennia, throughout the Holocene: see also Figure 24 and the extended discussion in \cite{Scafetta2012c}.

A 115-year oscillation can be observed in proxy temperature models going back for 2000 years \citep[e.g.:][]{Ogurtsov,Qian}, and can be correlated with grand-solar minima such as the Maunder, Dalton and the solar minimum around 1910 and other grand solar minima during the last 1000 years.  The 115-year oscillation is projected to reach a minimum in 2030-2040. Figure 24A suggests that this oscillations may be characterized by a temperature variation between 0.05 and 0.15 $^oC$. This cycle can be approximated as

\begin{equation}\label{}
  h_{115}(t)=0.05 \cos(2\pi(t-1980)/115).
\end{equation}

A great millennial oscillation is observed throughout the Holocene \citep{Bond,Kerr,Ogurtsov,Kirkby,Steinhilber,Scafetta2012c} and  was responsible for the Roman Warm Period, the Dark Age Cold Period, the Medieval Warm Period, the Little Ice Age and the Current Warm Period, and would peak around 2060: see Figure 24A.   As evident from poleoclimatic temperature proxy models, the millennial oscillation is skewed by other multisecular harmonics  with a likely minimum around 1680 during the Maunder Solar Minimum: see also \cite{Humlum} and Figures 23 and 24.   Assuming that the millennial temperature oscillation presents a variation of about $0.7\pm0.3$ $^oC$, as approximately shown in \cite{Ljungqvist2010} and in \cite{Moberg} (see Figures 23 and 24),  it may be approximately   modeled from  1680 to 2060 as

\begin{equation}\label{}
  h_{983}(t)=0.35 \cos(2\pi(t-2060)/760),
\end{equation}
where the adoption of the shorter period of  760 years  takes into account the skewness of the millennial cycle with a minimum in the middle of the Maunder solar minimum around 1680 and  a predicted maximum in 2060. Note that if the cloud system is modulated by these solar/astronomical cycles, the GST could be modulated by these oscillations relatively quickly with a time lags spanning from a few months to just a  few years as the frequency decreases \citep{Scafetta2008,Scafetta2009}.

  Figure 24A shows the proposed solar model versus   the extra-tropical Norther Hemisphere temperature reconstruction by \cite{Ljungqvist2010} (black). Note the two synchronous quasi millennial cycles and the common 115-year oscillation modulation, which is also highlighted at the bottom of the figure with an appropriate filtering of the temperature record. The blue curve at the bottom is the function $h_{115}(t)$. The figure highlights also the Roman Warm period (RWP), Dark Ages Cold Period (DACP), Medieval Warm Period (MWP), Little Ice Age (LIA) and Current Warm Period (CWP). Figure 24B depicts the  solar model  (red) versus  HadCRUT4 (annual smooth: black) merged in 1850-1900 with the proxy temperature model by \cite{Moberg} (blue). Note the synchronous occurrence of both the colder periods during the Maunder and Dalton modeled solar minima, and the quasi 61-year modulation from 1850 to 2010. The solar model predicts a $\sim$61-year oscillation from 1850 to 2150 whose maxima are highlighted by the black circles. Before 1850, the 61-oscillation weakens. The Sun may be entering into a (minor) grand minimum centered in the 2030s.   Indeed, sunspot cycles 19-23 (1955-2008) resemble sunspot cycles 1-4 (1755-1798) that preceded the Dalton solar minimum (1790-1830), and sunspot cycle 24 (2008-2021?) is approximately replicating the low  sunspot cycle 5 \citep[][Fig. 10]{Scafetta2012c}. The millennial modulation, $h_{983}(t)$, is also highlighted in the figure (green).

\begin{figure*}[!t]
\begin{center}
 \includegraphics[angle=-90,width=0.9\textwidth]{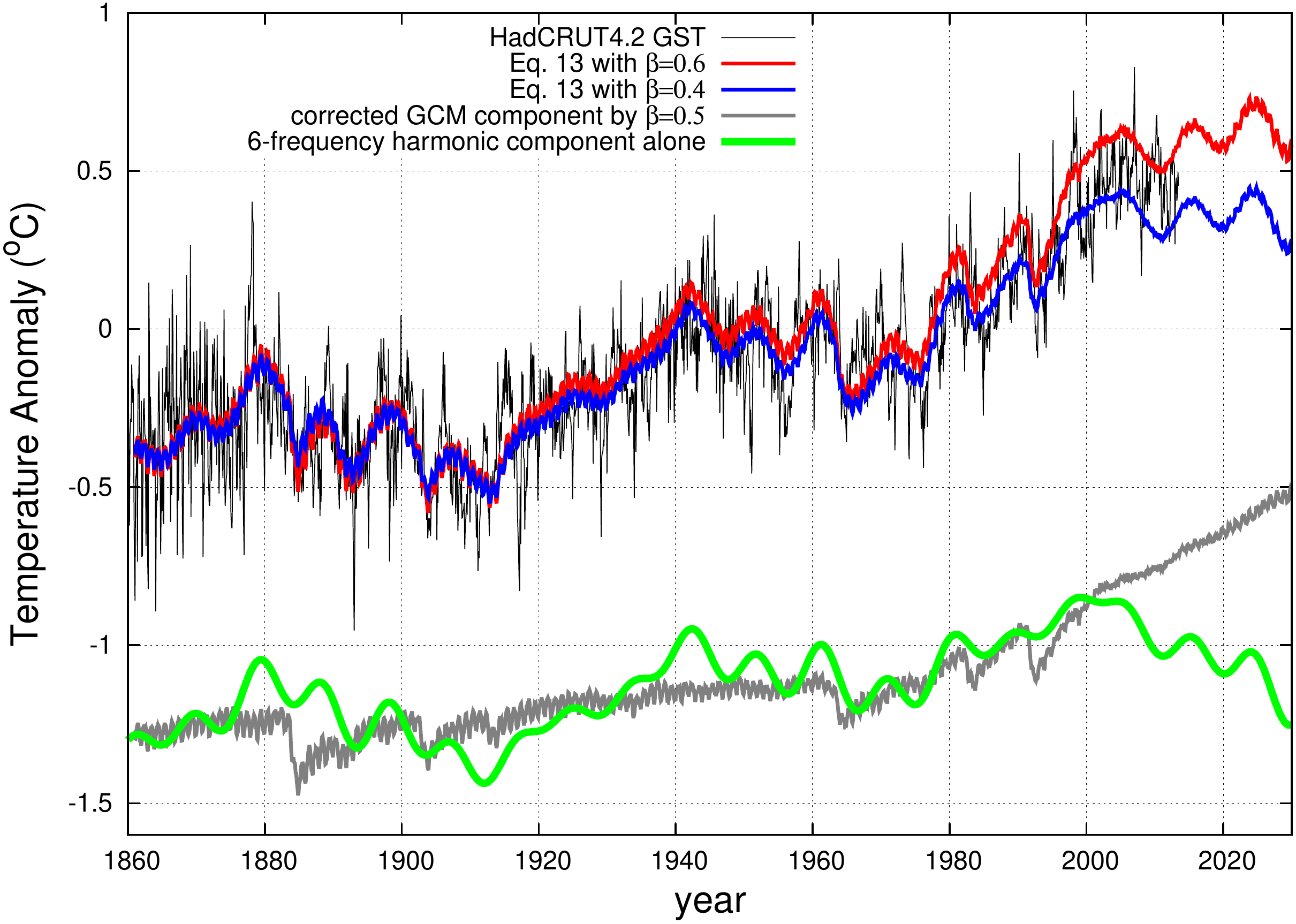}
\end{center}
 \caption{The semi-empirical model, Eq. \ref{eqg33}, using $\beta=0.4$ (blue) and $\beta=0.6$ (red) attenuation of the CMIP5 ensemble mean simulation vs.  HadCRUT4 GST record since 1860. The upward warming trending and all decadal and multidecadal patterns are well reconstructed. The bottom depicts the 6-frequency harmonic component that models the harmonic natural variability (green) and the GCM radiative component corrected by  $\beta=0.5$  (grey), respectively. }
\end{figure*}

\subsection{The six-harmonic astronomical/solar model for climate change}

With the above information a first approximation six-harmonic astronomical/solar model for climate change can be constructed, which phenomenologically simulates the  corresponding natural oscillations that the GCMs are currently not able to reproduced. The additional radiative forcing component (e.g. GHG, aerosol, volcano effects) can to a first approximation be simulated by using the CMIP5 mean projections reduced by a given factor $\beta$. Thus, the semi-empirical model is given by the equation:

\begin{equation}\label{eqg33}
\begin{split}
    H(t)=h_{983}(t)+ h_{115}(t) + h_{60}(t) + h_{20}(t)+h_{10.4}(t)\\+h_{9.1}(t)+ \beta*m(t) + const,
    \end{split}
\end{equation}
where the function $m(t)$ is a CMIP5  ensemble mean simulation depicted in Figure 2.

 Figure 25 shows that to  reproduce  accurately the HadCRUT4 GST warming trend since 1850 it is necessary to use  $\beta=0.5\pm0.1$. This result is reasonably  compatible with that found in \cite{Scafetta2012b} where a value of $\beta=0.45\pm0.05$ was chosen using the CMIP3 models and the HadCRUT3 GST, with a slight lower secular global warming than the HadCRUT4 GST record. \cite{Scafetta2012b} determined the coefficient $\beta$ only using an argument based on the GST residual from the harmonic model during the  period 1970-2000. Indeed, because from 1970 to 2000 the HadCRUT4 GST warms a little bit more than the HadCRUT3 record, the same argument used in \cite{Scafetta2012b}  but applied to the HadCRUT4 record  yields  $\beta\approx 0.5$. The fact that with $\beta=0.5$ Eq. \ref{eqg33} simultaneously  reconstructs both the 1970-2000 calibrating period and the entire 1850-2013 period validates the model with a hindcast test.

 The result implies that the climate sensitivity to radiative forcing has been overestimated by the CMIP5 GCMs by  about a factor of 2. Thus, the climate sensitivity to $CO_2$ doubling should be reduced from the \cite{IPCC} claimed 2.0-4.5 $^oC$ range to a 1.0-2.3 $^oC$ range with a likely median of $\sim1.5$ $^oC$ instead of $\sim3.0$ $^oC$.

\begin{figure*}[!t]
\begin{center}
 \includegraphics[angle=-90,width=1\textwidth]{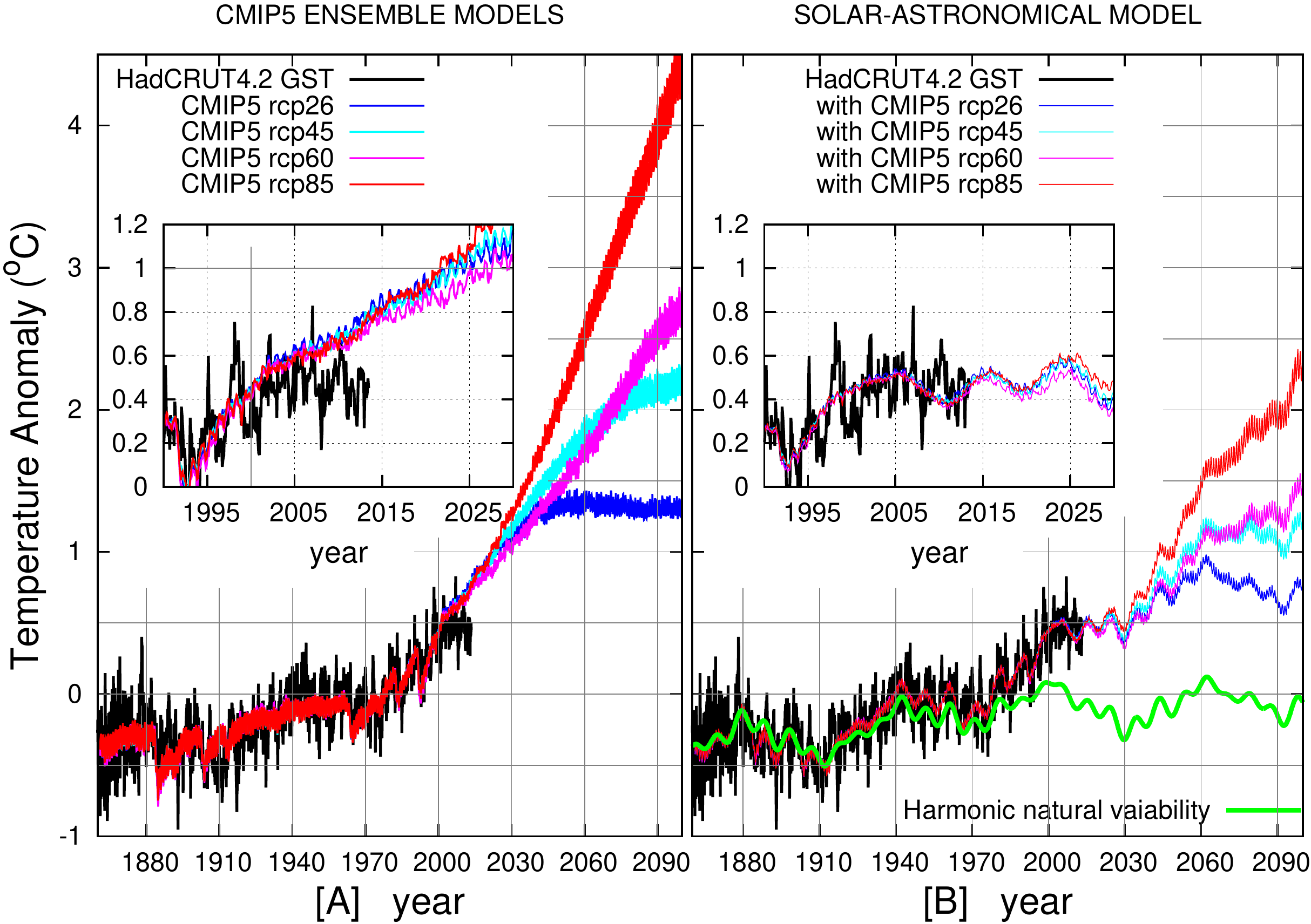}
\end{center}
 \caption{ [A] The four CMIP5 ensemble average projections versus the HadCRUT4 GST record. [B] The solar-astronomical semi-empirical  model, Eq. \ref{eqg33} with $\beta=0.5$, against the HadCRUT4 GST record: a common 1900-2000 baseline is used. The figure highlights the better performance of the solar-astronomical semi-empirical  model versus the CMIP5 models, which is particularly evident since 2000 as shown in the inserts.}
\end{figure*}

 A reduction by about half of the climate sensitivity to radiative forcing is  compatible with the results discussed in the subsection 6.1 and shown in Figure 23B concerning the interpretation of modern paleoclimatic temperature reconstructions with a larger pre-industrial variability with a MWP as warm as the 1950-2000 period. The result is also consistent with those  by \cite{Chylek2008} who found a climate sensitivity between 1.3 $^oC$ and 2.3 $^oC$ due to doubling of atmospheric concentration of $CO_2$, by \cite{Ring} who found a climate sensitivity in the range from about 1.5 $^oC$ to 2.0 $^oC$,  and by  \cite{Lewis} who found a climate sensitivity range from 1.2 $^oC$ to 2.2 $^oC$ (median 1.6 $^oC$). The result is also consistent with \cite{Zhou} and \cite{Chylek2013}, who found that the anthropogenic warming has been overestimated by about a factor of 2.

The quasi 60-year harmonic component of the model may be sufficiently valid for the period 1880-2150 because \cite{Scafetta2012c} three-frequency solar model   predicts that during this period the major natural patterns in the solar dynamics may be described by a quasi 60-year oscillation slightly  modulated by the 115-year oscillation, while the millennial oscillation is at its maximum.

The author notes  that the CMIP5 models contain also a solar signature, but it is very small because the CMIP5 adopted TSI  \citep{Lean} has a smaller secular trend than the TSI records adopted for the CMIP3 GCMs in the \cite{IPCC}. Herein,  this correction is ignored because  within the 20\% error associated to the $\beta=0.5$ factor, although the presence of a solar signature in $m(t)$ may slightly amplify the decadal cycle in $H(t)$. Note that Eq. \ref{eqg33} is not optimized for periods before 1850 because no GST are available before 1850, and the climate response to the deep grand solar minima (Dalton, Maunder etc.) could necessitate additional frequencies such as a 80-90 year oscillation \citep{ScafettaW2013a} and probably the inclusion of nonlinear dynamical effect.   The 80-90 year cycle  modulates the ~60 year cycle by producing a beat with a period of ~210 years known as the Suess (a.k.a. de Vries) solar cycle.

\begin{figure*}[!t]
\begin{center}
 \includegraphics[angle=-90,width=1\textwidth]{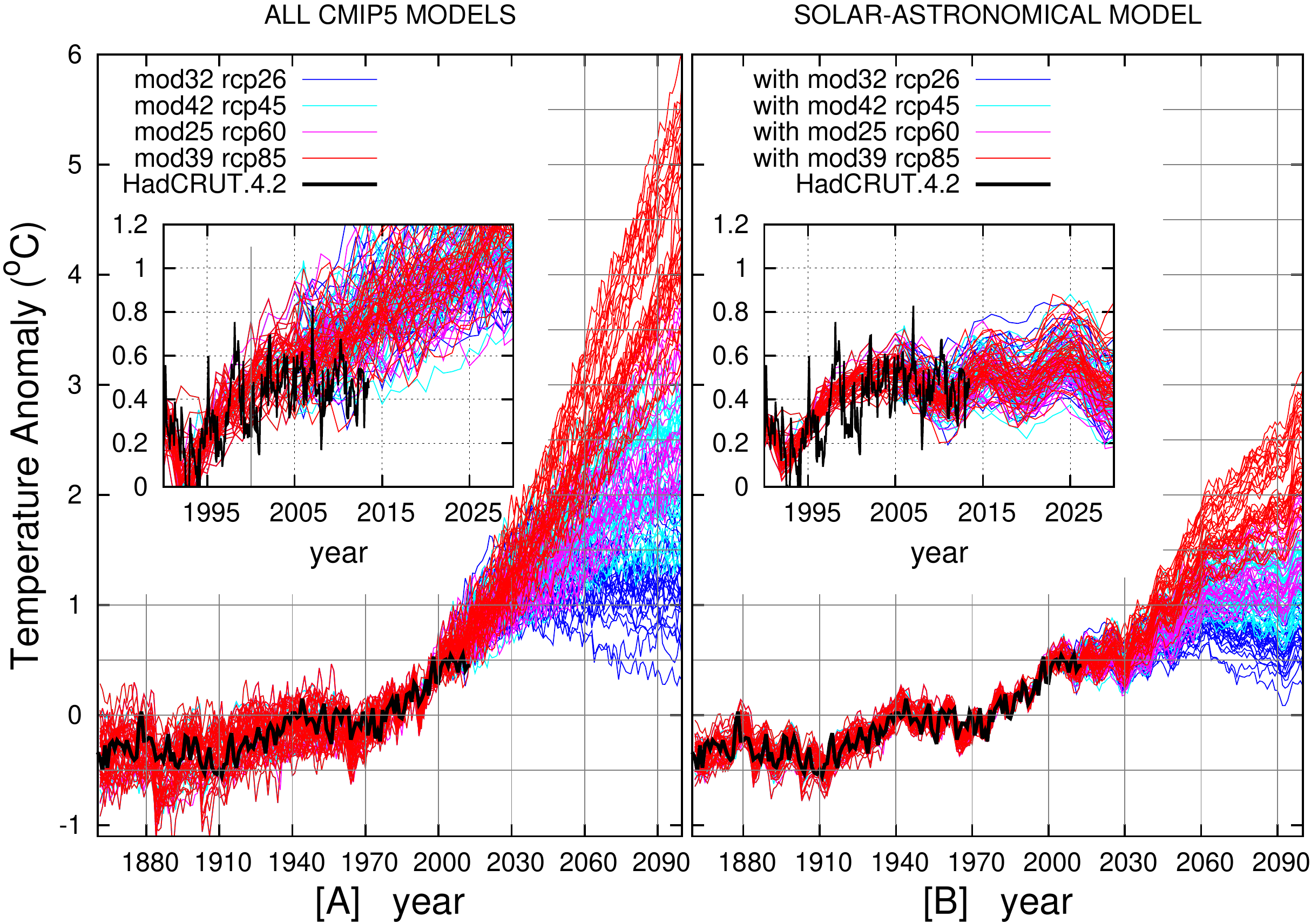}
\end{center}
 \caption{[A] All CMIP5 model projections versus the HadCRUT4 GST record. [B] The solar-astronomical semi-empirical  model, Eq. \ref{eqg33} with $\beta=0.5$ , against the HadCRUT4 GST record: a common 1980-2000 baseline and annually resolved records are used in the large figures while the monthly GST record is used in the inserts. The figure highlights the better performance of the solar-astronomical semi-empirical  model versus the CMIP5 models.}
\end{figure*}

Figure 26 compares the four CMIP5 ensemble average projections  (panel A) and the solar-astronomical semi-empirical  model using $\beta=0.5$ in Eq. \ref{eqg33} (panel B) against the HadCRUT4 GST record: a common 1900-2000 baseline is used. The figure highlights the superior performance of the solar-astronomical semi-empirical  model versus the CMIP5 ensemble mean models. Eq. \ref{eqg33} reconstructs  all major decadal and multidecadal temperature patterns observed since 1860. The temperature plateau observed since 2000 is better reconstructed by the semi-empirical model than by the GCM mean projections. Thus, the 2000-2013 GST plateau appears to be due to the cooling phase of a natural quasi 60-year oscillation that has balanced a strongly reduced projected anthropogenic warming trend. The four adopted CMIP5 GCM mean projections display a 2000-2100 warming between a minimum of about 1 $^oC$ to a maximum of about 4 $^oC$ (see figure 26A). However, as Figure 26B shows, the four corrected projections  predict a 2000-2100 warming between a minimum of about 0.3 $^oC$ to a maximum of about 1.8 $^oC$ using the same anthropogenic scenarios. The inserts magnify the period 1990-2030 to highlight  the strong mismatch between the GST and the GCM simulations since 2000 that  is resolved with the solar-astronomical semi-empirical model.

\begin{figure*}[!t]
\begin{center}
 \includegraphics[angle=-90,width=0.9\textwidth]{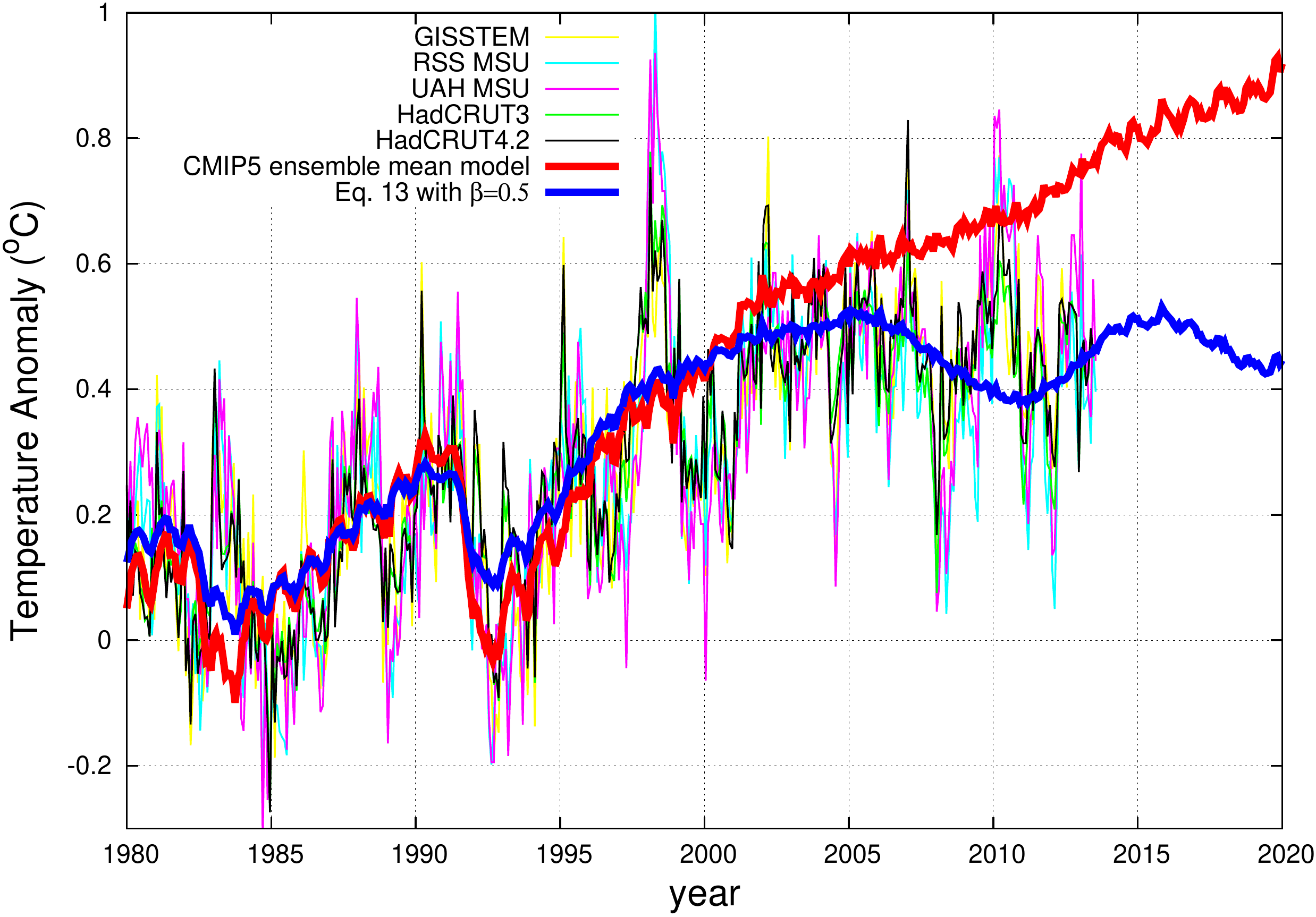}
\end{center}
 \caption{  Eq. \ref{eqg33} with $\beta=0.5$ (blue) and the original CMIP5 ensemble mean model (red) against six global temperature estimates (HadCRUT3, HadCRUT4, UAH MSU, RSS MSU, GISS and NCDC), which were  base-lined with HadCRUT4 from Jan/1980 to Dec/1999.  Temperature data from: {http://www.ncdc.noaa.gov}, {http://www.metoffice.gov.uk}, {http://data.giss.nasa.gov}, {http://www.remss.com/}, {http://vortex.nsstc.uah.edu/}
}
\end{figure*}

Figure 27 is similar to Figure 26 but with all CMIP5 model individual simulations used in Eq. \ref{eqg33} with $\beta=0.5$. Again, the figure highlights the superior performance of the solar-astronomical semi-empirical  model versus the CMIP5 model simulations. The  solar-astronomical semi-empirical  model also causes a reduction of the range among the simulation by a factor of two.
The root-mean-square deviation (RMSD) value between the 49-month average smooth GST and the full empirical model is about 0.04 $^oC$, while for the RMSD for the GCM model mean is twice, 0.09 $^oC$, and for the single GCM runs  is normally larger (see  Table 2).

Figure 28 magnifies the period 1980-2020 and uses six global temperature estimates (HadCRUT3, HadCRUT4, UAH MSU, RSS MSU, GISS and NCDC) that appear to agree sufficiently well with each other and all of them disagree with the CMIP5 simulations after 2000. The temperature may slightly increase from 2013 to 2016, and decrease from 2016 to 2020 because of the two decadal cycles. However, fast ENSO temperature fluctuations at scales shorter than 7 years may mask the result.

 In conclusion, because the temperature patterns appears well correlated with solar/astronomical oscillations at multiple scales and that semi-empirical models based on these oscillations  reconstruct and hindcast GST variations significantly better than the current GCMs,  it is very likely that the observed GST oscillations   have an astronomical origin whose mechanisms are not implemented in the GCMs yet.
 To better appreciate the finding, it is important to stress that the harmonic constituents of the proposed model, such as the frequencies and the phases, are coherent with major astronomical oscillations. Although in the future the empirical model may be improved with a better understanding of these oscillations, the proposed semi-empirical model already appears to outperform  all CMIP5 GCMs.

\section{Conclusion}

As for the CMIP3 GCMs used by the IPCC 2007 \citep{Scafetta2012b}, the upgraded  CMIP5 GCMs  to be used in the IPCC Fifth Assessment Report (AR5, 2013) do not reproduce the detectable decadal and multidecadal climate oscillations observed in the GST records since 1850.   Multiple analyses suggest that these GCMs  overestimate the anthropogenic warming effect by about 50\%.  This would also imply that the climate sensitivity to the radiative forcing should be significantly lowered by half. It may be $\sim1.5$  $^oC$ (or less if part of the warming is spurious, for example due to uncorrected UHI effects), and it may possibly range between 1 $^oC$ and 2.3 $^oC$ instead of the \cite{IPCC} proposed range from 2 $^oC$ to 4.5 $^oC$.  Very important physical mechanisms necessary for reproducing multiple climatic oscillations, which are  responsible  for about half  of the 1850-2010 warming appear to be still missing in the GCMs.

\begin{figure*}[!t]
\begin{center}
\includegraphics[width=0.9\textwidth]{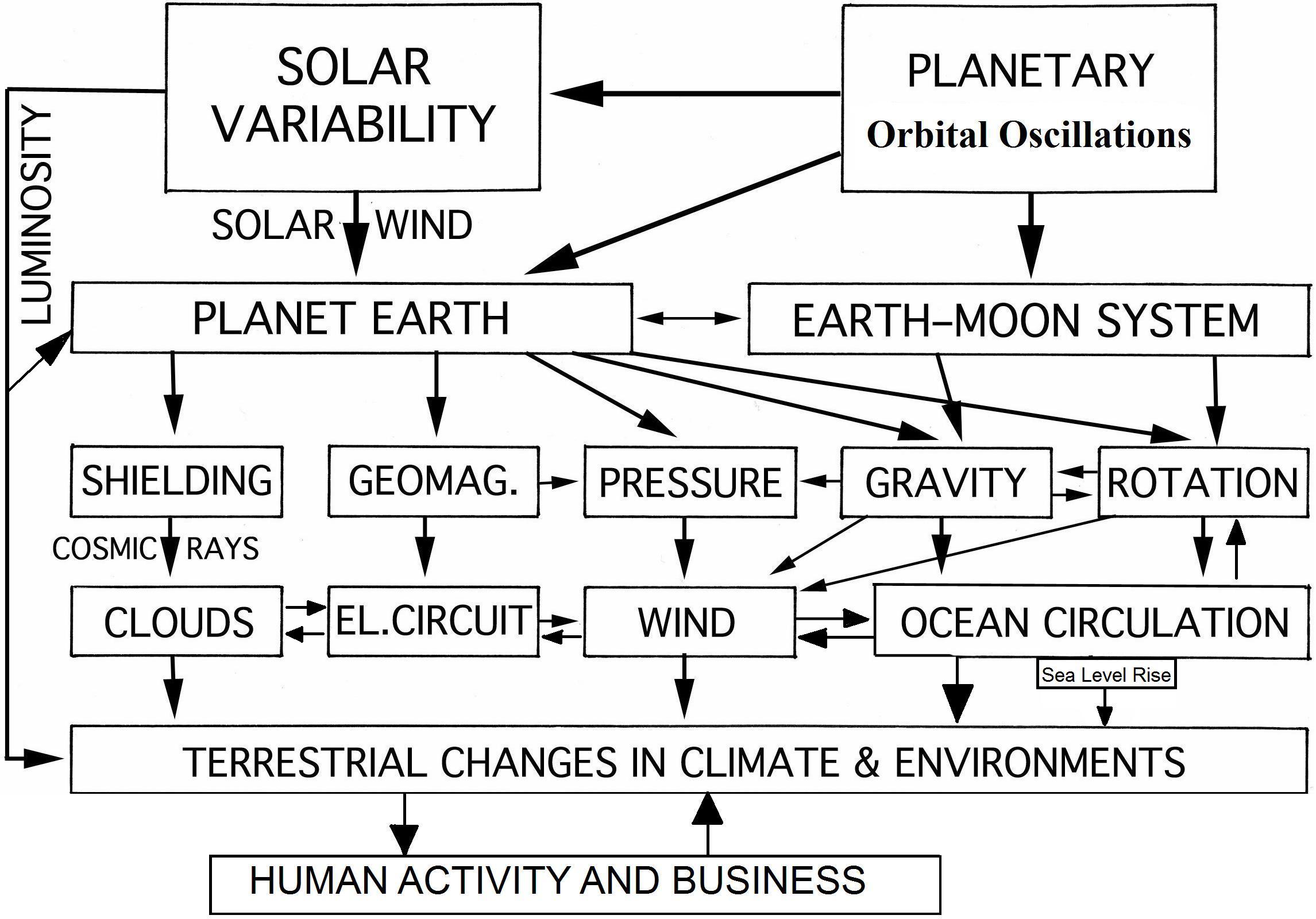}
\end{center}
\caption{Network of the possible physical interaction between planetary harmonics,
solar variability and climate and environments changes on planet Earth
(adapted with permission after \citet{Morner2012}, see also \cite{Scafetta2013}).}
\end{figure*}

The physical origin of the detected climatic oscillations is currently uncertain, but  in this paper it has been argued that they  may  be astronomically induced. This conclusion  derives from the coherence found among astronomical and climate oscillations from the decadal to the millennial time scales. This harmonic component  cannot be ignored in properly interpreting and forecasting climate change.  I have shown that an empirical model based on specific major astronomical harmonics  plus a correction of the outputs of the current CMIP5 GCM simulations simultaneously reconstructs the decadal, multidecadal and secular  patterns observed in the GST records since 1850.

The most reasonable conclusion is that the climate system is  synchronized to the natural oscillations found in the solar system and that this harmonic dynamics constitutes an important component of the Earth's climate. The proposed semi-empirical model may  produce more reliable projections for the 21st century, which are far less alarmist than the current CMIP5 projections. Under the same anthropogenic emission scenarios, the model projects a possible 2000-2100 warming ranging from 0.3 $^oC$ to 1.8 $^oC$. This range is significantly below the original CMIP5 GCM ensemble mean projections spanning from about 1 $^oC$ to 4 $^oC$.

In conclusion, multiple statistical tests suggest that the proposed semi-empirical GST model based on astronomically induced harmonics plus an     anthropogenic plus volcano contribution reduced by about 50\% from the CMIP5 GCM prediction   outperforms all CMIP5 GCMs in reconstructing and interpreting the GST patterns observed since 1850. The model would be compatible also with modern paleoclimatic reconstructions showing a larger pre-industrial variability with a medieval period as warm as the 1950-2000 global surface temperatures.

Future research should  investigate space-climate coupling
mechanisms in order to develop more advanced analytical and semi-empirical climate
models. Figure 29 schematically represents a possible network of physical interactions causing climatic changes.

\section*{Appendix}

The proposed harmonic model, Eq. \ref{eqg33}, uses a specific set of harmonics first determined in \cite{Scafetta2010}. \cite{Scafetta2013a} also proposed a model with slightly different decadal and multidecadal frequencies: harmonics with period of 10.2 years, 21 years and 61 years were used  instead of harmonics with period of 10.4 years, 20 years, and 60 years. \cite{Scafetta2013a} choice was justified by spectral analysis results and by the possibility that the observed harmonics could be induced by multiple closed astronomical frequencies with a different physical origin.

 For example, the quasi 20-year GST oscillation could be induced by a combination of the 18.6-year lunar nutation cycle, of the 19.85-year  oscillation of the speed of the wobbling sun related to the conjunction of Jupiter and Saturn and of the quasi 22-year Hale solar magnetic cycle. Indeed, \cite{Chylek2011} found evidence in multisecular ice-core data for a prominent near 20-year time-scale oscillation that beats and appears to be composed of three close frequencies. Similarly, the quasi 60-year GST oscillation could be caused by the 59.6 years cycle in the speed of the wobbling sun and by the 61-year tidal beat between Jupiter orbit (11.86 years) and Jupiter and Saturn's spring tide (9.93 years).
It is evident that if the planets are modulating solar activity and, directly or indirectly, the climate of the Earth,  numerous harmonics may be involved in the process, as also found in \cite{ScafettaW2013a,ScafettaW2013b}.

\begin{figure*}[!t]
\begin{center}
\includegraphics[angle=-90,width=0.9\textwidth]{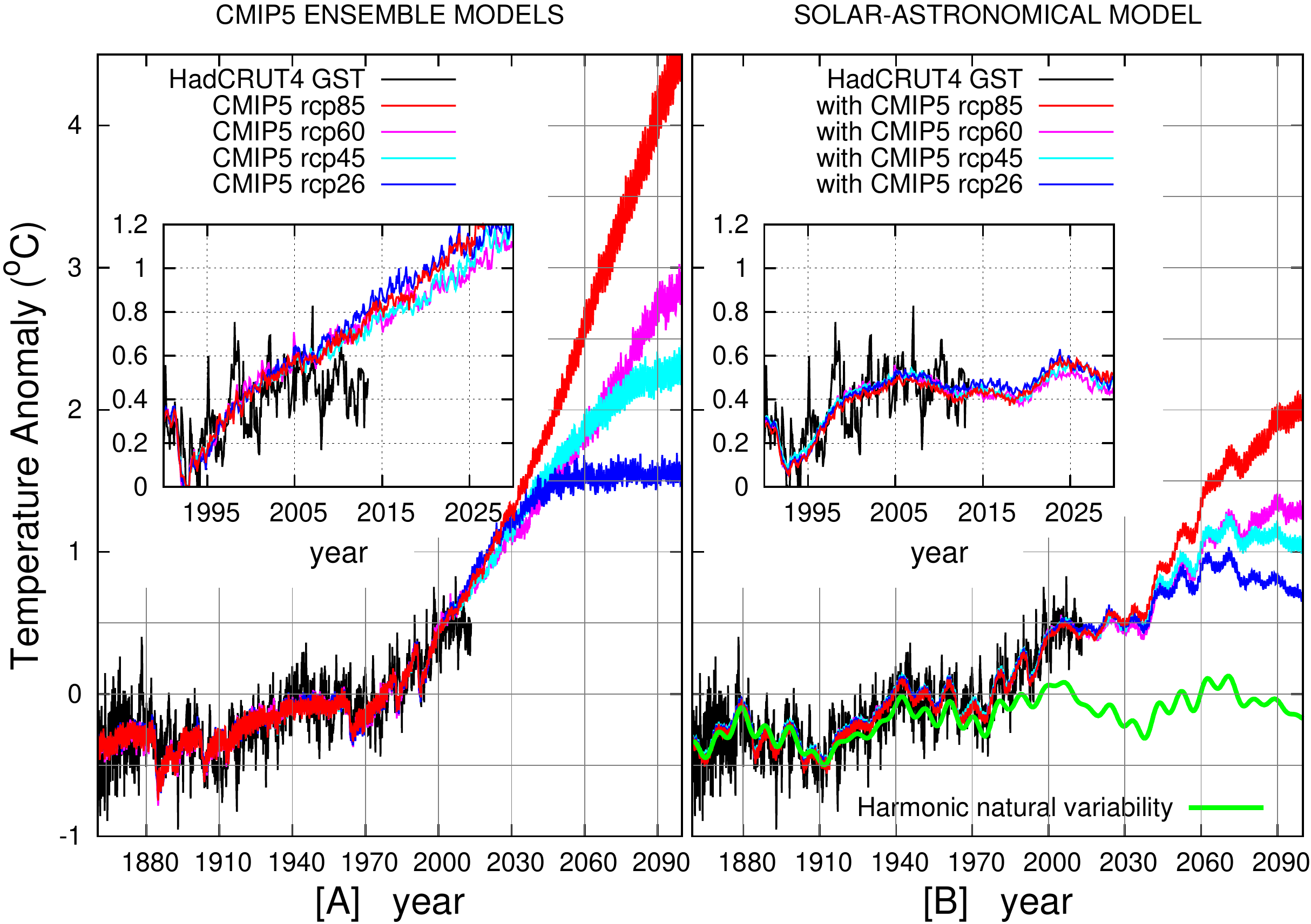}
\end{center}
\caption{Reproduction of Figure 26 with the alternative model and data proposed in \cite{Scafetta2013a} that uses slight different decadal and multidecadal harmonics.}
\end{figure*}

By analogy, the harmonic constituent model currently used to predict the ocean tides  \citep{Kelvin,Ehret} uses 30-40 harmonics whose frequencies are deduced from the orbits of the sun and of the moon. Many of the tidal harmonics are closely clustered (see \url{http://en.wikipedia.org/wiki/Theory_of_tides#Tidal_constituents}). For example, the semi-diurnal tidal oscillation is modeled using 8 frequencies spanning from 11.6 h (the shallow water semidiurnal cycle) to 12.9 h (the lunar elliptical semidiurnal second-order cycle).

Thus, the proposed harmonic model, Eq. \ref{eqg33}, which is based on a choice of just six harmonics spanning from the decadal to the millennial scales  should be interpreted as a minimal first approximation model. Future research should better define the true physical harmonics involved in the process which may be numerous. In any case, from a practical point of view the alternative model proposed in \cite{Scafetta2013a} produces  similar results  to Eq. \ref{eqg33}, as shown in Figure 30 that reproduces Figure 26 with the alternative model and data.

\onecolumn

\newpage

{\tiny
\begin{longtable}{|llr|c|c|c|c|c|c|c|c|}
\caption{\normalsize See Section 3.1.
The regression coefficients are evaluated using Eq. \ref{eqg} for all available 162 GCM simulations and for their ensemble mean (which is
numbered -0- in the list). The table reports the $\chi^2$ test values, which were calculated
using Eq. \ref{eqg33}, and the root-mean-square deviation (RMSD) values for each
simulations from the harmonic model Eq. \ref{eqf}, which were calculated using a 49-month running mean of the
original time sequences. See also Figure 12.}\\
\hline
\#	& Model&	Simulation	&	$\alpha$			&	$\alpha$			&	$\alpha$			&	$\alpha$			&	$\beta$			&	$\gamma$			&	$\chi^2$	&	RMSD	\\	
	& &	\#	&	(60-years)			&	(20-years)			&	(10.4=years)			&	(9.1-years)			&	(upward trend)			&	(bias)			&		&		\\	 \hline
	&	GST	&		&	1	$\pm$	0.05	&	1	$\pm$	0.12	&	1	$\pm$	0.17	&	1	$\pm$	0.11	&	1	$\pm$	0.02	&	0.0	$\pm$	0.01	&	0	&	0.05	\\ \hline
0	&	GCM mean	&		&	0.63	$\pm$	0.03	&	0.73	$\pm$	0.08	&	0.44	$\pm$	0.12	&	0.40	$\pm$	0.08	&	1.04	$\pm$	0.01	&	-0.55	$\pm$	0.00	&	14.3	&	0.09	\\ \hline
1	&	ACCESS1-0	&	0	&	0.39	$\pm$	0.05	&	0.92	$\pm$	0.14	&	0.88	$\pm$	0.20	&	0.70	$\pm$	0.13	&	0.84	$\pm$	0.02	&	0.12	$\pm$	0.01	&	21.4	&	0.14	\\ \hline
2	&	ACCESS1-3	&	0	&	0.46	$\pm$	0.05	&	0.89	$\pm$	0.12	&	1.05	$\pm$	0.17	&	0.02	$\pm$	0.12	&	0.58	$\pm$	0.02	&	0.08	$\pm$	0.00	&	69.4	&	0.14	\\
3	&		&	1	&	0.67	$\pm$	0.05	&	0.81	$\pm$	0.12	&	0.23	$\pm$	0.17	&	-0.07	$\pm$	0.12	&	0.51	$\pm$	0.02	&	0.07	$\pm$	0.00	&	81.7	&	0.15	\\
4	&		&	2	&	0.75	$\pm$	0.04	&	1.33	$\pm$	0.11	&	0.08	$\pm$	0.16	&	-0.03	$\pm$	0.11	&	0.39	$\pm$	0.02	&	0.06	$\pm$	0.00	&	123.6	&	0.15	\\ \hline
5	&	bcc-csm1-1	&	0	&	1.29	$\pm$	0.05	&	0.62	$\pm$	0.14	&	0.19	$\pm$	0.19	&	0.55	$\pm$	0.13	&	1.42	$\pm$	0.02	&	0.17	$\pm$	0.01	&	51.4	&	0.14	\\
6	&		&	1	&	0.64	$\pm$	0.06	&	0.41	$\pm$	0.14	&	0.59	$\pm$	0.20	&	1.03	$\pm$	0.14	&	1.72	$\pm$	0.02	&	0.22	$\pm$	0.01	&	129.9	&	0.19	\\
7	&		&	2	&	0.77	$\pm$	0.06	&	0.69	$\pm$	0.15	&	0.60	$\pm$	0.22	&	0.57	$\pm$	0.15	&	1.50	$\pm$	0.02	&	0.19	$\pm$	0.01	&	58.3	&	0.17	\\ \hline
8	&	bcc-csm1-1-m	&	0	&	0.40	$\pm$	0.06	&	0.83	$\pm$	0.15	&	0.81	$\pm$	0.22	&	0.48	$\pm$	0.15	&	1.44	$\pm$	0.02	&	0.20	$\pm$	0.01	&	55.7	&	0.16	\\
9	&		&	1	&	0.31	$\pm$	0.07	&	0.73	$\pm$	0.17	&	1.19	$\pm$	0.25	&	1.09	$\pm$	0.17	&	1.46	$\pm$	0.03	&	0.18	$\pm$	0.01	&	54.1	&	0.18	\\
10	&		&	2	&	0.25	$\pm$	0.07	&	1.07	$\pm$	0.19	&	0.68	$\pm$	0.26	&	1.01	$\pm$	0.18	&	1.65	$\pm$	0.03	&	0.23	$\pm$	0.01	&	86.6	&	0.22	\\ \hline
11	&	BNU-ESM	&	0	&	0.82	$\pm$	0.06	&	1.12	$\pm$	0.16	&	0.19	$\pm$	0.23	&	-0.18	$\pm$	0.16	&	1.78	$\pm$	0.02	&	0.27	$\pm$	0.01	&	131.5	&	0.20	\\ \hline
12	&	CanESM2	&	0	&	0.88	$\pm$	0.07	&	0.85	$\pm$	0.17	&	0.14	$\pm$	0.24	&	0.04	$\pm$	0.17	&	1.17	$\pm$	0.03	&	0.17	$\pm$	0.01	&	12.3	&	0.15	\\
13	&		&	1	&	1.17	$\pm$	0.06	&	1.39	$\pm$	0.16	&	0.51	$\pm$	0.23	&	0.66	$\pm$	0.15	&	1.05	$\pm$	0.02	&	0.15	$\pm$	0.01	&	3.5	&	0.13	\\
14	&		&	2	&	0.89	$\pm$	0.06	&	0.48	$\pm$	0.16	&	0.60	$\pm$	0.22	&	0.83	$\pm$	0.15	&	1.10	$\pm$	0.02	&	0.16	$\pm$	0.01	&	4.5	&	0.13	\\
15	&		&	3	&	0.85	$\pm$	0.06	&	1.00	$\pm$	0.16	&	0.78	$\pm$	0.22	&	-0.22	$\pm$	0.15	&	1.16	$\pm$	0.02	&	0.17	$\pm$	0.01	&	14.6	&	0.13	\\
16	&		&	4	&	1.17	$\pm$	0.06	&	1.20	$\pm$	0.16	&	1.11	$\pm$	0.23	&	0.11	$\pm$	0.15	&	1.19	$\pm$	0.02	&	0.17	$\pm$	0.01	&	12.7	&	0.14	\\ \hline
17	&	CCSM4	&	0	&	0.74	$\pm$	0.06	&	0.75	$\pm$	0.15	&	1.03	$\pm$	0.22	&	0.13	$\pm$	0.15	&	1.58	$\pm$	0.02	&	0.24	$\pm$	0.01	&	81.5	&	0.17	\\
18	&		&	1	&	0.74	$\pm$	0.05	&	0.68	$\pm$	0.14	&	-0.02	$\pm$	0.20	&	0.47	$\pm$	0.13	&	1.64	$\pm$	0.02	&	0.25	$\pm$	0.01	&	107.7	&	0.16	\\
19	&		&	2	&	0.63	$\pm$	0.06	&	0.27	$\pm$	0.15	&	0.83	$\pm$	0.21	&	0.82	$\pm$	0.14	&	1.55	$\pm$	0.02	&	0.24	$\pm$	0.01	&	78.0	&	0.15	\\
20	&		&	3	&	0.59	$\pm$	0.06	&	0.60	$\pm$	0.15	&	-0.06	$\pm$	0.22	&	0.98	$\pm$	0.15	&	1.59	$\pm$	0.02	&	0.24	$\pm$	0.01	&	85.4	&	0.17	\\
21	&		&	4	&	0.58	$\pm$	0.06	&	0.45	$\pm$	0.15	&	0.73	$\pm$	0.21	&	0.28	$\pm$	0.14	&	1.60	$\pm$	0.02	&	0.24	$\pm$	0.01	&	92.8	&	0.17	\\
22	&		&	5	&	0.56	$\pm$	0.05	&	1.16	$\pm$	0.14	&	0.83	$\pm$	0.20	&	0.75	$\pm$	0.14	&	1.59	$\pm$	0.02	&	0.25	$\pm$	0.01	&	92.6	&	0.16	\\ \hline
23	&	CESM1-BGC	&	0	&	0.67	$\pm$	0.05	&	0.45	$\pm$	0.14	&	1.09	$\pm$	0.20	&	1.07	$\pm$	0.14	&	1.50	$\pm$	0.02	&	0.23	$\pm$	0.01	&	64.6	&	0.15	\\ \hline
24	&	CESM1-CAM5	&	0	&	0.29	$\pm$	0.05	&	0.66	$\pm$	0.12	&	0.30	$\pm$	0.17	&	0.78	$\pm$	0.12	&	0.88	$\pm$	0.02	&	0.13	$\pm$	0.00	&	27.9	&	0.11	\\
25	&		&	1	&	1.01	$\pm$	0.05	&	1.01	$\pm$	0.12	&	0.80	$\pm$	0.17	&	0.58	$\pm$	0.12	&	0.79	$\pm$	0.02	&	0.11	$\pm$	0.00	&	13.7	&	0.10	\\
26	&		&	2	&	0.99	$\pm$	0.05	&	0.55	$\pm$	0.13	&	1.41	$\pm$	0.18	&	0.48	$\pm$	0.12	&	0.92	$\pm$	0.02	&	0.13	$\pm$	0.00	&	5.7	&	0.10	\\ \hline
27	&	CESM1-CAM5-1-FV2	&	0	&	0.87	$\pm$	0.06	&	0.04	$\pm$	0.16	&	1.32	$\pm$	0.23	&	0.63	$\pm$	0.16	&	0.71	$\pm$	0.02	&	0.09	$\pm$	0.01	&	22.2	&	0.15	\\
28	&		&	1	&	1.05	$\pm$	0.06	&	0.34	$\pm$	0.14	&	1.48	$\pm$	0.20	&	0.08	$\pm$	0.14	&	0.72	$\pm$	0.02	&	0.10	$\pm$	0.01	&	27.1	&	0.11	\\
29	&		&	2	&	0.77	$\pm$	0.05	&	-0.08	$\pm$	0.14	&	0.11	$\pm$	0.20	&	-0.05	$\pm$	0.14	&	0.85	$\pm$	0.02	&	0.12	$\pm$	0.01	&	23.8	&	0.11	\\
30	&		&	3	&	0.61	$\pm$	0.06	&	0.64	$\pm$	0.16	&	-0.06	$\pm$	0.22	&	0.16	$\pm$	0.15	&	0.81	$\pm$	0.02	&	0.11	$\pm$	0.01	&	20.3	&	0.13	\\ \hline
31	&	CESM1-FASTCHEM	&	0	&	0.62	$\pm$	0.06	&	0.79	$\pm$	0.15	&	0.25	$\pm$	0.21	&	0.92	$\pm$	0.14	&	1.72	$\pm$	0.02	&	0.26	$\pm$	0.01	&	122.8	&	0.19	\\
32	&		&	1	&	0.77	$\pm$	0.06	&	0.97	$\pm$	0.14	&	0.72	$\pm$	0.20	&	0.52	$\pm$	0.14	&	1.66	$\pm$	0.02	&	0.25	$\pm$	0.01	&	106.8	&	0.16	\\
33	&		&	2	&	0.58	$\pm$	0.06	&	0.86	$\pm$	0.14	&	0.60	$\pm$	0.21	&	1.15	$\pm$	0.14	&	1.82	$\pm$	0.02	&	0.28	$\pm$	0.01	&	161.8	&	0.20	\\ \hline
34	&	CESM1-WACCM	&	0	&	0.65	$\pm$	0.06	&	0.59	$\pm$	0.15	&	-0.47	$\pm$	0.22	&	0.89	$\pm$	0.15	&	1.55	$\pm$	0.02	&	0.23	$\pm$	0.01	&	76.0	&	0.17	\\ \hline
35	&	CMCC-CESM	&	0	&	0.50	$\pm$	0.07	&	0.01	$\pm$	0.18	&	-0.15	$\pm$	0.25	&	-0.11	$\pm$	0.17	&	0.49	$\pm$	0.03	&	0.07	$\pm$	0.01	&	67.4	&	0.16	\\ \hline
36	&	CMCC-CM	&	0	&	0.56	$\pm$	0.05	&	-0.55	$\pm$	0.12	&	1.04	$\pm$	0.17	&	0.09	$\pm$	0.11	&	0.89	$\pm$	0.02	&	0.14	$\pm$	0.00	&	35.0	&	0.12	\\ \hline
37	&	CMCC-CMS	&	0	&	0.66	$\pm$	0.06	&	0.49	$\pm$	0.15	&	0.98	$\pm$	0.21	&	-0.17	$\pm$	0.14	&	0.63	$\pm$	0.02	&	0.09	$\pm$	0.01	&	44.1	&	0.16	\\ \hline
38	&	CNRM-CM5	&	0	&	0.42	$\pm$	0.06	&	0.96	$\pm$	0.15	&	1.13	$\pm$	0.21	&	0.39	$\pm$	0.14	&	0.92	$\pm$	0.02	&	0.14	$\pm$	0.01	&	15.6	&	0.13	\\
39	&		&	1	&	0.67	$\pm$	0.05	&	1.24	$\pm$	0.12	&	0.72	$\pm$	0.17	&	0.81	$\pm$	0.12	&	0.99	$\pm$	0.02	&	0.14	$\pm$	0.00	&	5.6	&	0.08	\\
40	&		&	2	&	0.70	$\pm$	0.05	&	1.27	$\pm$	0.13	&	0.54	$\pm$	0.19	&	0.64	$\pm$	0.13	&	0.91	$\pm$	0.02	&	0.14	$\pm$	0.00	&	7.5	&	0.11	\\
41	&		&	3	&	0.92	$\pm$	0.05	&	0.57	$\pm$	0.13	&	0.35	$\pm$	0.18	&	0.94	$\pm$	0.12	&	1.20	$\pm$	0.02	&	0.18	$\pm$	0.00	&	13.9	&	0.11	\\
42	&		&	4	&	0.56	$\pm$	0.05	&	0.93	$\pm$	0.14	&	1.01	$\pm$	0.20	&	0.55	$\pm$	0.13	&	0.85	$\pm$	0.02	&	0.14	$\pm$	0.01	&	13.7	&	0.12	\\
43	&		&	5	&	0.45	$\pm$	0.05	&	0.98	$\pm$	0.13	&	0.98	$\pm$	0.19	&	0.59	$\pm$	0.13	&	1.10	$\pm$	0.02	&	0.18	$\pm$	0.00	&	15.3	&	0.11	\\
44	&		&	6	&	1.20	$\pm$	0.05	&	0.97	$\pm$	0.13	&	0.56	$\pm$	0.19	&	0.70	$\pm$	0.13	&	0.95	$\pm$	0.02	&	0.14	$\pm$	0.00	&	3.5	&	0.10	\\
45	&		&	7	&	0.36	$\pm$	0.05	&	0.59	$\pm$	0.13	&	-0.01	$\pm$	0.19	&	0.36	$\pm$	0.13	&	1.18	$\pm$	0.02	&	0.19	$\pm$	0.00	&	31.0	&	0.13	\\
46	&		&	8	&	0.71	$\pm$	0.05	&	0.21	$\pm$	0.12	&	0.73	$\pm$	0.18	&	0.86	$\pm$	0.12	&	0.74	$\pm$	0.02	&	0.12	$\pm$	0.00	&	27.1	&	0.11	\\
47	&		&	9	&	1.07	$\pm$	0.05	&	0.88	$\pm$	0.13	&	-0.07	$\pm$	0.18	&	0.62	$\pm$	0.12	&	1.15	$\pm$	0.02	&	0.17	$\pm$	0.00	&	11.2	&	0.11	\\ \hline
48	&	CSIRO-Mk3-6-0	&	0	&	0.53	$\pm$	0.06	&	1.32	$\pm$	0.15	&	-0.42	$\pm$	0.22	&	0.64	$\pm$	0.15	&	0.70	$\pm$	0.02	&	0.11	$\pm$	0.01	&	34.7	&	0.16	\\
49	&		&	1	&	0.98	$\pm$	0.05	&	0.35	$\pm$	0.14	&	1.07	$\pm$	0.20	&	-0.47	$\pm$	0.13	&	0.60	$\pm$	0.02	&	0.10	$\pm$	0.01	&	56.1	&	0.16	\\
50	&		&	2	&	1.12	$\pm$	0.05	&	0.66	$\pm$	0.13	&	1.21	$\pm$	0.18	&	0.00	$\pm$	0.12	&	0.78	$\pm$	0.02	&	0.12	$\pm$	0.00	&	20.9	&	0.12	\\
51	&		&	3	&	0.88	$\pm$	0.06	&	1.39	$\pm$	0.15	&	1.05	$\pm$	0.21	&	0.30	$\pm$	0.14	&	0.45	$\pm$	0.02	&	0.06	$\pm$	0.01	&	75.6	&	0.17	\\
52	&		&	4	&	0.80	$\pm$	0.06	&	0.74	$\pm$	0.17	&	-0.07	$\pm$	0.24	&	0.34	$\pm$	0.16	&	0.70	$\pm$	0.02	&	0.10	$\pm$	0.01	&	24.5	&	0.17	\\
53	&		&	5	&	0.60	$\pm$	0.06	&	0.77	$\pm$	0.14	&	0.54	$\pm$	0.20	&	0.35	$\pm$	0.14	&	0.69	$\pm$	0.02	&	0.11	$\pm$	0.01	&	31.8	&	0.15	\\
54	&		&	6	&	0.52	$\pm$	0.05	&	0.22	$\pm$	0.13	&	0.29	$\pm$	0.18	&	0.09	$\pm$	0.12	&	0.76	$\pm$	0.02	&	0.12	$\pm$	0.00	&	35.8	&	0.13	\\
55	&		&	7	&	1.07	$\pm$	0.06	&	0.54	$\pm$	0.14	&	0.21	$\pm$	0.20	&	0.55	$\pm$	0.14	&	0.63	$\pm$	0.02	&	0.09	$\pm$	0.01	&	37.1	&	0.15	\\
56	&		&	8	&	0.36	$\pm$	0.05	&	-0.31	$\pm$	0.13	&	0.43	$\pm$	0.19	&	0.31	$\pm$	0.13	&	0.77	$\pm$	0.02	&	0.13	$\pm$	0.00	&	44.4	&	0.14	\\
57	&		&	9	&	0.39	$\pm$	0.05	&	0.63	$\pm$	0.14	&	-0.22	$\pm$	0.19	&	0.01	$\pm$	0.13	&	0.80	$\pm$	0.02	&	0.13	$\pm$	0.01	&	36.0	&	0.14	\\ \hline
58	&	EC-EARTH	&	0	&	0.48	$\pm$	0.05	&	0.75	$\pm$	0.12	&	0.51	$\pm$	0.17	&	0.37	$\pm$	0.11	&	1.41	$\pm$	0.02	&	0.19	$\pm$	0.00	&	63.8	&	0.14	\\
59	&		&	1	&	-0.13	$\pm$	0.05	&	1.08	$\pm$	0.12	&	0.94	$\pm$	0.16	&	0.06	$\pm$	0.11	&	1.54	$\pm$	0.02	&	0.21	$\pm$	0.00	&	147.3	&	0.16	\\
60	&		&	2	&	0.53	$\pm$	0.05	&	0.40	$\pm$	0.12	&	0.83	$\pm$	0.17	&	-0.04	$\pm$	0.12	&	1.35	$\pm$	0.02	&	0.19	$\pm$	0.00	&	54.7	&	0.13	\\
61	&		&	3	&	0.14	$\pm$	0.04	&	0.84	$\pm$	0.12	&	0.77	$\pm$	0.16	&	0.28	$\pm$	0.11	&	1.40	$\pm$	0.02	&	0.23	$\pm$	0.00	&	83.2	&	0.14	\\
62	&		&	4	&	0.38	$\pm$	0.05	&	0.41	$\pm$	0.12	&	0.09	$\pm$	0.17	&	0.69	$\pm$	0.11	&	1.54	$\pm$	0.02	&	0.20	$\pm$	0.00	&	105.9	&	0.15	\\
63	&		&	5	&	0.77	$\pm$	0.04	&	1.04	$\pm$	0.11	&	0.18	$\pm$	0.16	&	0.18	$\pm$	0.11	&	1.47	$\pm$	0.02	&	0.18	$\pm$	0.00	&	75.3	&	0.14	\\
64	&		&	6	&	0.31	$\pm$	0.05	&	1.03	$\pm$	0.12	&	0.51	$\pm$	0.17	&	-0.01	$\pm$	0.11	&	1.44	$\pm$	0.02	&	0.19	$\pm$	0.00	&	86.2	&	0.14	\\
65	&		&	7	&	0.71	$\pm$	0.04	&	0.07	$\pm$	0.11	&	0.51	$\pm$	0.16	&	0.97	$\pm$	0.11	&	1.38	$\pm$	0.02	&	0.18	$\pm$	0.00	&	54.6	&	0.11	\\
66	&		&	8	&	0.59	$\pm$	0.05	&	1.03	$\pm$	0.12	&	0.90	$\pm$	0.17	&	0.86	$\pm$	0.12	&	1.38	$\pm$	0.02	&	0.18	$\pm$	0.00	&	47.1	&	0.13	\\ \hline
67	&	FGOALS-g2	&	0	&	0.37	$\pm$	0.04	&	0.26	$\pm$	0.10	&	0.09	$\pm$	0.15	&	-0.01	$\pm$	0.10	&	1.00	$\pm$	0.02	&	0.15	$\pm$	0.00	&	36.1	&	0.10	\\
68	&		&	1	&	0.72	$\pm$	0.04	&	0.10	$\pm$	0.10	&	0.01	$\pm$	0.14	&	-0.03	$\pm$	0.10	&	0.95	$\pm$	0.01	&	0.10	$\pm$	0.00	&	25.4	&	0.08	\\
69	&		&	2	&	0.71	$\pm$	0.04	&	-0.07	$\pm$	0.11	&	0.05	$\pm$	0.16	&	-0.05	$\pm$	0.11	&	0.95	$\pm$	0.02	&	0.14	$\pm$	0.00	&	25.6	&	0.10	\\
70	&		&	3	&	0.49	$\pm$	0.04	&	0.47	$\pm$	0.11	&	0.15	$\pm$	0.15	&	0.08	$\pm$	0.10	&	0.84	$\pm$	0.02	&	0.10	$\pm$	0.00	&	32.6	&	0.11	\\
71	&		&	4	&	0.51	$\pm$	0.04	&	0.14	$\pm$	0.10	&	-0.04	$\pm$	0.15	&	-0.04	$\pm$	0.10	&	0.95	$\pm$	0.02	&	0.12	$\pm$	0.00	&	32.3	&	0.10	\\ \hline
72	&	FIO-ESM	&	0	&	0.31	$\pm$	0.05	&	0.26	$\pm$	0.14	&	0.57	$\pm$	0.20	&	0.02	$\pm$	0.14	&	1.33	$\pm$	0.02	&	0.21	$\pm$	0.01	&	53.8	&	0.13	\\
73	&		&	1	&	0.35	$\pm$	0.05	&	0.69	$\pm$	0.13	&	0.85	$\pm$	0.18	&	0.12	$\pm$	0.13	&	1.23	$\pm$	0.02	&	0.19	$\pm$	0.00	&	36.7	&	0.10	\\
74	&		&	2	&	0.36	$\pm$	0.05	&	-0.11	$\pm$	0.13	&	-0.40	$\pm$	0.19	&	0.19	$\pm$	0.13	&	1.46	$\pm$	0.02	&	0.23	$\pm$	0.00	&	89.0	&	0.13	\\ \hline
75	&	GFDL-CM3	&	0	&	1.60	$\pm$	0.06	&	1.49	$\pm$	0.16	&	-0.07	$\pm$	0.23	&	0.57	$\pm$	0.16	&	0.49	$\pm$	0.02	&	0.05	$\pm$	0.01	&	69.2	&	0.18	\\
76	&		&	1	&	1.39	$\pm$	0.06	&	1.01	$\pm$	0.15	&	1.10	$\pm$	0.21	&	0.47	$\pm$	0.14	&	0.56	$\pm$	0.02	&	0.07	$\pm$	0.01	&	52.2	&	0.15	\\
77	&		&	2	&	0.77	$\pm$	0.06	&	0.95	$\pm$	0.16	&	1.24	$\pm$	0.22	&	0.34	$\pm$	0.15	&	0.52	$\pm$	0.02	&	0.07	$\pm$	0.01	&	52.4	&	0.17	\\
78	&		&	3	&	1.12	$\pm$	0.07	&	1.13	$\pm$	0.17	&	0.57	$\pm$	0.25	&	0.26	$\pm$	0.17	&	0.63	$\pm$	0.03	&	0.08	$\pm$	0.01	&	29.1	&	0.18	\\
79	&		&	4	&	1.27	$\pm$	0.07	&	1.11	$\pm$	0.17	&	0.12	$\pm$	0.24	&	0.98	$\pm$	0.17	&	0.27	$\pm$	0.03	&	0.02	$\pm$	0.01	&	105.1	&	0.21	\\ \hline
80	&	GFDL-ESM2G	&	0	&	0.85	$\pm$	0.07	&	1.42	$\pm$	0.17	&	0.14	$\pm$	0.24	&	0.14	$\pm$	0.16	&	0.85	$\pm$	0.03	&	0.11	$\pm$	0.01	&	11.4	&	0.16	\\
81	&		&	1	&	0.19	$\pm$	0.05	&	1.27	$\pm$	0.14	&	0.80	$\pm$	0.19	&	0.10	$\pm$	0.13	&	0.89	$\pm$	0.02	&	0.13	$\pm$	0.01	&	33.6	&	0.12	\\
82	&		&	2	&	0.33	$\pm$	0.06	&	0.80	$\pm$	0.15	&	-0.33	$\pm$	0.21	&	0.55	$\pm$	0.14	&	1.02	$\pm$	0.02	&	0.14	$\pm$	0.01	&	21.9	&	0.13	\\ \hline
83	&	GFDL-ESM2M	&	0	&	0.22	$\pm$	0.06	&	0.48	$\pm$	0.16	&	0.18	$\pm$	0.23	&	0.40	$\pm$	0.15	&	0.95	$\pm$	0.02	&	0.13	$\pm$	0.01	&	24.6	&	0.14	\\ \hline
84	&	GISS-E2-H p1	&	0	&	0.29	$\pm$	0.04	&	0.50	$\pm$	0.11	&	0.25	$\pm$	0.15	&	0.43	$\pm$	0.10	&	1.37	$\pm$	0.02	&	0.23	$\pm$	0.00	&	72.2	&	0.13	\\
85	&		&	1	&	0.33	$\pm$	0.04	&	0.52	$\pm$	0.10	&	0.29	$\pm$	0.15	&	0.67	$\pm$	0.10	&	1.34	$\pm$	0.02	&	0.21	$\pm$	0.00	&	62.5	&	0.12	\\
86	&		&	2	&	0.35	$\pm$	0.04	&	0.22	$\pm$	0.11	&	0.51	$\pm$	0.16	&	0.67	$\pm$	0.11	&	1.30	$\pm$	0.02	&	0.21	$\pm$	0.00	&	53.1	&	0.12	\\
87	&		&	3	&	0.41	$\pm$	0.04	&	0.53	$\pm$	0.11	&	0.34	$\pm$	0.15	&	0.13	$\pm$	0.10	&	1.32	$\pm$	0.02	&	0.21	$\pm$	0.00	&	57.9	&	0.12	\\
88	&		&	4	&	0.40	$\pm$	0.04	&	0.76	$\pm$	0.10	&	0.90	$\pm$	0.14	&	0.46	$\pm$	0.10	&	1.33	$\pm$	0.01	&	0.21	$\pm$	0.00	&	56.2	&	0.11	\\
89	&		&	5	&	0.40	$\pm$	0.04	&	0.73	$\pm$	0.10	&	0.41	$\pm$	0.15	&	0.65	$\pm$	0.10	&	1.24	$\pm$	0.02	&	0.16	$\pm$	0.00	&	38.7	&	0.11	\\ \hline
90	&	GISS-E2-H p2	&	0	&	0.54	$\pm$	0.05	&	0.52	$\pm$	0.12	&	0.48	$\pm$	0.17	&	0.35	$\pm$	0.12	&	1.05	$\pm$	0.02	&	0.17	$\pm$	0.00	&	15.6	&	0.12	\\
91	&		&	1	&	0.25	$\pm$	0.05	&	0.94	$\pm$	0.12	&	0.30	$\pm$	0.17	&	0.75	$\pm$	0.11	&	1.06	$\pm$	0.02	&	0.18	$\pm$	0.00	&	28.0	&	0.12	\\
92	&		&	2	&	0.43	$\pm$	0.04	&	0.91	$\pm$	0.11	&	0.16	$\pm$	0.15	&	0.54	$\pm$	0.10	&	1.12	$\pm$	0.02	&	0.18	$\pm$	0.00	&	23.9	&	0.11	\\
93	&		&	3	&	0.40	$\pm$	0.05	&	0.58	$\pm$	0.12	&	0.47	$\pm$	0.17	&	0.44	$\pm$	0.11	&	1.03	$\pm$	0.02	&	0.17	$\pm$	0.00	&	21.0	&	0.12	\\
94	&		&	4	&	0.45	$\pm$	0.04	&	0.57	$\pm$	0.10	&	0.63	$\pm$	0.15	&	0.57	$\pm$	0.10	&	1.07	$\pm$	0.02	&	0.18	$\pm$	0.00	&	20.0	&	0.10	\\ \hline
95	&	GISS-E2-H p3	&	0	&	0.36	$\pm$	0.05	&	0.50	$\pm$	0.12	&	0.96	$\pm$	0.18	&	0.26	$\pm$	0.12	&	1.36	$\pm$	0.02	&	0.22	$\pm$	0.00	&	57.0	&	0.14	\\
96	&		&	1	&	0.32	$\pm$	0.04	&	0.55	$\pm$	0.11	&	0.20	$\pm$	0.15	&	0.70	$\pm$	0.10	&	1.66	$\pm$	0.02	&	0.26	$\pm$	0.00	&	161.4	&	0.17	\\
97	&		&	2	&	0.41	$\pm$	0.05	&	0.31	$\pm$	0.12	&	0.34	$\pm$	0.17	&	0.37	$\pm$	0.11	&	1.47	$\pm$	0.02	&	0.24	$\pm$	0.00	&	85.2	&	0.15	\\
98	&		&	3	&	0.65	$\pm$	0.04	&	0.72	$\pm$	0.11	&	0.45	$\pm$	0.15	&	0.44	$\pm$	0.10	&	1.42	$\pm$	0.02	&	0.23	$\pm$	0.00	&	63.6	&	0.13	\\
99	&		&	4	&	0.31	$\pm$	0.05	&	1.01	$\pm$	0.12	&	0.72	$\pm$	0.17	&	0.28	$\pm$	0.11	&	1.36	$\pm$	0.02	&	0.22	$\pm$	0.00	&	62.1	&	0.14	\\
100	&		&	5	&	0.45	$\pm$	0.05	&	0.99	$\pm$	0.12	&	0.28	$\pm$	0.17	&	0.19	$\pm$	0.11	&	1.29	$\pm$	0.02	&	0.17	$\pm$	0.00	&	44.9	&	0.13	\\ \hline
101	&	GISS-E2-H-CC p1	&	0	&	0.37	$\pm$	0.04	&	0.77	$\pm$	0.11	&	0.43	$\pm$	0.16	&	0.50	$\pm$	0.11	&	1.60	$\pm$	0.02	&	0.23	$\pm$	0.00	&	127.7	&	0.16	\\ \hline
102	&	GISS-E2-R p1	&	0	&	0.19	$\pm$	0.04	&	0.21	$\pm$	0.11	&	0.30	$\pm$	0.15	&	0.19	$\pm$	0.10	&	0.97	$\pm$	0.02	&	0.15	$\pm$	0.00	&	43.9	&	0.11	\\
103	&		&	1	&	0.31	$\pm$	0.04	&	1.03	$\pm$	0.11	&	0.76	$\pm$	0.15	&	0.32	$\pm$	0.10	&	1.10	$\pm$	0.02	&	0.17	$\pm$	0.00	&	30.2	&	0.10	\\
104	&		&	2	&	0.51	$\pm$	0.04	&	0.61	$\pm$	0.10	&	0.20	$\pm$	0.15	&	0.43	$\pm$	0.10	&	1.02	$\pm$	0.02	&	0.16	$\pm$	0.00	&	18.4	&	0.09	\\
105	&		&	3	&	0.49	$\pm$	0.04	&	0.71	$\pm$	0.11	&	0.86	$\pm$	0.16	&	0.54	$\pm$	0.11	&	1.09	$\pm$	0.02	&	0.17	$\pm$	0.00	&	17.0	&	0.10	\\
106	&		&	4	&	0.27	$\pm$	0.04	&	1.05	$\pm$	0.11	&	0.78	$\pm$	0.16	&	0.36	$\pm$	0.11	&	1.15	$\pm$	0.02	&	0.18	$\pm$	0.00	&	33.9	&	0.11	\\
107	&		&	5	&	0.32	$\pm$	0.05	&	0.83	$\pm$	0.12	&	0.46	$\pm$	0.17	&	0.58	$\pm$	0.11	&	0.96	$\pm$	0.02	&	0.15	$\pm$	0.00	&	23.3	&	0.12	\\ \hline
108	&	GISS-E2-R p2	&	0	&	0.42	$\pm$	0.04	&	0.48	$\pm$	0.11	&	0.26	$\pm$	0.16	&	0.41	$\pm$	0.11	&	0.88	$\pm$	0.02	&	0.14	$\pm$	0.00	&	26.5	&	0.11	\\
109	&		&	1	&	0.63	$\pm$	0.04	&	-0.52	$\pm$	0.11	&	0.23	$\pm$	0.16	&	0.50	$\pm$	0.11	&	0.72	$\pm$	0.02	&	0.12	$\pm$	0.00	&	49.7	&	0.13	\\
110	&		&	2	&	0.17	$\pm$	0.04	&	0.34	$\pm$	0.11	&	1.15	$\pm$	0.16	&	0.61	$\pm$	0.11	&	0.78	$\pm$	0.02	&	0.13	$\pm$	0.00	&	50.0	&	0.12	\\
111	&		&	3	&	0.58	$\pm$	0.05	&	0.68	$\pm$	0.12	&	0.80	$\pm$	0.17	&	0.55	$\pm$	0.12	&	0.65	$\pm$	0.02	&	0.10	$\pm$	0.00	&	43.2	&	0.14	\\
112	&		&	4	&	0.21	$\pm$	0.04	&	0.86	$\pm$	0.11	&	0.47	$\pm$	0.16	&	0.33	$\pm$	0.11	&	0.72	$\pm$	0.02	&	0.12	$\pm$	0.00	&	55.4	&	0.13	\\
113	&		&	5	&	0.39	$\pm$	0.04	&	0.80	$\pm$	0.11	&	0.50	$\pm$	0.15	&	0.36	$\pm$	0.10	&	0.71	$\pm$	0.02	&	0.11	$\pm$	0.00	&	49.7	&	0.12	\\ \hline
114	&	GISS-E2-R p3	&	0	&	0.31	$\pm$	0.05	&	0.78	$\pm$	0.12	&	0.46	$\pm$	0.17	&	0.28	$\pm$	0.11	&	0.95	$\pm$	0.02	&	0.15	$\pm$	0.00	&	27.0	&	0.12	\\
115	&		&	1	&	0.24	$\pm$	0.04	&	0.56	$\pm$	0.11	&	0.51	$\pm$	0.16	&	0.88	$\pm$	0.11	&	1.19	$\pm$	0.02	&	0.18	$\pm$	0.00	&	39.8	&	0.11	\\
116	&		&	2	&	0.56	$\pm$	0.05	&	0.73	$\pm$	0.12	&	0.74	$\pm$	0.18	&	0.70	$\pm$	0.12	&	1.24	$\pm$	0.02	&	0.19	$\pm$	0.00	&	24.4	&	0.12	\\
117	&		&	3	&	0.45	$\pm$	0.04	&	1.29	$\pm$	0.12	&	0.63	$\pm$	0.16	&	0.01	$\pm$	0.11	&	1.17	$\pm$	0.02	&	0.19	$\pm$	0.00	&	31.2	&	0.11	\\
118	&		&	4	&	0.17	$\pm$	0.05	&	1.02	$\pm$	0.12	&	0.33	$\pm$	0.17	&	0.01	$\pm$	0.12	&	1.08	$\pm$	0.02	&	0.17	$\pm$	0.00	&	40.4	&	0.12	\\
119	&		&	5	&	0.37	$\pm$	0.04	&	1.17	$\pm$	0.12	&	0.61	$\pm$	0.16	&	0.84	$\pm$	0.11	&	1.14	$\pm$	0.02	&	0.18	$\pm$	0.00	&	24.0	&	0.11	\\ \hline
120	&	GISS-E2-R-CC p1	&	0	&	0.27	$\pm$	0.04	&	0.52	$\pm$	0.11	&	0.45	$\pm$	0.15	&	0.38	$\pm$	0.10	&	1.09	$\pm$	0.02	&	0.15	$\pm$	0.00	&	33.8	&	0.11	\\ \hline
121	&	HadGEM2-AO	&	0	&	1.22	$\pm$	0.05	&	1.56	$\pm$	0.14	&	1.56	$\pm$	0.20	&	0.38	$\pm$	0.13	&	0.80	$\pm$	0.02	&	0.10	$\pm$	0.01	&	16.9	&	0.13	\\ \hline
122	&	HadGEM2-CC	&	0	&	0.98	$\pm$	0.05	&	0.77	$\pm$	0.14	&	0.21	$\pm$	0.20	&	0.49	$\pm$	0.13	&	0.30	$\pm$	0.02	&	0.03	$\pm$	0.01	&	120.9	&	0.19	\\ \hline
123	&	HadGEM2-ES	&	0	&	1.03	$\pm$	0.05	&	0.65	$\pm$	0.14	&	-0.16	$\pm$	0.20	&	0.53	$\pm$	0.13	&	0.48	$\pm$	0.02	&	0.06	$\pm$	0.01	&	73.3	&	0.16	\\
124	&		&	1	&	0.76	$\pm$	0.06	&	1.09	$\pm$	0.15	&	0.06	$\pm$	0.22	&	0.65	$\pm$	0.15	&	0.49	$\pm$	0.02	&	0.06	$\pm$	0.01	&	63.0	&	0.19	\\
125	&		&	2	&	1.04	$\pm$	0.05	&	1.46	$\pm$	0.13	&	0.70	$\pm$	0.18	&	0.33	$\pm$	0.12	&	0.65	$\pm$	0.02	&	0.08	$\pm$	0.00	&	37.5	&	0.13	\\
126	&		&	3	&	0.40	$\pm$	0.06	&	0.94	$\pm$	0.15	&	0.59	$\pm$	0.21	&	-0.44	$\pm$	0.14	&	0.68	$\pm$	0.02	&	0.09	$\pm$	0.01	&	49.7	&	0.17	\\ \hline
127	&	inmcm4	&	0	&	0.36	$\pm$	0.03	&	0.17	$\pm$	0.09	&	0.21	$\pm$	0.12	&	0.01	$\pm$	0.08	&	1.02	$\pm$	0.01	&	0.16	$\pm$	0.00	&	41.9	&	0.08	\\ \hline
128	&	IPSL-CM5A-LR	&	0	&	0.89	$\pm$	0.06	&	0.25	$\pm$	0.15	&	0.73	$\pm$	0.21	&	0.21	$\pm$	0.14	&	1.60	$\pm$	0.02	&	0.25	$\pm$	0.01	&	89.3	&	0.18	\\
129	&		&	1	&	1.06	$\pm$	0.05	&	0.67	$\pm$	0.13	&	-0.33	$\pm$	0.19	&	-0.11	$\pm$	0.13	&	1.65	$\pm$	0.02	&	0.26	$\pm$	0.00	&	122.3	&	0.18	\\
130	&		&	2	&	1.12	$\pm$	0.05	&	0.93	$\pm$	0.13	&	0.68	$\pm$	0.19	&	0.37	$\pm$	0.13	&	1.52	$\pm$	0.02	&	0.24	$\pm$	0.00	&	71.4	&	0.15	\\
131	&		&	3	&	0.53	$\pm$	0.06	&	0.94	$\pm$	0.14	&	0.61	$\pm$	0.20	&	0.24	$\pm$	0.14	&	1.67	$\pm$	0.02	&	0.26	$\pm$	0.01	&	116.9	&	0.18	\\
132	&		&	4	&	0.65	$\pm$	0.06	&	0.19	$\pm$	0.15	&	0.89	$\pm$	0.21	&	0.57	$\pm$	0.14	&	1.55	$\pm$	0.02	&	0.25	$\pm$	0.01	&	78.5	&	0.17	\\
133	&		&	5	&	0.92	$\pm$	0.05	&	1.35	$\pm$	0.14	&	-0.06	$\pm$	0.19	&	0.08	$\pm$	0.13	&	1.36	$\pm$	0.02	&	0.21	$\pm$	0.01	&	43.1	&	0.14	\\ \hline
134	&	IPSL-CM5A-MR	&	0	&	0.55	$\pm$	0.05	&	0.96	$\pm$	0.13	&	0.65	$\pm$	0.18	&	0.42	$\pm$	0.12	&	1.30	$\pm$	0.02	&	0.21	$\pm$	0.00	&	35.7	&	0.12	\\
135	&		&	1	&	0.63	$\pm$	0.05	&	0.91	$\pm$	0.14	&	0.62	$\pm$	0.19	&	-0.02	$\pm$	0.13	&	1.54	$\pm$	0.02	&	0.24	$\pm$	0.01	&	85.2	&	0.16	\\
136	&		&	2	&	0.86	$\pm$	0.05	&	0.47	$\pm$	0.13	&	0.44	$\pm$	0.18	&	0.16	$\pm$	0.12	&	1.66	$\pm$	0.02	&	0.26	$\pm$	0.00	&	127.1	&	0.17	\\ \hline
137	&	IPSL-CM5B-LR	&	0	&	0.04	$\pm$	0.05	&	0.10	$\pm$	0.13	&	0.88	$\pm$	0.19	&	0.17	$\pm$	0.13	&	1.43	$\pm$	0.02	&	0.24	$\pm$	0.00	&	91.9	&	0.16	\\ \hline
138	&	MIROC5	&	0	&	0.88	$\pm$	0.06	&	0.01	$\pm$	0.16	&	1.25	$\pm$	0.22	&	0.15	$\pm$	0.15	&	0.70	$\pm$	0.02	&	0.07	$\pm$	0.01	&	29.4	&	0.14	\\
139	&		&	1	&	0.62	$\pm$	0.06	&	0.80	$\pm$	0.14	&	0.36	$\pm$	0.20	&	0.61	$\pm$	0.14	&	0.81	$\pm$	0.02	&	0.10	$\pm$	0.01	&	15.9	&	0.13	\\
140	&		&	2	&	0.83	$\pm$	0.05	&	0.95	$\pm$	0.14	&	0.89	$\pm$	0.20	&	0.25	$\pm$	0.13	&	0.88	$\pm$	0.02	&	0.10	$\pm$	0.01	&	8.4	&	0.11	\\
141	&		&	3	&	0.52	$\pm$	0.06	&	0.24	$\pm$	0.15	&	1.06	$\pm$	0.21	&	0.24	$\pm$	0.14	&	0.68	$\pm$	0.02	&	0.07	$\pm$	0.01	&	37.4	&	0.14	\\
142	&		&	4	&	0.59	$\pm$	0.06	&	0.20	$\pm$	0.14	&	0.41	$\pm$	0.20	&	0.72	$\pm$	0.14	&	0.79	$\pm$	0.02	&	0.10	$\pm$	0.01	&	21.3	&	0.14	\\ \hline
143	&	MIROC-ESM	&	0	&	0.84	$\pm$	0.05	&	0.49	$\pm$	0.13	&	0.35	$\pm$	0.18	&	0.80	$\pm$	0.12	&	1.15	$\pm$	0.02	&	0.18	$\pm$	0.00	&	10.8	&	0.11	\\
144	&		&	1	&	0.28	$\pm$	0.05	&	-0.21	$\pm$	0.14	&	-0.70	$\pm$	0.19	&	0.16	$\pm$	0.13	&	1.12	$\pm$	0.02	&	0.18	$\pm$	0.01	&	45.5	&	0.15	\\
145	&		&	2	&	0.62	$\pm$	0.05	&	0.77	$\pm$	0.14	&	-0.04	$\pm$	0.19	&	0.37	$\pm$	0.13	&	1.06	$\pm$	0.02	&	0.16	$\pm$	0.01	&	12.7	&	0.13	\\ \hline
146	&	MIROC-ESM-CHEM	&	0	&	0.49	$\pm$	0.04	&	1.36	$\pm$	0.11	&	-0.52	$\pm$	0.16	&	0.72	$\pm$	0.11	&	1.05	$\pm$	0.02	&	0.16	$\pm$	0.00	&	22.1	&	0.10	\\ \hline
147	&	MPI-ESM-LR	&	0	&	0.11	$\pm$	0.06	&	0.11	$\pm$	0.15	&	0.17	$\pm$	0.22	&	0.70	$\pm$	0.15	&	1.44	$\pm$	0.02	&	0.22	$\pm$	0.01	&	76.1	&	0.18	\\
148	&		&	1	&	0.09	$\pm$	0.05	&	0.41	$\pm$	0.14	&	0.57	$\pm$	0.20	&	1.07	$\pm$	0.13	&	1.54	$\pm$	0.02	&	0.24	$\pm$	0.01	&	105.2	&	0.16	\\
149	&		&	2	&	0.41	$\pm$	0.06	&	0.88	$\pm$	0.14	&	0.92	$\pm$	0.20	&	-0.81	$\pm$	0.14	&	1.31	$\pm$	0.02	&	0.20	$\pm$	0.01	&	56.3	&	0.14	\\ \hline
150	&	MPI-ESM-MR	&	0	&	0.58	$\pm$	0.05	&	0.41	$\pm$	0.13	&	-0.42	$\pm$	0.19	&	-0.51	$\pm$	0.13	&	1.40	$\pm$	0.02	&	0.21	$\pm$	0.00	&	72.2	&	0.15	\\
151	&		&	1	&	0.10	$\pm$	0.05	&	1.13	$\pm$	0.13	&	0.86	$\pm$	0.19	&	0.77	$\pm$	0.13	&	1.40	$\pm$	0.02	&	0.21	$\pm$	0.00	&	73.1	&	0.14	\\
152	&		&	2	&	0.54	$\pm$	0.05	&	0.72	$\pm$	0.12	&	-0.02	$\pm$	0.18	&	0.60	$\pm$	0.12	&	1.39	$\pm$	0.02	&	0.20	$\pm$	0.00	&	54.3	&	0.14	\\ \hline
153	&	MPI-ESM-P	&	0	&	0.39	$\pm$	0.05	&	0.71	$\pm$	0.13	&	1.17	$\pm$	0.19	&	0.05	$\pm$	0.13	&	1.46	$\pm$	0.02	&	0.23	$\pm$	0.00	&	75.1	&	0.15	\\
154	&		&	1	&	0.33	$\pm$	0.06	&	0.67	$\pm$	0.14	&	-0.13	$\pm$	0.20	&	0.34	$\pm$	0.14	&	1.47	$\pm$	0.02	&	0.22	$\pm$	0.01	&	75.7	&	0.17	\\ \hline
155	&	MRI-CGCM3	&	0	&	0.13	$\pm$	0.05	&	0.29	$\pm$	0.12	&	0.33	$\pm$	0.17	&	0.17	$\pm$	0.12	&	0.72	$\pm$	0.02	&	0.11	$\pm$	0.00	&	63.8	&	0.12	\\
156	&		&	1	&	0.42	$\pm$	0.05	&	0.60	$\pm$	0.13	&	0.39	$\pm$	0.19	&	0.46	$\pm$	0.13	&	0.52	$\pm$	0.02	&	0.08	$\pm$	0.00	&	76.3	&	0.15	\\
157	&		&	2	&	0.52	$\pm$	0.05	&	0.36	$\pm$	0.13	&	0.16	$\pm$	0.18	&	0.45	$\pm$	0.13	&	0.70	$\pm$	0.02	&	0.11	$\pm$	0.00	&	39.7	&	0.13	\\ \hline
158	&	MRI-ESM1	&	0	&	0.54	$\pm$	0.05	&	0.56	$\pm$	0.14	&	-0.14	$\pm$	0.19	&	0.52	$\pm$	0.13	&	0.89	$\pm$	0.02	&	0.13	$\pm$	0.01	&	17.7	&	0.11	\\ \hline
159	&	NorESM1-M	&	0	&	0.72	$\pm$	0.05	&	0.43	$\pm$	0.12	&	0.00	$\pm$	0.18	&	0.39	$\pm$	0.12	&	0.82	$\pm$	0.02	&	0.13	$\pm$	0.00	&	20.2	&	0.10	\\
160	&		&	1	&	0.42	$\pm$	0.05	&	0.63	$\pm$	0.12	&	-0.02	$\pm$	0.17	&	0.80	$\pm$	0.12	&	0.87	$\pm$	0.02	&	0.13	$\pm$	0.00	&	23.4	&	0.11	\\
161	&		&	2	&	0.32	$\pm$	0.05	&	0.49	$\pm$	0.13	&	0.76	$\pm$	0.18	&	0.30	$\pm$	0.12	&	0.86	$\pm$	0.02	&	0.13	$\pm$	0.00	&	29.3	&	0.11	\\ \hline
162	&	NorESM1-ME	&	0	&	0.55	$\pm$	0.05	&	1.03	$\pm$	0.14	&	0.58	$\pm$	0.20	&	0.64	$\pm$	0.13	&	0.87	$\pm$	0.02	&	0.12	$\pm$	0.01	&	13.4	&	0.11	\\ \hline
	&	average	&		&	0.58	$\pm$	0.30	&	0.68	$\pm$	0.42	&	0.48	$\pm$	0.48	&	0.39	$\pm$	0.37	&	1.09	$\pm$	0.36	&	0.16	$\pm$	0.06	&	51.30	&	0.14	\\ \hline
\end{longtable} }

\newpage

{\tiny
\begin{longtable}{|llr|c|c|c|c|c|c|c|}
\caption{\normalsize See Sections 3.2 and 3.3.
The regression coefficients $\phi$  using Eq. \ref{eq23355} for band-pass frequency scales S1, S2, S3 and S4 for all available 162 GCM simulations and for their ensemble mean (which is numbered
-0- in the list). The  $\chi^2$ test values  were calculated
using Eq. \ref{phi}. Last two columns report the  correlation coefficients between the GST power spectrum (periodogram and MEM) and the correspondent GCM power spectra. See also Figures 13, 14 and 15.}
\\ \hline
\#	&	model	&	simulation	&	S1	&	S2	&	S3	&	S4	&	$\chi^2$	&	Corr. Coeff.	&	Corr. Coeff	\\
	&		&	\#	&	0.5-7 years	&	7-14 years	&	14-28 years	&	28-104 years	&		&	Periodogram	&	MEM	\\ \hline
0	&	ensemble mean	&		&	0.05	&	0.36	&	0.81	&	0.63	&	8.6	&	0.79	&	-0.02	\\ \hline
1	&	ACCESS1-0	&	0	&	-0.05	&	0.53	&	0.91	&	0.51	&	6.9	&	0.34	&	0.08	\\ \hline
2	&	ACCESS1-3	&	0	&	0.06	&	0.12	&	1.22	&	0.55	&	15.3	&	0.48	&	-0.07	\\
3	&		&	1	&	-0.03	&	-0.11	&	1.14	&	0.77	&	19.4	&	0.63	&	0.12	\\
4	&		&	2	&	0.03	&	-0.03	&	1.22	&	0.78	&	17.1	&	0.81	&	0.28	\\ \hline
5	&	bcc-csm1-1	&	0	&	0.03	&	0.37	&	1.21	&	1.08	&	6.7	&	0.97	&	0.36	\\
6	&		&	1	&	0.05	&	0.63	&	0.92	&	0.79	&	2.7	&	0.79	&	0.19	\\
7	&		&	2	&	0.06	&	0.62	&	1.33	&	0.70	&	5.1	&	0.93	&	0.32	\\ \hline
8	&	bcc-csm1-1-m	&	0	&	0.02	&	0.57	&	1.18	&	0.23	&	12.0	&	0.25	&	-0.11	\\
9	&		&	1	&	0.09	&	1.00	&	0.91	&	0.63	&	2.1	&	0.21	&	-0.06	\\
10	&		&	2	&	0.00	&	0.83	&	1.75	&	0.20	&	18.3	&	0.50	&	-0.08	\\ \hline
11	&	BNU-ESM	&	0	&	0.02	&	0.29	&	1.56	&	0.78	&	12.8	&	0.84	&	-0.03	\\ \hline
12	&	CanESM2	&	0	&	-0.06	&	0.24	&	1.22	&	0.75	&	10.2	&	0.79	&	0.02	\\
13	&		&	1	&	-0.02	&	0.57	&	1.12	&	1.11	&	3.1	&	0.87	&	0.09	\\
14	&		&	2	&	-0.03	&	0.60	&	0.57	&	0.87	&	5.3	&	0.79	&	0.00	\\
15	&		&	3	&	0.05	&	0.07	&	0.86	&	0.90	&	13.3	&	0.77	&	-0.04	\\
16	&		&	4	&	-0.01	&	0.55	&	0.82	&	1.11	&	3.7	&	0.89	&	0.25	\\ \hline
17	&	CCSM4	&	0	&	0.08	&	0.38	&	1.26	&	0.68	&	8.1	&	0.93	&	0.01	\\
18	&		&	1	&	-0.01	&	0.48	&	1.13	&	0.61	&	6.5	&	0.94	&	0.42	\\
19	&		&	2	&	0.10	&	0.57	&	0.67	&	0.49	&	8.2	&	0.93	&	0.11	\\
20	&		&	3	&	0.13	&	0.69	&	0.74	&	0.52	&	5.8	&	0.88	&	-0.08	\\
21	&		&	4	&	0.09	&	0.30	&	0.97	&	0.48	&	11.3	&	0.94	&	0.13	\\
22	&		&	5	&	0.03	&	0.41	&	1.50	&	0.49	&	12.8	&	0.84	&	0.04	\\ \hline
23	&	CESM1-BGC	&	0	&	0.02	&	0.72	&	0.88	&	0.51	&	4.9	&	0.96	&	-0.03	\\ \hline
24	&	CESM1-CAM5	&	0	&	0.04	&	0.47	&	0.68	&	0.25	&	14.0	&	0.81	&	0.40	\\
25	&		&	1	&	0.13	&	0.57	&	0.91	&	0.97	&	2.9	&	0.97	&	0.03	\\
26	&		&	2	&	0.03	&	0.53	&	0.58	&	1.11	&	5.9	&	0.87	&	0.06	\\ \hline
27	&	CESM1-CAM5-1-FV2	&	0	&	0.09	&	0.37	&	0.41	&	0.77	&	11.8	&	0.85	&	-0.11	\\
28	&		&	1	&	0.05	&	0.28	&	0.57	&	1.01	&	10.5	&	0.97	&	-0.02	\\
29	&		&	2	&	0.09	&	-0.06	&	0.26	&	0.72	&	26.0	&	0.98	&	0.45	\\
30	&		&	3	&	0.00	&	0.34	&	0.69	&	0.59	&	10.3	&	0.64	&	0.09	\\ \hline
31	&	CESM1-FASTCHEM	&	0	&	0.05	&	0.53	&	1.30	&	0.56	&	7.5	&	0.87	&	0.06	\\
32	&		&	1	&	0.06	&	0.63	&	0.89	&	0.71	&	3.4	&	0.93	&	0.01	\\
33	&		&	2	&	0.03	&	0.76	&	1.24	&	0.44	&	6.3	&	0.91	&	0.01	\\ \hline
34	&	CESM1-WACCM	&	0	&	-0.01	&	0.63	&	1.43	&	0.50	&	8.4	&	0.90	&	0.29	\\ \hline
35	&	CMCC-CESM	&	0	&	-0.04	&	0.03	&	-0.16	&	0.41	&	38.9	&	0.71	&	0.13	\\ \hline
36	&	CMCC-CM	&	0	&	0.04	&	0.38	&	-0.53	&	0.40	&	45.9	&	0.34	&	0.03	\\ \hline
37	&	CMCC-CMS	&	0	&	-0.05	&	-0.19	&	0.11	&	0.36	&	38.8	&	0.01	&	-0.04	\\ \hline
38	&	CNRM-CM5	&	0	&	0.02	&	0.62	&	1.02	&	0.45	&	6.7	&	0.15	&	0.01	\\
39	&		&	1	&	-0.04	&	0.62	&	0.98	&	0.70	&	3.4	&	0.88	&	-0.01	\\
40	&		&	2	&	0.04	&	0.41	&	1.07	&	0.78	&	5.9	&	0.69	&	-0.01	\\
41	&		&	3	&	0.08	&	0.60	&	0.96	&	0.99	&	2.4	&	0.87	&	0.44	\\
42	&		&	4	&	0.09	&	0.64	&	1.18	&	0.53	&	5.7	&	0.78	&	0.06	\\
43	&		&	5	&	0.11	&	0.52	&	0.97	&	0.64	&	5.4	&	0.46	&	0.13	\\
44	&		&	6	&	0.04	&	0.54	&	1.03	&	1.04	&	3.2	&	0.98	&	0.63	\\
45	&		&	7	&	0.00	&	0.29	&	0.97	&	0.54	&	10.6	&	0.25	&	-0.10	\\
46	&		&	8	&	0.07	&	0.55	&	0.30	&	0.63	&	12.2	&	0.94	&	0.19	\\
47	&		&	9	&	0.03	&	0.30	&	0.92	&	1.10	&	7.6	&	0.88	&	0.46	\\ \hline
48	&	CSIRO-Mk3-6-0	&	0	&	0.04	&	0.70	&	1.07	&	0.80	&	2.0	&	0.22	&	-0.12	\\
49	&		&	1	&	0.05	&	0.26	&	0.66	&	1.15	&	10.2	&	0.57	&	-0.10	\\
50	&		&	2	&	0.08	&	0.35	&	0.66	&	1.10	&	8.1	&	0.90	&	0.46	\\
51	&		&	3	&	-0.02	&	0.32	&	1.48	&	1.10	&	10.4	&	0.57	&	-0.06	\\
52	&		&	4	&	-0.03	&	0.14	&	0.96	&	1.22	&	11.8	&	0.42	&	-0.10	\\
53	&		&	5	&	0.07	&	0.11	&	0.75	&	0.85	&	13.1	&	0.34	&	-0.09	\\
54	&		&	6	&	0.08	&	0.26	&	0.43	&	0.54	&	16.0	&	0.50	&	-0.03	\\
55	&		&	7	&	0.04	&	0.60	&	0.61	&	1.17	&	5.0	&	0.66	&	-0.07	\\
56	&		&	8	&	0.04	&	0.15	&	0.07	&	0.48	&	27.5	&	0.17	&	0.01	\\
57	&		&	9	&	-0.06	&	0.31	&	0.28	&	0.62	&	16.9	&	0.16	&	-0.08	\\ \hline
58	&	EC-EARTH	&	0	&	0.03	&	0.34	&	0.91	&	0.45	&	11.1	&	0.76	&	-0.05	\\
59	&		&	1	&	0.07	&	0.47	&	0.98	&	0.02	&	18.5	&	0.68	&	0.23	\\
60	&		&	2	&	0.01	&	0.05	&	0.89	&	0.53	&	16.7	&	0.87	&	0.08	\\
61	&		&	3	&	0.04	&	0.34	&	0.90	&	0.13	&	17.8	&	0.77	&	-0.08	\\
62	&		&	4	&	0.10	&	0.59	&	0.61	&	0.34	&	11.3	&	0.81	&	0.11	\\
63	&		&	5	&	0.01	&	0.06	&	1.01	&	0.81	&	13.8	&	0.94	&	0.01	\\
64	&		&	6	&	0.05	&	0.21	&	0.84	&	0.48	&	13.6	&	0.83	&	-0.04	\\
65	&		&	7	&	0.05	&	0.57	&	0.60	&	0.63	&	7.2	&	0.81	&	0.14	\\
66	&		&	8	&	0.01	&	0.75	&	1.30	&	0.57	&	5.1	&	0.89	&	-0.04	\\ \hline
67	&	FGOALS-g2	&	0	&	0.00	&	0.06	&	0.19	&	0.21	&	32.0	&	0.48	&	-0.07	\\
68	&		&	1	&	-0.03	&	-0.04	&	0.34	&	0.56	&	25.2	&	0.90	&	0.74	\\
69	&		&	2	&	-0.01	&	-0.05	&	-0.06	&	0.46	&	37.2	&	0.78	&	0.86	\\
70	&		&	3	&	-0.01	&	0.20	&	0.24	&	0.33	&	24.7	&	0.83	&	-0.03	\\
71	&		&	4	&	-0.02	&	0.03	&	0.26	&	0.35	&	28.4	&	0.65	&	0.44	\\ \hline
72	&	FIO-ESM	&	0	&	0.03	&	0.04	&	0.26	&	0.19	&	31.4	&	0.68	&	-0.12	\\
73	&		&	1	&	0.01	&	0.31	&	0.35	&	0.17	&	23.6	&	0.44	&	-0.07	\\
74	&		&	2	&	0.05	&	0.12	&	-0.07	&	0.21	&	37.7	&	0.53	&	0.01	\\ \hline
75	&	GFDL-CM3	&	0	&	-0.03	&	0.33	&	1.06	&	1.63	&	12.5	&	0.87	&	-0.06	\\
76	&		&	1	&	0.02	&	0.70	&	0.82	&	1.27	&	2.9	&	0.98	&	0.11	\\
77	&		&	2	&	-0.03	&	0.62	&	0.52	&	1.00	&	5.5	&	0.55	&	-0.12	\\
78	&		&	3	&	0.10	&	0.44	&	0.94	&	1.28	&	5.7	&	0.58	&	0.02	\\
79	&		&	4	&	0.07	&	0.32	&	0.83	&	1.41	&	9.8	&	0.76	&	0.14	\\ \hline
80	&	GFDL-ESM2G	&	0	&	0.05	&	0.09	&	1.26	&	0.81	&	13.8	&	0.63	&	0.18	\\
81	&		&	1	&	-0.02	&	0.16	&	1.19	&	0.31	&	18.0	&	-0.02	&	-0.02	\\
82	&		&	2	&	0.07	&	0.25	&	0.86	&	0.39	&	14.2	&	0.11	&	-0.03	\\ \hline
83	&	GFDL-ESM2M	&	0	&	0.15	&	0.35	&	0.79	&	0.29	&	14.3	&	0.29	&	-0.11	\\ \hline
84	&	GISS-E2-H p1	&	0	&	0.01	&	0.37	&	0.70	&	0.41	&	12.3	&	0.33	&	0.06	\\
85	&		&	1	&	0.05	&	0.49	&	0.82	&	0.36	&	10.2	&	0.47	&	0.28	\\
86	&		&	2	&	0.04	&	0.52	&	0.62	&	0.41	&	10.7	&	0.37	&	-0.08	\\
87	&		&	3	&	0.10	&	0.17	&	0.66	&	0.42	&	16.9	&	0.72	&	0.16	\\
88	&		&	4	&	0.05	&	0.53	&	0.77	&	0.42	&	9.2	&	0.83	&	-0.06	\\
89	&		&	5	&	0.07	&	0.47	&	0.95	&	0.45	&	8.8	&	0.69	&	0.00	\\ \hline
90	&	GISS-E2-H p2	&	0	&	0.03	&	0.45	&	0.88	&	0.57	&	7.5	&	0.45	&	-0.11	\\
91	&		&	1	&	0.06	&	0.57	&	0.74	&	0.37	&	9.6	&	0.06	&	-0.14	\\
92	&		&	2	&	0.04	&	0.39	&	1.04	&	0.57	&	8.3	&	0.50	&	0.07	\\
93	&		&	3	&	0.04	&	0.36	&	1.09	&	0.57	&	9.0	&	0.29	&	-0.13	\\
94	&		&	4	&	0.02	&	0.53	&	0.80	&	0.54	&	7.1	&	0.53	&	-0.03	\\ \hline
95	&	GISS-E2-H p3	&	0	&	0.03	&	0.49	&	0.91	&	0.56	&	6.9	&	0.57	&	-0.07	\\
96	&		&	1	&	0.06	&	0.53	&	0.82	&	0.43	&	8.6	&	0.54	&	0.07	\\
97	&		&	2	&	0.04	&	0.38	&	0.85	&	0.55	&	9.1	&	0.35	&	0.37	\\
98	&		&	3	&	0.04	&	0.51	&	0.84	&	0.72	&	5.1	&	0.81	&	0.41	\\
99	&		&	4	&	0.05	&	0.40	&	1.10	&	0.47	&	9.8	&	0.50	&	-0.07	\\
100	&		&	5	&	0.06	&	0.40	&	1.03	&	0.63	&	7.3	&	0.48	&	-0.06	\\ \hline
101	&	GISS-E2-H-CC p1	&	0	&	0.05	&	0.45	&	0.98	&	0.42	&	9.4	&	0.66	&	0.22	\\ \hline
102	&	GISS-E2-R p1	&	0	&	0.06	&	0.25	&	0.41	&	0.24	&	22.1	&	0.35	&	-0.07	\\
103	&		&	1	&	0.03	&	0.41	&	0.90	&	0.31	&	12.3	&	0.44	&	-0.07	\\
104	&		&	2	&	0.03	&	0.29	&	0.83	&	0.45	&	12.4	&	0.69	&	0.16	\\
105	&		&	3	&	0.03	&	0.39	&	0.81	&	0.52	&	9.5	&	0.72	&	-0.05	\\
106	&		&	4	&	0.03	&	0.43	&	0.95	&	0.23	&	13.7	&	0.90	&	-0.07	\\
107	&		&	5	&	0.06	&	0.38	&	0.85	&	0.45	&	10.5	&	0.37	&	-0.08	\\ \hline
108	&	GISS-E2-R p2	&	0	&	0.04	&	0.36	&	0.79	&	0.53	&	10.0	&	0.44	&	-0.09	\\
109	&		&	1	&	0.01	&	0.23	&	0.10	&	0.62	&	23.1	&	0.45	&	0.25	\\
110	&		&	2	&	0.03	&	0.62	&	0.49	&	0.28	&	13.8	&	0.05	&	-0.14	\\
111	&		&	3	&	0.06	&	0.54	&	0.88	&	0.74	&	4.3	&	0.52	&	0.13	\\
112	&		&	4	&	-0.01	&	0.25	&	0.84	&	0.35	&	15.0	&	0.25	&	-0.13	\\
113	&		&	5	&	-0.03	&	0.43	&	0.80	&	0.41	&	10.5	&	0.59	&	0.05	\\ \hline
114	&	GISS-E2-R p3	&	0	&	0.05	&	0.32	&	0.97	&	0.40	&	12.2	&	0.26	&	0.01	\\
115	&		&	1	&	0.03	&	0.52	&	0.88	&	0.24	&	12.2	&	0.62	&	0.09	\\
116	&		&	2	&	0.01	&	0.69	&	1.08	&	0.56	&	4.4	&	0.91	&	-0.06	\\
117	&		&	3	&	0.09	&	0.41	&	1.15	&	0.49	&	9.5	&	0.60	&	0.46	\\
118	&		&	4	&	0.12	&	0.21	&	0.94	&	0.25	&	17.7	&	-0.20	&	-0.17	\\
119	&		&	5	&	0.05	&	0.60	&	1.06	&	0.26	&	10.6	&	0.67	&	0.01	\\ \hline
120	&	GISS-E2-R-CC p1	&	0	&	0.07	&	0.45	&	0.71	&	0.31	&	12.7	&	0.32	&	-0.07	\\ \hline
121	&	HadGEM2-AO	&	0	&	0.07	&	0.36	&	1.27	&	1.32	&	8.6	&	0.80	&	0.40	\\ \hline
122	&	HadGEM2-CC	&	0	&	0.00	&	0.15	&	0.45	&	1.01	&	15.3	&	0.80	&	0.29	\\ \hline
123	&	HadGEM2-ES	&	0	&	0.12	&	0.24	&	0.54	&	1.02	&	11.7	&	0.75	&	-0.11	\\
124	&		&	1	&	0.04	&	0.30	&	1.08	&	0.99	&	7.3	&	0.40	&	-0.11	\\
125	&		&	2	&	0.02	&	0.13	&	1.06	&	1.13	&	11.6	&	0.67	&	-0.04	\\
126	&		&	3	&	0.10	&	-0.10	&	0.95	&	0.58	&	20.7	&	0.28	&	-0.11	\\ \hline
127	&	inmcm4	&	0	&	-0.05	&	0.01	&	0.26	&	0.34	&	29.0	&	0.95	&	0.63	\\ \hline
128	&	IPSL-CM5A-LR	&	0	&	0.00	&	0.48	&	0.33	&	0.97	&	10.7	&	0.78	&	0.02	\\
129	&		&	1	&	0.09	&	0.30	&	1.13	&	0.96	&	7.5	&	0.84	&	0.49	\\
130	&		&	2	&	0.02	&	0.65	&	1.10	&	1.07	&	2.0	&	0.91	&	0.67	\\
131	&		&	3	&	0.05	&	0.64	&	0.83	&	0.47	&	6.6	&	0.75	&	0.08	\\
132	&		&	4	&	0.04	&	0.45	&	0.64	&	0.63	&	8.4	&	0.68	&	0.02	\\
133	&		&	5	&	0.00	&	0.15	&	1.30	&	0.73	&	13.0	&	0.84	&	0.28	\\ \hline
134	&	IPSL-CM5A-MR	&	0	&	0.00	&	0.58	&	1.07	&	0.59	&	5.3	&	0.74	&	-0.06	\\
135	&		&	1	&	0.04	&	0.33	&	1.08	&	0.71	&	7.9	&	0.66	&	0.05	\\
136	&		&	2	&	0.00	&	0.04	&	0.65	&	0.82	&	15.9	&	0.77	&	0.65	\\ \hline
137	&	IPSL-CM5B-LR	&	0	&	0.04	&	0.48	&	0.21	&	0.15	&	23.8	&	0.67	&	-0.11	\\ \hline
138	&	MIROC5	&	0	&	0.10	&	0.59	&	-0.05	&	1.03	&	18.7	&	0.71	&	0.50	\\
139	&		&	1	&	-0.03	&	0.37	&	0.81	&	0.73	&	7.5	&	0.52	&	-0.06	\\
140	&		&	2	&	-0.03	&	0.18	&	0.81	&	0.94	&	10.5	&	0.76	&	-0.01	\\
141	&		&	3	&	0.08	&	0.44	&	0.27	&	0.58	&	15.2	&	0.49	&	-0.08	\\
142	&		&	4	&	0.02	&	0.67	&	0.58	&	0.79	&	4.9	&	0.35	&	-0.13	\\ \hline
143	&	MIROC-ESM	&	0	&	0.02	&	0.63	&	0.72	&	0.92	&	3.3	&	0.82	&	0.60	\\
144	&		&	1	&	0.02	&	-0.07	&	0.35	&	0.55	&	26.4	&	0.28	&	-0.12	\\
145	&		&	2	&	0.03	&	0.21	&	1.10	&	0.78	&	10.2	&	0.51	&	-0.08	\\ \hline
146	&	MIROC-ESM-CHEM	&	0	&	0.02	&	0.36	&	1.39	&	0.52	&	11.7	&	0.65	&	0.04	\\ \hline
147	&	MPI-ESM-LR	&	0	&	0.03	&	0.58	&	0.96	&	-0.05	&	19.0	&	0.84	&	-0.03	\\
148	&		&	1	&	0.09	&	0.30	&	0.81	&	0.21	&	17.1	&	0.67	&	0.07	\\
149	&		&	2	&	0.04	&	0.42	&	0.95	&	0.35	&	11.3	&	0.81	&	-0.10	\\ \hline
150	&	MPI-ESM-MR	&	0	&	0.05	&	-0.01	&	0.77	&	0.62	&	18.2	&	0.85	&	0.02	\\
151	&		&	1	&	0.12	&	0.82	&	0.90	&	0.14	&	11.6	&	0.05	&	-0.10	\\
152	&		&	2	&	0.01	&	0.34	&	0.51	&	0.53	&	13.3	&	0.75	&	0.03	\\ \hline
153	&	MPI-ESM-P	&	0	&	0.04	&	0.23	&	0.97	&	0.47	&	13.1	&	0.62	&	0.45	\\
154	&		&	1	&	0.10	&	0.38	&	0.87	&	0.22	&	14.9	&	0.64	&	-0.12	\\ \hline
155	&	MRI-CGCM3	&	0	&	0.00	&	0.15	&	0.34	&	0.20	&	26.6	&	0.35	&	0.38	\\
156	&		&	1	&	-0.04	&	0.08	&	0.54	&	0.55	&	18.7	&	0.50	&	-0.04	\\
157	&		&	2	&	0.03	&	0.19	&	0.49	&	0.65	&	15.3	&	0.39	&	-0.13	\\ \hline
158	&	MRI-ESM1	&	0	&	0.05	&	0.12	&	0.82	&	0.62	&	14.1	&	0.57	&	-0.08	\\ \hline
159	&	NorESM1-M	&	0	&	0.04	&	0.40	&	0.44	&	0.68	&	11.4	&	0.90	&	-0.01	\\
160	&		&	1	&	-0.04	&	0.40	&	0.67	&	0.43	&	11.9	&	0.76	&	-0.01	\\
161	&		&	2	&	0.00	&	0.14	&	0.59	&	0.40	&	18.9	&	0.58	&	-0.02	\\ \hline
162	&	NorESM1-ME	&	0	&	0.09	&	0.69	&	0.93	&	0.55	&	4.5	&	0.71	&	-0.05	\\ \hline
	&	average	&		&	0.03 $\pm$0.04	&	0.38 $\pm$0.22	&	0.80	$\pm$0.36&	0.61	$\pm$0.30&	12.6	$\pm$ 8.0&	0.63$\pm$0.25	&	0.08$\pm$	0.21\\ \hline
\end{longtable} }

\end{document}